\documentclass[twocolumn,showpacs,superscriptaddress,floatfix,prx]{revtex4-2}
\usepackage{graphicx,amsfonts,amssymb,amsmath,hyperref,hypcap,enumerate}
\usepackage{braket}
\usepackage{lipsum} 
\usepackage{amsthm}
\usepackage{multirow}
\usepackage{graphicx}
\usepackage{dcolumn}   
\usepackage{bm}        
\usepackage{amssymb}   
\usepackage{amsmath}
\usepackage{braket}
\usepackage{epstopdf}
\usepackage{color}
\usepackage{subfigure}
\usepackage{diagbox}

\usepackage{multirow}
\usepackage{makecell}
\usepackage{bbold}
\usepackage{bbding}

\DeclareMathAlphabet{\mathitb}{OT1}{cmr}{bx}{sl}
\usepackage[utf8]{inputenc}

\newcommand{\mg}[1]{\langle g_{#1} \rangle}
\newcommand{\Tr}{{\rm Tr}}
\newcommand{\h}{\mathcal{H}}

\begin{document}

\title{Anisotropic Topological Anderson Transitions in Chiral Symmetry Classes}

\author{Zhenyu Xiao}
\affiliation{International Center for Quantum Materials, Peking University, Beijing 100871, China}

\author{Kohei Kawabata}
\affiliation{Department of Physics, Princeton University, Princeton, New Jersey 08544, USA}
\affiliation{Institute for Solid State Physics, University of Tokyo, Kashiwa, Chiba 277-8581, Japan}

\author{Xunlong Luo}
\affiliation{Science and Technology on Surface Physics and Chemistry Laboratory, Mianyang 621907, China}

\author{Tomi Ohtsuki}
\affiliation{Physics Division, Sophia University, Chiyoda-ku, Tokyo 102-8554, Japan}

\author{Ryuichi Shindou}
\email{rshindou@pku.edu.cn}
\affiliation{International Center for Quantum Materials, Peking University, Beijing 100871, China}

\date{\today}

\begin{abstract}
We study quantum phase transitions of three-dimensional disordered systems in the chiral classes (AIII and BDI) with and without weak topological indices. 
We show that the systems with a nontrivial weak topological index universally exhibit an emergent thermodynamic phase where wave functions are delocalized along one spatial direction but exponentially localized in the other two spatial directions, which we call the quasi-localized phase.
Our extensive numerical study clarifies that the critical exponent of the Anderson transition between the metallic and quasi-localized phases, as well as that between the quasi-localized and localized phases, are different from that with no weak topological index, signaling the new universality classes induced by topology. 
The quasi-localized phase and concomitant topological Anderson transition manifest themselves in the anisotropic transport phenomena of disordered weak topological insulators and nodal-line semimetals, which exhibit the metallic behavior in one direction but the insulating behavior in the other directions.
\end{abstract}

\maketitle

\textit{Introduction}---The last decades have seen remarkable discoveries of topological materials~\cite{Hasan10, qi2011,chiu16}.
The interplay of disorder and topology leads to new types of quantum phase transitions, including the quantum Hall plateau transitions~\cite{klitzing80, chalker88, pruisken88, huckestein95, bhaseen00, Slevin09, prodan10, zhu19, puschmann19, dresselhaus22}. 
The universality classes of the disorder-driven metal-insulator transitions, known as the Anderson transitions, are characterized by the critical exponents and scaling functions, which are commonly believed to be determined solely by symmetry and spatial dimensions~\cite{Evers08}.
Many theories investigated whether topology can change the universality classes of the Anderson transitions~\cite{Asada02, Asada05, onoda07, Obuse07, ryu07, nomura07, mirlin10, Konig2012, Ringel12, Fu12, Slevin16, Roy17, Luo18QMCT, Yoshioka18, song21, son21, pan21, Wang21b, luo21unifying, supplemental}. 
Still, the role of topology in the Anderson transitions has been elusive.

Prime examples of three-dimensional (3D) topological materials include nodal-line semimetals characterized by the weak topological invariant~\cite{Burkov-11, Schnyder2015, fang2016, Armitage18}.
Several recent experiments realized nodal-line semimetals in solid states~\cite{Bian16, Schoop16, Chen22}, as well as synthetic materials of ultracold atoms~\cite{Bo2019} and photonic~\cite{Gao2018, xia2019} and phononic~\cite{Deng2019} systems.
Despite the significant interest in the physics of nodal-line semimetals~\cite{nandkishore2016, sur2016, syzranov2017, goncalves2020, Luo20}, their unique transport signatures have remained largely unexplored.

\begin{figure}[b]
    \centering
	\subfigure[topological models]{
		\begin{minipage}[t]{0.5\linewidth}
			\centering
			\includegraphics[width=1\linewidth]{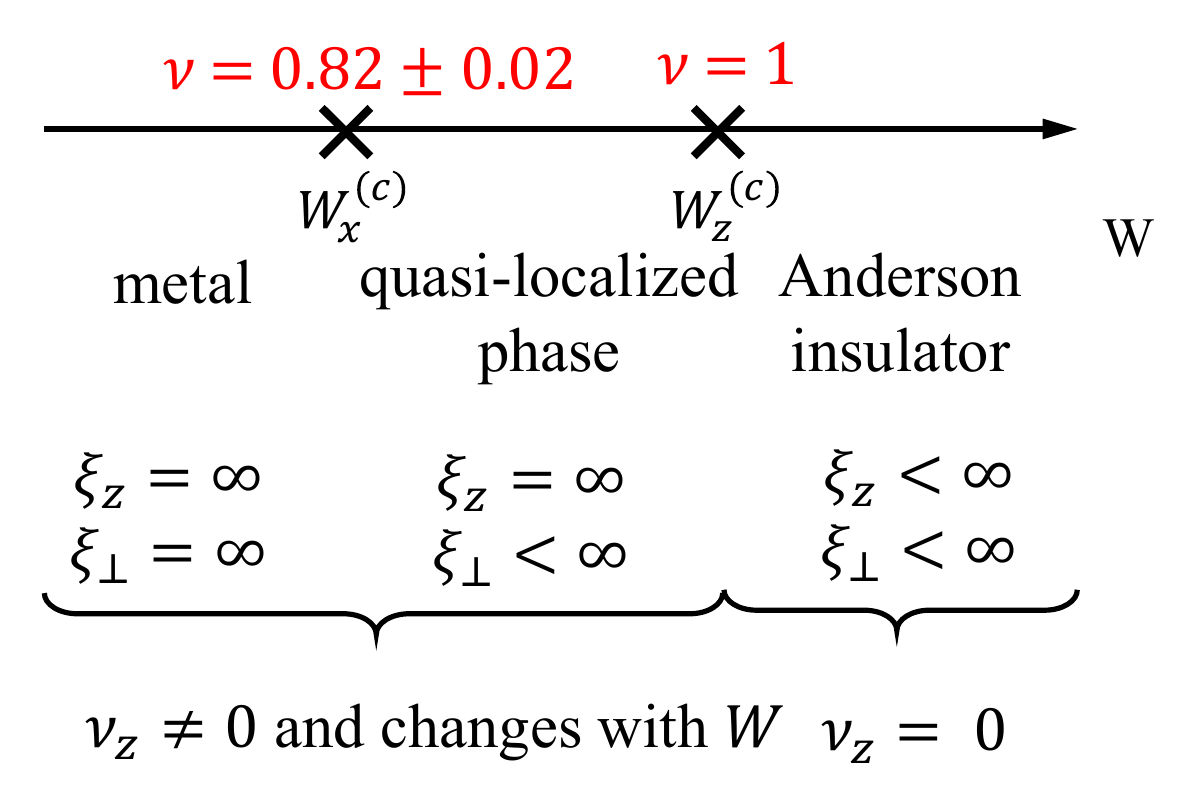}
        \label{t_pd}
		\end{minipage}%
	}%
    \subfigure[nontopological models]{
		\begin{minipage}[t]{0.5\linewidth}
			\centering
			\includegraphics[width=1\linewidth]{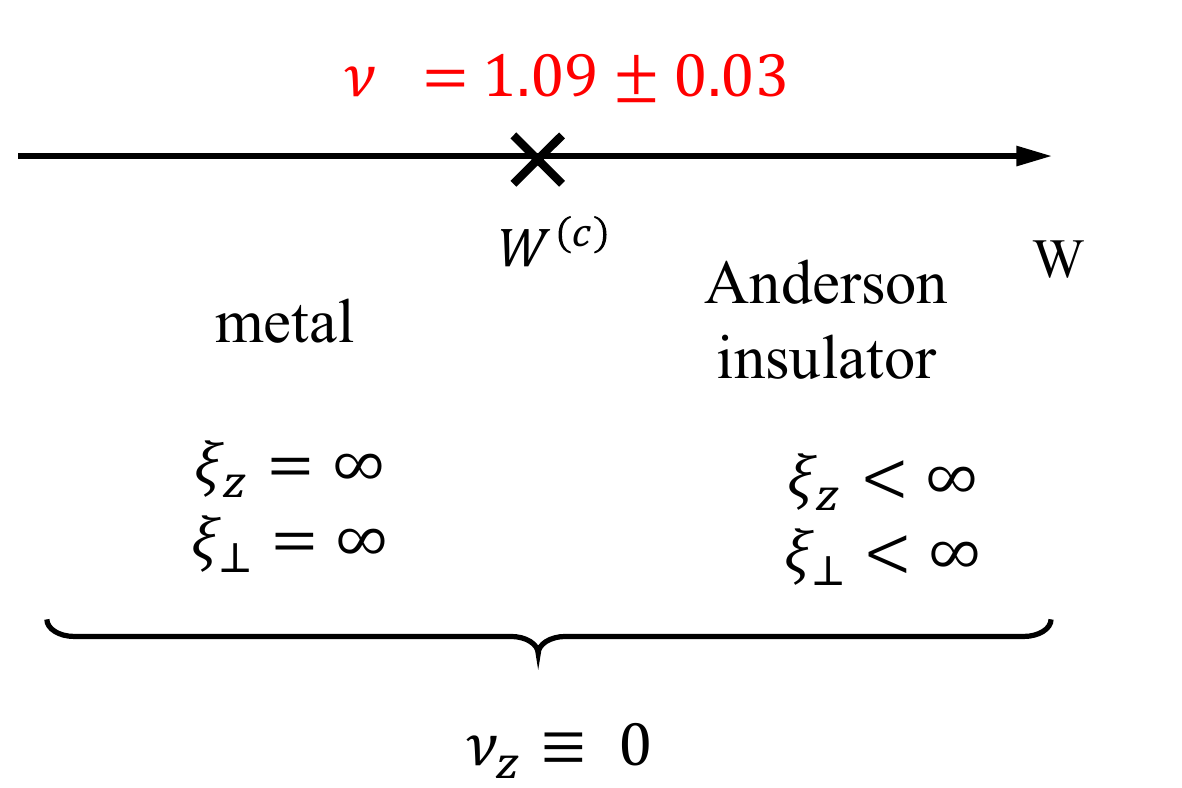}
		\label{nt_pd}
		\end{minipage}%
	}%

	\caption{Phase diagrams of 3D disordered Hamiltonians in the chiral symmetry classes~(a) with and (b) without the weak topological index $\nu_z$. 
    The critical exponents $\nu$ and localization lengths $\xi_{z}, \xi_{\perp} \!\ (\perp = x,y)$ along different directions
    are shown for different phases.
    The nontrivial critical exponents $\nu = 0.82 \pm 0.02$ and $\nu = 1.09 \pm 0.03$ are obtained for class BDI.
    }
    \label{pd}
\end{figure} 

In this Letter, we elucidate that the weak topological indices induce a novel thermodynamic phase in 3D disordered systems, including topological nodal-line semimetals, in the chiral classes.
There, 3D wave functions are delocalized along one spatial direction and exponentially localized along the other two spatial directions---quasi-localized phase [Fig.~\ref{pd}(a)].   
From extensive numerical calculations, we evaluate correlation-length critical exponents of the Anderson transitions among the metallic, quasi-localized, and localized phases [Fig.~\ref{pd}(a)] and find that they are distinct from the critical exponent in topologically trivial systems [Fig.~\ref{pd}(b)], signaling new universality classes induced by the topological indices. 
Notably, our quasi-localized phase and concomitant topological Anderson transition are of direct experimental relevance in the anisotropic transport that exhibits the metallic behavior in one direction but the insulating behavior in the other directions. 
While such anisotropic transport has played an important role in condensed matter physics~\cite{cohen1974, lilly1999, du1999, ando2002, borzi2007, hinkov2008, chu2010, fradkin2010}, our results provide its new universal mechanism induced by the interplay of disorder and topology.


\textit{Lyapunov exponents and topological indices}---We study disorder-induced quantum phase transitions of 3D chiral-symmetric Hamiltonians $\mathcal{H}$.
The localization properties along the $\mu$ direction ($\mu = x, y, z$) are efficiently captured
by the Lyapunov exponents (LEs) along the $\mu$ direction in the limit $L \rightarrow \infty$, which are eigenvalues of~\cite{MacKinnon81,MacKinnon83}
\begin{equation}
    \lim_{L_{\mu} \rightarrow \infty} \log {(M^{\dagger}M)^{\frac{1}{2L_{\mu}}}}.
\end{equation}
Here, $M \equiv M_{L_{\mu}} M_{L_{\mu} - 1} \cdots M_1$ is the product of transfer 
matrices along the $\mu$ direction.
The smallest positive LE gives the inverse of 
the localization length along the $\mu$ direction~\cite{Slevin14}. 
In the limit $L \to \infty$, the LEs of ${\cal H}$ form several continuous spectra~\cite{markos1995}. 
If the spectra do not include zero, the wave function is localized along the $\mu$ direction.
By contrast, if the spectra include zero,
the localization length diverges, 
which means the delocalization of the wave function. 
The finite (infinite) localization length leads to the vanishing 
(nonvanishing) conductance in the same direction, as shown in the Supplemental Material~\cite{supplemental}.

Symmetries of Hamiltonians give constraints on the spectrum of 
the LEs. For example, because of Hermiticity of ${\cal H}$, 
the LEs come in opposite-sign pairs.
Moreover, in the presence of chiral symmetry, ${\cal H}$ can be brought into the block off-diagonal structure,
\begin{align}
{\cal H}=
\begin{pmatrix}
0& h\\
h^{\dag}&0\\
\end{pmatrix},
    \label{eq: chiral basis}
\end{align}
where the off-diagonal part $h$ is assumed to be a square matrix.
Because of chiral symmetry, the LEs of ${\cal H}$ reduce to the LEs of $h$ and $h^{\dag}$, 
which come in opposite-sign pairs, as shown in the Supplemental Material~\cite{supplemental}.
Consequently, we only need to calculate the product of the transfer matrices of $h$.

We demonstrate that a weak topological index $\nu_{\mu}$ imposes another constraint on the spectrum of the LEs and plays a vital role in the emergence of the quasi-localized phase in disordered chiral-symmetric systems. 
To introduce $\nu_{\mu}$ along the $\mu$ direction in the presence of disorder, let us insert a magnetic flux $\phi_{\mu}$ through a closed loop along the $\mu$ direction.
Then, the weak topological index $\nu_{\mu}$ is given by the winding of $\det h \left( \phi_{\mu} \right)$ in Eq.~(\ref{eq: chiral basis}) under an adiabatic insertion of a unit flux~\cite{Mondragon-Shem14, Altland14, claes21weak}:
\begin{align}
    \nu_{\mu} 
    \equiv \frac{\rm i}{L^2}\int^{2\pi}_{0}  \frac{ d\phi_{\mu}}{2\pi } 
    \partial_{\phi_{\mu}} {\rm Tr}\big[\log\big[h(\phi_{\mu})]\big] \, , 
\label{topological} 
\end{align}
where $L^2$ is the system size within the two directions 
perpendicular
to the $\mu$ direction. 
Here, $\nu_{\mu}$ is not necessarily quantized and takes an arbitrary real number.
Notably, the weak topological index $\nu_{\mu}$ and LEs of $h$ are related to each other by~\cite{supplemental,molinari2003} 
\begin{align}
    \nu_{\mu} 
    =\frac{1}{2L^2} (N_{+,\mu}-N_{-,\mu})\, ,
    \label{winding_LE-main}
\end{align}
where $N_{+,\mu}$ and $N_{-,\mu}$ are the numbers of positive and negative LEs of 
$h$ along the $\mu$ direction, respectively. 

Suppose ${\cal H}$ has a mobility gap around $E = 0$ and its zero-energy state is characterized by 
the weak topological indices $\nu_x = \nu_y = 0$, $\nu_z = 1$.
From Eq.~(\ref{winding_LE-main}), a finite gap exists between 
the smallest positive LE and the largest negative LE 
such that $N_{+,z} - N_{-,z} = 2L^2$. 
By contrast, when disorder is strong enough, the zero-energy state is in a topologically-trivial localized phase with $N_{+,z} = N_{-,z}$. 
Between the two localized phases, $L^2$ 
positive LEs of $h$ cross zero, and 
$\nu_z$ continuously changes from $1$ to $0$ with 
respect to the disorder strength, where  
the localization length $\xi_z$ along the $z$ direction always diverges. 
Within this finite range with divergent $\xi_z$, the zero-energy state undergoes the Anderson transitions along the $x$ and $y$ directions, and thus a quasi-localized phase with divergent $\xi_z$ and finite $\xi_{x}$ and $\xi_y$ emerges. 
Below, we clarify its nature, obtain the critical exponents of the Anderson transitions among the metallic, quasi-localized, and localized phases, and demonstrate the existence of new universality classes. 

\textit{Model}---As a prototypical example, we study a two-orbital tight-binding model on a 3D cubic lattice~\cite{Luo20}
\begin{align}
{\cal H}&= \sum_{\bm r=(r_x,r_y,r_z)} \left\{ \epsilon_{\bm{r}} c^{\dagger}_{\bm{r}} \sigma_z c_{\bm{r}} +  \left[ \sum_{\mu = x,y} \left(  t_{\perp}c^{\dagger}_{\bm r+\bm{e_{\mu}}} \sigma_z c_{\bm{r}}  \right) \right. \right. \nonumber \\
 & \left. \left. -{\rm i} t_{\|}   c^{\dagger}_{\bm r+\bm{ e_z}}  \sigma_y c_{\bm{r}} + t_{\|}^{\prime}   c^{\dagger}_{\bm r+\bm{ e_z}}  \sigma_z c_{\bm{r}} +\text{H.c.} \right]  \right\}.
    \label{NDSM_H}
\end{align}
Here, $c_{\bm{r}}$ is a two-component annihilation operator at the cubic lattice site 
${\bm r}$, $\sigma_{\mu}$ $(\mu = x, y,z)$ are Pauli matrices,
$t_{\perp}$, $t_{\|}$, $t_{\|}^{\prime}$ are real-valued parameters, 
and $\epsilon_{\bm{r}}$ is a random potential 
that distributes uniformly in $[-W/2,W/2]$. 
We assume $t_{\perp}, t_{\|} >0 $ for simplicity. 
This Hamiltonian respects time-reversal symmetry ${\cal H} = {\cal H}^*$ and 
chiral symmetry ${\cal H} = -\sigma_x{\cal H} \sigma_x$, 
and hence belongs to class BDI~\cite{Altland97, Evers08, chiu16}.
In addition, the ensemble of Hamiltonians
is statistically invariant 
under the combination of time reversal and 
reflection with respect to the $xy$ plane, 
which requires $N_{+,x}=N_{-,x}$, $N_{+,y}=N_{-,y}$ and 
$\nu_x = \nu_y = 0$, as shown in the Supplemental Material~\cite{supplemental}, while $\nu_z$ can be nonzero. 
In the clean limit, the Hamiltonian has an energy gap around $E=0$ with 
$\nu_z=1$ for $4 t_{\perp} < 2|t_{\|}^{\prime}|$.  
For $4 t_{\perp} > 2|t_{\|}^{\prime}|$, 
by contrast, the zero-energy state forms a nodal line in 
momentum space, resulting in $0 < \nu_z < 1$. 
In the following, we focus on 
the nodal-line-semimetal phase for  
$t_{\|} = t_{\|}^{\prime} = 1/2$, $t_{\perp} = 1$ 
and study the Anderson transitions of the zero modes 
along all the directions. 
Still, we stress that the weak topological invariant $\nu_\mu$, rather than a nodal line itself, is the main ingredient for the quasi-localized phase.


\begin{figure}[bt]
    \centering
    \includegraphics[width=1.0\linewidth]{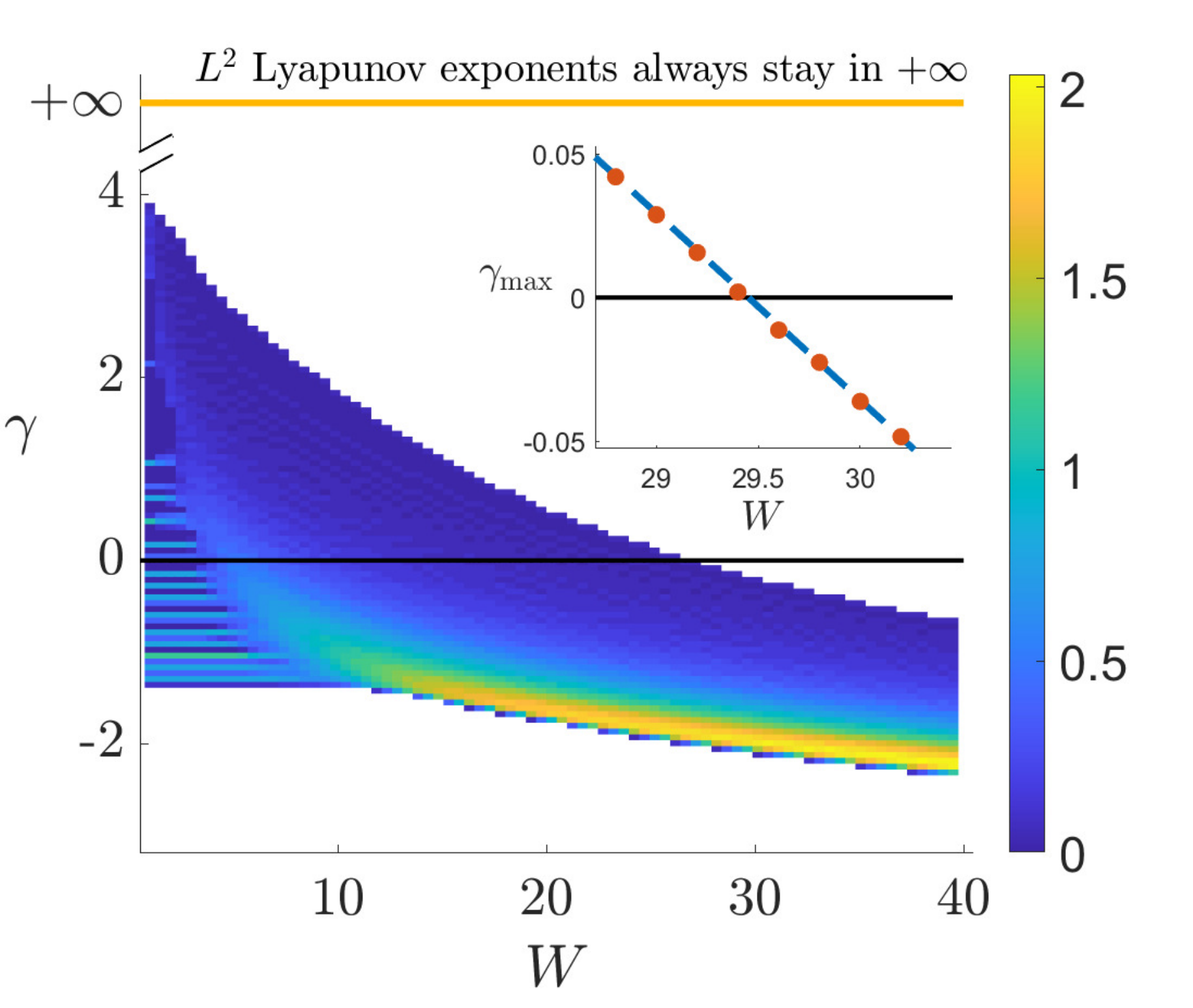}
    \caption{$2L^2$ Lyapunov exponents (LEs) of the right-upper part $h$ of the 3D nodal-line-semimetal model ${\cal H}$ along the $z$ direction with the quasi-1D geometry $L\times L \times L_z$ ($L = 18$, $L_z = 2\times 10^6$), plotted as a function of the disorder strength $W$.
    The color scale stands for the density $\rho(\gamma)$ of the LEs with the normalization  $\int  \rho(\gamma) d \gamma = 1$.
    The LEs of ${\cal H}$ are composed of the LEs of $h$ and $h^{\dag}$.
    Inset: the largest LE $\gamma_{\rm max}(W,L)$ among the smaller $L^2$ LEs as a function of $W$ in the limit $L\rightarrow \infty$, 
    obtained by a finite-size scaling fit. 
    The error bars are smaller than the marks.  
    The plot crosses zero linearly at $W^{(z)}_{c}=29.45 \pm 0.05$; 
    $\xi_z \sim (W - W_c^{(z)})^{-\nu^{\prime}}$ with $\nu^{\prime} = 1$ for $W>W_c^{(z)}$.}
\label{LE_z_one_way}
\end{figure}

\textit{Localization length $\xi_z$}---Figure~\ref{LE_z_one_way} shows the  distribution of 
LEs $\gamma$ of 
$h$ in Eq.~(\ref{eq: chiral basis}) 
for the nodal-line-semimetal model ${\cal H}$ in Eq.~(\ref{NDSM_H})
along the $z$ direction 
in the quasi-1D geometry $L\times L \times L_z$. 
The distribution consists of  
two separate spectra, each of which contains $L^2$ LEs. 
The upper spectrum
is always $\gamma=+\infty$~\cite{supplemental}
and irrelevant to the Anderson transitions. 
For $W \leq W_c^{(z)} \approx 29$, the lower spectrum includes zero $\gamma=0$.  
Every positive LE in the lower spectrum 
for $W<W_c^{(z)}$ crosses zero when we increase $W$. 
At each crossing point, $N_{-,z}$ changes by one.  
For $L \rightarrow \infty$, the crossing points become dense and  
$\nu_z = 1 - N_{-,z}/L^2$ changes continuously with $W$. 
For $W > W_c^{(z)}$, all the LEs in the lower spectrum are negative (i.e., $N_{-,z}=L^2$), and 
the system is in a localized phase with no weak topological index 
$\nu_z = 0 $. At $W = W_c^{(z)}$, the maximal LE in the lower spectrum crosses zero. 
Notably, $W_c^{(z)}$ for $L\rightarrow \infty$ cannot be determined by fitting
$\xi_z/L$ with a standard scaling function [e.g., see Eq.~(\ref{fit1-main})] because 
$\xi_z$ with finite $L$ diverges at 
some $W<W_c^{(z)}$. Instead, we map the non-Hermitian matrix $h$ into 
a well-localized Hermitian matrix by a similarity 
transformation~\cite{supplemental, Hatano96},
where the localization length 
obeys a scaling form in the strong disorder limit~\cite{Asada04}. 
Then, we obtain the scaling form of the largest LE $\gamma_{\rm max}(W,L)$, 
\begin{equation}
    \gamma_{\rm max}(W,L) = a/L +\gamma_{\rm max}(W,L = \infty)  \, .
    \label{LE_z_scaling}
\end{equation}
We numerically verify this scaling and determine the critical disorder strength $W^{(z)}_c=29.45\pm 0.05$ (inset of Fig.~\ref{LE_z_one_way}).

\begin{figure}[bt]
    \centering
    \includegraphics[width=1\linewidth]{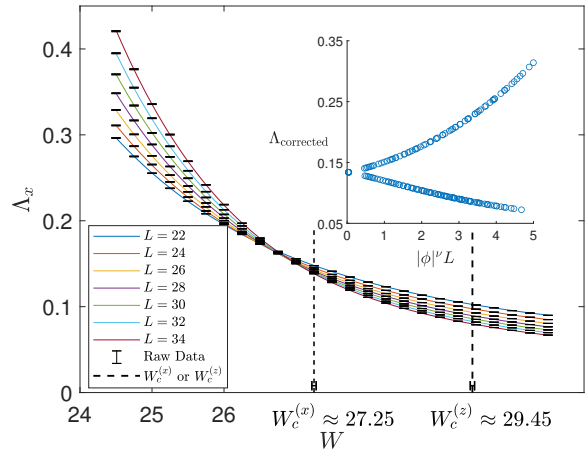}
    \caption{Normalized localization length $\Lambda_x \equiv \xi_x/L$ along 
    the $x$ direction
    as a function of 
    the disorder strength $W$ in the nodal-line-semimetal model in Eq.~(\ref{NDSM_H}) with the quasi-1D geometry $L\times L\times L_x$.  
    The black points are the raw data with the error bars. 
    The solid lines for different $L$ and the dashed vertical line $W_c^{(x)}$ with the error bars are the results of the fitting according to Eq.~(\ref{fit1-main}) with $n = 3$.
    The dashed line $W_c^{(z)}$ is evaluated by the fitting of the Lyapunov exponent along the 
    $z$ direction by Eq.~(\ref{LE_z_scaling}). 
    Inset: single-parameter scaling function of $\Lambda_x$. $\Lambda_\text{corrected}$ is $\Lambda_x$ subtracted by a contribution of the irrelevant scaling variable $c$ in Eq.~(\ref{fit1-main}), and
    $\phi$ 
    is the relevant scaling variable.
    }
    \label{Lambda_x_one_way}
\end{figure}

\textit{Localization length $\xi_x$, $\xi_y$}---The statistical symmetries mentioned above 
require LEs of $h$ along the $x$ and $y$ directions to 
come in opposite-sign pairs.
Thus, the localization length $\xi_x$ along the $x$ direction 
is always finite in the quasi-1D geometry with finite $L$. 
As shown in Fig.~\ref{Lambda_x_one_way}, the normalized localization length 
$\Lambda_x(W,L) \equiv \xi_x(W,L)/L$ shows 
scale-invariant behavior at a certain 
disorder strength $W^{(x)}_{c}$ well below $W=W^{(z)}_{c}$, indicating a quantum phase transition at 
$W = W^{(x)}_c <W^{(z)}_{c}$. To determine 
$W^{(x)}_c$ and the critical exponent $\nu$, we 
use a finite-size scaling function and its polynomial 
expansion~\cite{Slevin99,Slevin14}. The scaling function for 
$\Lambda_x(W,L)$ is Taylor-expanded with respect to the relevant 
scaling variable $\phi(w)$ and the least irrelevant scaling 
variable $c$ up to the $n$th order and first order, respectively, 
\begin{align}
\Lambda_x(W,L) = \sum^n_{i=0} \sum^1_{j=0} a_{i,j} 
\big(\phi(w)L^{1/\nu}\big)^i \big(c L^{-y}\big)^j, \label{fit1-main}
\end{align}
with $w \equiv (W-W^{(x)}_c)/W^{(x)}_c$ and the scaling dimension $-y$ $(<0)$ of the least irrelevant scaling variable around a saddle-point fixed point. 
The fitting is carried out by the $\chi^2$ fitting method, and  
the confidence error bars for the optimal parameters are determined 
by the Monte Carlo method, as detailed in the Supplemental Material~\cite{supplemental}.

The first row in Table~\ref{fitting_table} shows the fitting results, where 
$W^{(x)}_{c}=27.24 \pm 0.05$ is significantly smaller than $W^{(z)}_{c}=29.45 \pm 0.05$ 
and the critical exponent at $W^{(x)}_{c}$ is evaluated as $\nu = 0.82 \pm 0.02$.
The two different critical disorder strengths illustrate the emergence 
of the three distinct phases as a function of the disorder strength $W$ [Fig.~\ref{t_pd}].
For $W<W^{(x)}_c$, the localization lengths diverge along all directions (metallic phase). For $W>W^{(z)}_c$, the localization lengths are finite along all directions (Anderson insulator phase). For $W^{(x)}_{c}<W<W^{(z)}_c$, the localization lengths are finite along the $x$ and $y$ directions but diverge along the $z$ direction (quasi-localized phase), 
and $\nu_z$ continuously changes as $W$ changes. 
Our extensive numerical calculations show that the quasi-localized phase 
with divergent $\xi_z$ but finite $\xi_x,\xi_y$ universally appears between metallic and localized phases in different 
models with nonzero $\nu_z$, as shown in the Supplemental Material~\cite{supplemental}.
The consistent critical exponent at $W=W^{(x)}_{c}$ was 
also obtained in Ref.~\cite{Luo20}, while 
a different critical exponent was obtained in
Ref.~\cite{Wang21}
even in the same class. 
In this Letter, we elucidate that this 
difference originates from the emergence of the quasi-localized phase, which was not identified previously.


\begin{table}[tb]
    \centering
    \caption{Critical disorder strength $W^{(\mu)}_c$ and critical exponent $\nu$ for the 3D chiral classes, 
    obtained by the polynomial fitting of the normalized localization length $\Lambda_{\mu} \equiv \xi_{\mu}/L$ along the $\mu$ direction ($\mu = x, y, z$) around critical points of different models with the quasi-one-dimensional geometry $L\times L\times L_{\mu}$. 
    In the column ``Topo'', ``$\surd$'' shows the nonzero weak topological index $\nu_{z}$ 
    around the critical point,
    and ``$\times$'' shows zero topological indices in all the directions.
    The square brackets denote the 95\% confidence interval.
    }
    \begin{tabular}{c|c|c
    |cc}
    \hline \hline
    Class & Topo & ~$\mu$~ 
    &$W^{(\mu)}_c$& $\nu$  \\ \hline
    BDI & $\surd$ & $x$ & 27.241[27.194,27.303] & 0.820[0.783,0.846] \\
    AIII  & $\surd$ & $x$ & 9.143[9.125,9.168] & 0.824[0.776,0.862] \\
    \hline 
    BDI & $\times$ & $z$ & 23.220[23.167,23.293] & 1.089[1.005,1.128] \\
    BDI & $\times$ & $x$ & 23.170[23.098,23.279] & 1.042[0.943,1.099] \\
    AIII& $\times$ & $z$  & 8.091[8.074,8.096] & 1.024[0.973,1.070] \\
    \hline
    \end{tabular}
    \label{fitting_table}
\end{table}

{\it Quasi-localized phase}---Now, we clarify the nature of the quasi-localized phase induced by the weak topological index $\nu_{\mu}$.  
Let $\Phi( {\bm r}) = \langle {\bm r}| \Phi \rangle$ be a normalized wave function. 
The wave function interacts with an effective disorder potential $V_{\rm eff} = \langle \Phi  |V| \Phi \rangle = \sum_{\bm r} V({\bm r}) |\Phi( {\bm r})|^2$, 
whose strength is given by $\langle V_{\rm eff}^2 \rangle =  W^2  P_2$ with the inverse participation ratio $P_2 \equiv \sum_{\bm r} |\Phi( {\bm r})|^4$.
Here, $\langle ... \rangle $ denotes the disorder average: 
$\langle V({\bm r}) V({\bm r}^{\prime}) \rangle = W^2 
\delta_{{\bm r},{\bm r}^{\prime}}$. 
As long as $W$ is finite, the following argument is applicable to general $V(\bm r)$, including the box disorder in $[-W/2,W/2]$ used for the numerical calculations.
Let us introduce the integrated weight of the wave function in the $z$th layer by $|\phi(z)|^2  = \sum_{x,y} |\Phi({\bm r})|^2$ and also the one-dimensional inverse 
participation ratio $P_2^z \equiv \sum_z |\phi(z)|^4$.    
$P_2^x, P_2^y$ can be defined in the same manner.  
$P_2^{\mu}$ measures the localization property of $\Phi({\bm r})$
along the $\mu$ direction, giving an upper bound of $P_2$: 
$P_2 \leq P_2^{\mu} \, (\mu = x,y,z)$~\cite{supplemental}.
If 
the wave function is extended along the $z$ direction 
(i.e., $P_{2}^{z}\sim L_{z}^{-1}$~\cite{Evers08}), 
$P_2$ and $\braket{V_{\rm eff}^2}$
should vanish for $L_z \rightarrow \infty$, and 
$\Phi({\bm r})$ must be extended along all the directions. 
If $P_2^z$ is finite even for $L_z \rightarrow \infty$, by contrast, $P_2^x$ and $P_2^y$ should also be finite for $L_x,L_y \rightarrow \infty$. 
Otherwise, $\Phi({\bm r})$ is extended within all the directions, which contradicts finite $P_2^z$. 
In the intermediate phase
discussed above,
we find that
$\xi_x$ is finite but $\xi_z$ diverges. 
While finite $\xi_x$ means finite $P_2$ and $P_2^z$, 
divergent $\xi_z$ with finite $P_2^z$ means 
that the wave function $\Phi({\bm r})$ must be quasi-localized 
along the $z$ direction. 
Thus, the wave function in the intermediate phase is localized within the 
$xy$ plane and delocalized only along the $z$ direction---quasi-localized phase. 
Here, $\Phi({\bm r})$ along the $z$ direction shares the same 
localization properties as 
wave functions of 1D chiral-symmetric 
systems at a topological phase transition, 
where the 1D topological index
changes~\cite{Evers08, Mondragon-Shem14, Altland14, Leon97, Mathur97, Brouwer98}.
The 3D system in the intermediate phase is effectively decoupled 
into 1D wires because of finite $\xi_{x,y}$. 

The emergence of the quasi-localized phase 
in 3D systems 
is a consequence of finite $P_2^{\rm 1D}$ at the 
topological phase transition of 1D chiral-symmetric systems. 
Generally, when a $d'$-dimensional wave function $\Phi({\bm R})$ 
in ${\bm R} \equiv ({\bm r},{\bm s})$ with ${\bm r}=(r_1,\cdots,r_d)$ and ${\bm s}=(s_1,\cdots, 
s_{d'-d})$ ($d<d'$) is made out of coupled $d$-dimensional wave functions $\psi({\bm r})$ 
at a critical point, 
$\Phi({\bm R})$ is more extended than 
$\psi({\bm r})$ 
along the ${\bm r}$ direction because of the 
interlayer 
coupling~\cite{supplemental}. 
Thus, 
the effective disorder strength for the $d'$-dimensional wave function 
$\Phi({\bm R})$ is bounded by the $d$-dimensional inverse participation 
ratio $P^{\psi(\bm r)}_2$ of $\psi(\bm r)$.
When the wave function $\psi({\bm r})$ has finite $P^{\psi({\bm r})}_2$ at the critical point, the effective disorder strength can be finite, and $\Phi({\bm R})$ can be 
either extended or localized within the ${\bm s}$ direction.
On the other hand, 
when $P^{\psi({\bm r})}_2$ is zero
at the critical point, e.g., 
2D critical wave functions at the quantum 
Hall plateau transition, 
the effective disorder strength  
is zero, and 
the 
$d'$-dimensional 
wave function should be always extended in 
both ${\bm r}$ and ${\bm s}$ directions. 
Notably, the 1D topological phase transitions in all the three chiral classes are characterized by finite $P_2$~\cite{Evers08}. 
In the following, we demonstrate the quasi-localized phases also in the 3D chiral unitary class, which is consistent with the above argument. 

{\it Model without time-reversal symmetry---}We add a 
time-reversal-breaking
but chiral-symmetric disorder $\Delta {\cal H}$ to the model ${\cal H}$ in Eq.~(\ref{NDSM_H}):
\begin{align}
  {\cal H}_1 = {\cal H} + \Delta {\cal H},\quad
  \Delta {\cal H} = \sum_{\bm r}  \epsilon^{\prime}_{\bm{r}}  c^{\dagger}_{\bm{r}} \sigma_y c_{\bm{r}},
  \label{TRS_breaking}
\end{align}
with the random potentials $\epsilon^{\prime}_{\bm{r}}$, where $(\epsilon_{\bm{r}},\epsilon^{\prime}_{\bm{r}}) 
= (V_{\bm r}\cos\theta_{\bm r},V_{\bm r}\sin\theta_{\bm r})$, and $\theta_{\bm r}$ and $V_{\bm r}$  
distribute uniformly in the range of $[0,2\pi)$ and $[0,W]$, respectively.
This model only respects chiral symmetry and belongs to class AIII, in which the weak topological indices are defined in the same manner. 
It shows a similar phase diagram as in the previous model in class BDI
with
$W^{(z)}_{c}=9.8 \pm 0.1$ and $W^{(x)}_{c}=9.14 \pm 0.01$ (see Fig.~\ref{pd} and 
Table~\ref{fitting_table}).
The critical exponents are the same as those in the models in class BDI,
which suggests possible super-universality in 3D systems in the chiral 
classes with the topological indices.


{\it Models with trivial topological indices---}To further clarify the role of the topological indices,
we also study a topologically trivial
model in class BDI with statistical symmetries. 
The statistical symmetry of time reversal combined with reflection 
with respect to the $xz$ or $yz$ plane makes all three topological 
indices vanish, as shown in the Supplemental Material~\cite{supplemental}. In addition, LEs of $h$ along any direction 
come in opposite-sign pairs, and the localization lengths along 
the $x$ and $y$ directions are the same. 
On increasing the disorder strength, the model undergoes the Anderson transition, where  
the normalized localization lengths $\Lambda_x$ and 
$\Lambda_z$ along the $x$ and $z$ directions both show scale-invariant behaviors. 
The critical disorder strengths and critical exponents determined from $\Lambda_x$ 
and $\Lambda_z$ 
are consistent with each other (see Table~\ref{fitting_table}), 
which suggests that the scale-invariant behavior of $\Lambda_x$ and $\Lambda_z$ 
comes from the same quantum phase transition [Fig.~\ref{pd}(b)]. 
The evaluated critical exponent $\nu = 1.089[1.005,1.128]$ is different from $\nu$ at $W=W^{(x)}_{c}$ of the topological model, and consistent with $\nu$ of the topologically trivial models in Ref.~\cite{Wang21}.
We also evaluate the critical exponent in 
the chiral unitary class without weak topological 
indices as $\nu = 1.024[0.973,1.070]$ (Table~\ref{fitting_table}), 
which is different from  $\nu$ of the topological models in the same
symmetry class and consistent with 
Refs.~\cite{Wang21,luo21unifying}. 

\textit{Summary and discussion}---In this Letter, we show that in 3D systems in the chiral classes, the weak topological indices induce a  
disorder-driven quasi-localized phase where wave functions are delocalized only along one direction and localized along the other two directions. 
The critical exponents of the Anderson transitions among metal, quasi-localized, and localized phases are all different (Fig.~\ref{pd}). 
We believe that these conclusions hold 
also in the chiral symplectic class (class CII). 
Our quasi-localized phase leads to the anisotropic transport phenomena of topological nodal-line semimetals~\cite{Schnyder2015, fang2016, Armitage18, Bian16, Schoop16, Chen22, Bo2019, Gao2018, xia2019, Deng2019}, 
where the conductance along the direction with the divergent localization length takes finite values with larger fluctuations, while it vanishes along the other directions in the thermodynamic limit, as shown in the Supplemental Material~\cite{supplemental}.
The quasi-localized phase may potentially find practical applications such as quantum devices that control the direction of currents.

Our results are also relevant to non-Hermitian physics~\cite{Konotop16, Feng17, El-Ganainy18}, where the interplay between disorder and dissipation has recently acquired renewed interest. 
In fact, all the disorder-driven phases and phase transitions in this Letter are characterized by the LEs of the off-diagonal part $h$ in Eq.~(\ref{eq: chiral basis}), which can be considered as a non-Hermitian Hamiltonian~\cite{luo21unifying}.
Anisotropy of ${\cal H}$ corresponds to nonreciprocity of $h$ and leads to transport phenomena unique to open systems.

3D chiral-symmetric systems also host a strong topological index~\cite{Hasan10, qi2011, chiu16}. By a similar numerical study, 
we find that the strong index does not lead to the quasi-localized phases, not influencing the universality classes of the Anderson transitions~\cite{unpublished}.  It also remains to be explored whether the quasi-localized phase appears and whether the topological indices change the universality classes of the Anderson transitions in 2D systems, as well as nodal-line semimetals protected by spatial symmetry.



\textit{Acknowledgement}---
Z.X. thanks Zhida Song, Lingxian Kong, and Yeyang Zhang for fruitful discussions.
Z.X. and R.S. were supported by the National Basic Research Programs of China (No.~2019YFA0308401) and by National Natural Science Foundation of China (No.~11674011 and No.~12074008). 
K.K. was supported by JSPS Overseas Research Fellowship, and Grant No.~GBMF8685 from the Gordon and Betty Moore Foundation toward the Princeton theory program. 
X.L. was supported by National Natural Science Foundation of China of Grant No.~12105253. 
T.O. was supported by JSPS KAKENHI Grant~19H00658 and 22H05114.

\bibliography{paper}

\begin{thebibliography}{85}%
\makeatletter
\providecommand \@ifxundefined [1]{%
 \@ifx{#1\undefined}
}%
\providecommand \@ifnum [1]{%
 \ifnum #1\expandafter \@firstoftwo
 \else \expandafter \@secondoftwo
 \fi
}%
\providecommand \@ifx [1]{%
 \ifx #1\expandafter \@firstoftwo
 \else \expandafter \@secondoftwo
 \fi
}%
\providecommand \natexlab [1]{#1}%
\providecommand \enquote  [1]{``#1''}%
\providecommand \bibnamefont  [1]{#1}%
\providecommand \bibfnamefont [1]{#1}%
\providecommand \citenamefont [1]{#1}%
\providecommand \href@noop [0]{\@secondoftwo}%
\providecommand \href [0]{\begingroup \@sanitize@url \@href}%
\providecommand \@href[1]{\@@startlink{#1}\@@href}%
\providecommand \@@href[1]{\endgroup#1\@@endlink}%
\providecommand \@sanitize@url [0]{\catcode `\\12\catcode `\$12\catcode
  `\&12\catcode `\#12\catcode `\^12\catcode `\_12\catcode `\%12\relax}%
\providecommand \@@startlink[1]{}%
\providecommand \@@endlink[0]{}%
\providecommand \url  [0]{\begingroup\@sanitize@url \@url }%
\providecommand \@url [1]{\endgroup\@href {#1}{\urlprefix }}%
\providecommand \urlprefix  [0]{URL }%
\providecommand \Eprint [0]{\href }%
\providecommand \doibase [0]{https://doi.org/}%
\providecommand \selectlanguage [0]{\@gobble}%
\providecommand \bibinfo  [0]{\@secondoftwo}%
\providecommand \bibfield  [0]{\@secondoftwo}%
\providecommand \translation [1]{[#1]}%
\providecommand \BibitemOpen [0]{}%
\providecommand \bibitemStop [0]{}%
\providecommand \bibitemNoStop [0]{.\EOS\space}%
\providecommand \EOS [0]{\spacefactor3000\relax}%
\providecommand \BibitemShut  [1]{\csname bibitem#1\endcsname}%
\let\auto@bib@innerbib\@empty
\bibitem [{\citenamefont {Hasan}\ and\ \citenamefont {Kane}(2010)}]{Hasan10}%
  \BibitemOpen
  \bibfield  {author} {\bibinfo {author} {\bibfnamefont {M.~Z.}\ \bibnamefont
  {Hasan}}\ and\ \bibinfo {author} {\bibfnamefont {C.~L.}\ \bibnamefont
  {Kane}},\ }\bibfield  {title} {\bibinfo {title} {{Colloquium: Topological
  insulators}},\ }\href {https://doi.org/10.1103/RevModPhys.82.3045} {\bibfield
   {journal} {\bibinfo  {journal} {Rev. Mod. Phys.}\ }\textbf {\bibinfo
  {volume} {82}},\ \bibinfo {pages} {3045} (\bibinfo {year}
  {2010})}\BibitemShut {NoStop}%
\bibitem [{\citenamefont {Qi}\ and\ \citenamefont {Zhang}(2011)}]{qi2011}%
  \BibitemOpen
  \bibfield  {author} {\bibinfo {author} {\bibfnamefont {X.-L.}\ \bibnamefont
  {Qi}}\ and\ \bibinfo {author} {\bibfnamefont {S.-C.}\ \bibnamefont {Zhang}},\
  }\bibfield  {title} {\bibinfo {title} {{Topological insulators and
  superconductors}},\ }\href {https://doi.org/10.1103/RevModPhys.83.1057}
  {\bibfield  {journal} {\bibinfo  {journal} {Rev. Mod. Phys.}\ }\textbf
  {\bibinfo {volume} {83}},\ \bibinfo {pages} {1057} (\bibinfo {year}
  {2011})}\BibitemShut {NoStop}%
\bibitem [{\citenamefont {Chiu}\ \emph {et~al.}(2016)\citenamefont {Chiu},
  \citenamefont {Teo}, \citenamefont {Schnyder},\ and\ \citenamefont
  {Ryu}}]{chiu16}%
  \BibitemOpen
  \bibfield  {author} {\bibinfo {author} {\bibfnamefont {C.-K.}\ \bibnamefont
  {Chiu}}, \bibinfo {author} {\bibfnamefont {J.~C.~Y.}\ \bibnamefont {Teo}},
  \bibinfo {author} {\bibfnamefont {A.~P.}\ \bibnamefont {Schnyder}},\ and\
  \bibinfo {author} {\bibfnamefont {S.}~\bibnamefont {Ryu}},\ }\bibfield
  {title} {\bibinfo {title} {{Classification of topological quantum matter with
  symmetries}},\ }\href {https://doi.org/10.1103/RevModPhys.88.035005}
  {\bibfield  {journal} {\bibinfo  {journal} {Rev. Mod. Phys.}\ }\textbf
  {\bibinfo {volume} {88}},\ \bibinfo {pages} {035005} (\bibinfo {year}
  {2016})}\BibitemShut {NoStop}%
\bibitem [{\citenamefont {Klitzing}\ \emph {et~al.}(1980)\citenamefont
  {Klitzing}, \citenamefont {Dorda},\ and\ \citenamefont
  {Pepper}}]{klitzing80}%
  \BibitemOpen
  \bibfield  {author} {\bibinfo {author} {\bibfnamefont {K.~v.}\ \bibnamefont
  {Klitzing}}, \bibinfo {author} {\bibfnamefont {G.}~\bibnamefont {Dorda}},\
  and\ \bibinfo {author} {\bibfnamefont {M.}~\bibnamefont {Pepper}},\
  }\bibfield  {title} {\bibinfo {title} {{New Method for High-Accuracy
  Determination of the Fine-Structure Constant Based on Quantized Hall
  Resistance}},\ }\href {https://doi.org/10.1103/PhysRevLett.45.494} {\bibfield
   {journal} {\bibinfo  {journal} {Phys. Rev. Lett.}\ }\textbf {\bibinfo
  {volume} {45}},\ \bibinfo {pages} {494} (\bibinfo {year} {1980})}\BibitemShut
  {NoStop}%
\bibitem [{\citenamefont {Chalker}\ and\ \citenamefont
  {Coddington}(1988)}]{chalker88}%
  \BibitemOpen
  \bibfield  {author} {\bibinfo {author} {\bibfnamefont {J.}~\bibnamefont
  {Chalker}}\ and\ \bibinfo {author} {\bibfnamefont {P.}~\bibnamefont
  {Coddington}},\ }\bibfield  {title} {\bibinfo {title} {{Percolation, quantum
  tunnelling and the integer Hall effect}},\ }\href
  {https://doi.org/10.1088/0022-3719/21/14/008} {\bibfield  {journal} {\bibinfo
   {journal} {J. Phys. C}\ }\textbf {\bibinfo {volume} {21}},\ \bibinfo {pages}
  {2665} (\bibinfo {year} {1988})}\BibitemShut {NoStop}%
\bibitem [{\citenamefont {Pruisken}(1988)}]{pruisken88}%
  \BibitemOpen
  \bibfield  {author} {\bibinfo {author} {\bibfnamefont {A.~M.~M.}\
  \bibnamefont {Pruisken}},\ }\bibfield  {title} {\bibinfo {title} {{Universal
  Singularities in the Integral Quantum Hall Effect}},\ }\href
  {https://doi.org/10.1103/PhysRevLett.61.1297} {\bibfield  {journal} {\bibinfo
   {journal} {Phys. Rev. Lett.}\ }\textbf {\bibinfo {volume} {61}},\ \bibinfo
  {pages} {1297} (\bibinfo {year} {1988})}\BibitemShut {NoStop}%
\bibitem [{\citenamefont {Huckestein}(1995)}]{huckestein95}%
  \BibitemOpen
  \bibfield  {author} {\bibinfo {author} {\bibfnamefont {B.}~\bibnamefont
  {Huckestein}},\ }\bibfield  {title} {\bibinfo {title} {{Scaling theory of the
  integer quantum Hall effect}},\ }\href
  {https://doi.org/10.1103/RevModPhys.67.357} {\bibfield  {journal} {\bibinfo
  {journal} {Rev. Mod. Phys.}\ }\textbf {\bibinfo {volume} {67}},\ \bibinfo
  {pages} {357} (\bibinfo {year} {1995})}\BibitemShut {NoStop}%
\bibitem [{\citenamefont {Bhaseen}\ \emph {et~al.}(2000)\citenamefont
  {Bhaseen}, \citenamefont {Kogan}, \citenamefont {Soloviev}, \citenamefont
  {Taniguchi},\ and\ \citenamefont {Tsvelik}}]{bhaseen00}%
  \BibitemOpen
  \bibfield  {author} {\bibinfo {author} {\bibfnamefont {M.~J.}\ \bibnamefont
  {Bhaseen}}, \bibinfo {author} {\bibfnamefont {I.~I.}\ \bibnamefont {Kogan}},
  \bibinfo {author} {\bibfnamefont {O.~A.}\ \bibnamefont {Soloviev}}, \bibinfo
  {author} {\bibfnamefont {N.}~\bibnamefont {Taniguchi}},\ and\ \bibinfo
  {author} {\bibfnamefont {A.~M.}\ \bibnamefont {Tsvelik}},\ }\bibfield
  {title} {\bibinfo {title} {{Towards a field theory of the plateau transitions
  in the integer quantum Hall effect}},\ }\href
  {https://doi.org/https://doi.org/10.1016/S0550-3213(00)00276-5} {\bibfield
  {journal} {\bibinfo  {journal} {Nucl. Phys. B}\ }\textbf {\bibinfo {volume}
  {580}},\ \bibinfo {pages} {688} (\bibinfo {year} {2000})}\BibitemShut
  {NoStop}%
\bibitem [{\citenamefont {Slevin}\ and\ \citenamefont
  {Ohtsuki}(2009)}]{Slevin09}%
  \BibitemOpen
  \bibfield  {author} {\bibinfo {author} {\bibfnamefont {K.}~\bibnamefont
  {Slevin}}\ and\ \bibinfo {author} {\bibfnamefont {T.}~\bibnamefont
  {Ohtsuki}},\ }\bibfield  {title} {\bibinfo {title} {{Critical exponent for
  the quantum Hall transition}},\ }\href
  {https://doi.org/10.1103/PhysRevB.80.041304} {\bibfield  {journal} {\bibinfo
  {journal} {Phys. Rev. B}\ }\textbf {\bibinfo {volume} {80}},\ \bibinfo
  {pages} {041304} (\bibinfo {year} {2009})}\BibitemShut {NoStop}%
\bibitem [{\citenamefont {Prodan}\ \emph {et~al.}(2010)\citenamefont {Prodan},
  \citenamefont {Hughes},\ and\ \citenamefont {Bernevig}}]{prodan10}%
  \BibitemOpen
  \bibfield  {author} {\bibinfo {author} {\bibfnamefont {E.}~\bibnamefont
  {Prodan}}, \bibinfo {author} {\bibfnamefont {T.~L.}\ \bibnamefont {Hughes}},\
  and\ \bibinfo {author} {\bibfnamefont {B.~A.}\ \bibnamefont {Bernevig}},\
  }\bibfield  {title} {\bibinfo {title} {{Entanglement Spectrum of a Disordered
  Topological Chern Insulator}},\ }\href
  {https://doi.org/10.1103/PhysRevLett.105.115501} {\bibfield  {journal}
  {\bibinfo  {journal} {Phys. Rev. Lett.}\ }\textbf {\bibinfo {volume} {105}},\
  \bibinfo {pages} {115501} (\bibinfo {year} {2010})}\BibitemShut {NoStop}%
\bibitem [{\citenamefont {Zhu}\ \emph {et~al.}(2019)\citenamefont {Zhu},
  \citenamefont {Wu}, \citenamefont {Bhatt},\ and\ \citenamefont
  {Wan}}]{zhu19}%
  \BibitemOpen
  \bibfield  {author} {\bibinfo {author} {\bibfnamefont {Q.}~\bibnamefont
  {Zhu}}, \bibinfo {author} {\bibfnamefont {P.}~\bibnamefont {Wu}}, \bibinfo
  {author} {\bibfnamefont {R.~N.}\ \bibnamefont {Bhatt}},\ and\ \bibinfo
  {author} {\bibfnamefont {X.}~\bibnamefont {Wan}},\ }\bibfield  {title}
  {\bibinfo {title} {{Localization-length exponent in two models of quantum
  Hall plateau transitions}},\ }\href
  {https://doi.org/10.1103/PhysRevB.99.024205} {\bibfield  {journal} {\bibinfo
  {journal} {Phys. Rev. B}\ }\textbf {\bibinfo {volume} {99}},\ \bibinfo
  {pages} {024205} (\bibinfo {year} {2019})}\BibitemShut {NoStop}%
\bibitem [{\citenamefont {Puschmann}\ \emph {et~al.}(2019)\citenamefont
  {Puschmann}, \citenamefont {Cain}, \citenamefont {Schreiber},\ and\
  \citenamefont {Vojta}}]{puschmann19}%
  \BibitemOpen
  \bibfield  {author} {\bibinfo {author} {\bibfnamefont {M.}~\bibnamefont
  {Puschmann}}, \bibinfo {author} {\bibfnamefont {P.}~\bibnamefont {Cain}},
  \bibinfo {author} {\bibfnamefont {M.}~\bibnamefont {Schreiber}},\ and\
  \bibinfo {author} {\bibfnamefont {T.}~\bibnamefont {Vojta}},\ }\bibfield
  {title} {\bibinfo {title} {{Integer quantum Hall transition on a
  tight-binding lattice}},\ }\href {https://doi.org/10.1103/PhysRevB.99.121301}
  {\bibfield  {journal} {\bibinfo  {journal} {Phys. Rev. B}\ }\textbf {\bibinfo
  {volume} {99}},\ \bibinfo {pages} {121301} (\bibinfo {year}
  {2019})}\BibitemShut {NoStop}%
\bibitem [{\citenamefont {Dresselhaus}\ \emph {et~al.}(2022)\citenamefont
  {Dresselhaus}, \citenamefont {Sbierski},\ and\ \citenamefont
  {Gruzberg}}]{dresselhaus22}%
  \BibitemOpen
  \bibfield  {author} {\bibinfo {author} {\bibfnamefont {E.~J.}\ \bibnamefont
  {Dresselhaus}}, \bibinfo {author} {\bibfnamefont {B.}~\bibnamefont
  {Sbierski}},\ and\ \bibinfo {author} {\bibfnamefont {I.~A.}\ \bibnamefont
  {Gruzberg}},\ }\bibfield  {title} {\bibinfo {title} {{Scaling Collapse of
  Longitudinal Conductance near the Integer Quantum Hall Transition}},\ }\href
  {https://doi.org/10.1103/PhysRevLett.129.026801} {\bibfield  {journal}
  {\bibinfo  {journal} {Phys. Rev. Lett.}\ }\textbf {\bibinfo {volume} {129}},\
  \bibinfo {pages} {026801} (\bibinfo {year} {2022})}\BibitemShut {NoStop}%
\bibitem [{\citenamefont {Evers}\ and\ \citenamefont {Mirlin}(2008)}]{Evers08}%
  \BibitemOpen
  \bibfield  {author} {\bibinfo {author} {\bibfnamefont {F.}~\bibnamefont
  {Evers}}\ and\ \bibinfo {author} {\bibfnamefont {A.~D.}\ \bibnamefont
  {Mirlin}},\ }\bibfield  {title} {\bibinfo {title} {Anderson transitions},\
  }\href {https://doi.org/https://doi.org/10.1103/RevModPhys.80.1355}
  {\bibfield  {journal} {\bibinfo  {journal} {Rev. Mod. Phys.}\ }\textbf
  {\bibinfo {volume} {80}},\ \bibinfo {pages} {1355} (\bibinfo {year}
  {2008})}\BibitemShut {NoStop}%
\bibitem [{\citenamefont {Asada}\ \emph {et~al.}(2002)\citenamefont {Asada},
  \citenamefont {Slevin},\ and\ \citenamefont {Ohtsuki}}]{Asada02}%
  \BibitemOpen
  \bibfield  {author} {\bibinfo {author} {\bibfnamefont {Y.}~\bibnamefont
  {Asada}}, \bibinfo {author} {\bibfnamefont {K.}~\bibnamefont {Slevin}},\ and\
  \bibinfo {author} {\bibfnamefont {T.}~\bibnamefont {Ohtsuki}},\ }\bibfield
  {title} {\bibinfo {title} {{Anderson Transition in Two-Dimensional Systems
  with Spin-Orbit Coupling}},\ }\href
  {https://doi.org/10.1103/PhysRevLett.89.256601} {\bibfield  {journal}
  {\bibinfo  {journal} {Phys. Rev. Lett.}\ }\textbf {\bibinfo {volume} {89}},\
  \bibinfo {pages} {256601} (\bibinfo {year} {2002})}\BibitemShut {NoStop}%
\bibitem [{\citenamefont {Asada}\ \emph {et~al.}(2005)\citenamefont {Asada},
  \citenamefont {Slevin},\ and\ \citenamefont {Ohtsuki}}]{Asada05}%
  \BibitemOpen
  \bibfield  {author} {\bibinfo {author} {\bibfnamefont {Y.}~\bibnamefont
  {Asada}}, \bibinfo {author} {\bibfnamefont {K.}~\bibnamefont {Slevin}},\ and\
  \bibinfo {author} {\bibfnamefont {T.}~\bibnamefont {Ohtsuki}},\ }\bibfield
  {title} {\bibinfo {title} {{Anderson Transition in the Three Dimensional
  Symplectic Universality Class}},\ }\href
  {https://doi.org/10.1143/JPSJS.74S.238} {\bibfield  {journal} {\bibinfo
  {journal} {J. Phys. Soc. Jpn.}\ }\textbf {\bibinfo {volume} {74}},\ \bibinfo
  {pages} {238} (\bibinfo {year} {2005})}\BibitemShut {NoStop}%
\bibitem [{\citenamefont {Onoda}\ \emph {et~al.}(2007)\citenamefont {Onoda},
  \citenamefont {Avishai},\ and\ \citenamefont {Nagaosa}}]{onoda07}%
  \BibitemOpen
  \bibfield  {author} {\bibinfo {author} {\bibfnamefont {M.}~\bibnamefont
  {Onoda}}, \bibinfo {author} {\bibfnamefont {Y.}~\bibnamefont {Avishai}},\
  and\ \bibinfo {author} {\bibfnamefont {N.}~\bibnamefont {Nagaosa}},\
  }\bibfield  {title} {\bibinfo {title} {{Localization in a Quantum Spin Hall
  System}},\ }\href {https://doi.org/10.1103/PhysRevLett.98.076802} {\bibfield
  {journal} {\bibinfo  {journal} {Phys. Rev. Lett.}\ }\textbf {\bibinfo
  {volume} {98}},\ \bibinfo {pages} {076802} (\bibinfo {year}
  {2007})}\BibitemShut {NoStop}%
\bibitem [{\citenamefont {Obuse}\ \emph {et~al.}(2007)\citenamefont {Obuse},
  \citenamefont {Furusaki}, \citenamefont {Ryu},\ and\ \citenamefont
  {Mudry}}]{Obuse07}%
  \BibitemOpen
  \bibfield  {author} {\bibinfo {author} {\bibfnamefont {H.}~\bibnamefont
  {Obuse}}, \bibinfo {author} {\bibfnamefont {A.}~\bibnamefont {Furusaki}},
  \bibinfo {author} {\bibfnamefont {S.}~\bibnamefont {Ryu}},\ and\ \bibinfo
  {author} {\bibfnamefont {C.}~\bibnamefont {Mudry}},\ }\bibfield  {title}
  {\bibinfo {title} {{Two-dimensional spin-filtered chiral network model for
  the ${\mathbb{Z}}_{2}$ quantum spin-Hall effect}},\ }\href
  {https://doi.org/10.1103/PhysRevB.76.075301} {\bibfield  {journal} {\bibinfo
  {journal} {Phys. Rev. B}\ }\textbf {\bibinfo {volume} {76}},\ \bibinfo
  {pages} {075301} (\bibinfo {year} {2007})}\BibitemShut {NoStop}%
\bibitem [{\citenamefont {Ryu}\ \emph {et~al.}(2007)\citenamefont {Ryu},
  \citenamefont {Mudry}, \citenamefont {Obuse},\ and\ \citenamefont
  {Furusaki}}]{ryu07}%
  \BibitemOpen
  \bibfield  {author} {\bibinfo {author} {\bibfnamefont {S.}~\bibnamefont
  {Ryu}}, \bibinfo {author} {\bibfnamefont {C.}~\bibnamefont {Mudry}}, \bibinfo
  {author} {\bibfnamefont {H.}~\bibnamefont {Obuse}},\ and\ \bibinfo {author}
  {\bibfnamefont {A.}~\bibnamefont {Furusaki}},\ }\bibfield  {title} {\bibinfo
  {title} {{${\mathbb{Z}}_{2}$ Topological Term, the Global Anomaly, and the
  Two-Dimensional Symplectic Symmetry Class of Anderson Localization}},\ }\href
  {https://doi.org/10.1103/PhysRevLett.99.116601} {\bibfield  {journal}
  {\bibinfo  {journal} {Phys. Rev. Lett.}\ }\textbf {\bibinfo {volume} {99}},\
  \bibinfo {pages} {116601} (\bibinfo {year} {2007})}\BibitemShut {NoStop}%
\bibitem [{\citenamefont {Nomura}\ \emph {et~al.}(2007)\citenamefont {Nomura},
  \citenamefont {Koshino},\ and\ \citenamefont {Ryu}}]{nomura07}%
  \BibitemOpen
  \bibfield  {author} {\bibinfo {author} {\bibfnamefont {K.}~\bibnamefont
  {Nomura}}, \bibinfo {author} {\bibfnamefont {M.}~\bibnamefont {Koshino}},\
  and\ \bibinfo {author} {\bibfnamefont {S.}~\bibnamefont {Ryu}},\ }\bibfield
  {title} {\bibinfo {title} {{Topological Delocalization of Two-Dimensional
  Massless Dirac Fermions}},\ }\href
  {https://doi.org/10.1103/PhysRevLett.99.146806} {\bibfield  {journal}
  {\bibinfo  {journal} {Phys. Rev. Lett.}\ }\textbf {\bibinfo {volume} {99}},\
  \bibinfo {pages} {146806} (\bibinfo {year} {2007})}\BibitemShut {NoStop}%
\bibitem [{\citenamefont {Mirlin}\ \emph {et~al.}(2010)\citenamefont {Mirlin},
  \citenamefont {Evers}, \citenamefont {Gornyi},\ and\ \citenamefont
  {Ostrovsky}}]{mirlin10}%
  \BibitemOpen
  \bibfield  {author} {\bibinfo {author} {\bibfnamefont {A.~D.}\ \bibnamefont
  {Mirlin}}, \bibinfo {author} {\bibfnamefont {F.}~\bibnamefont {Evers}},
  \bibinfo {author} {\bibfnamefont {I.~V.}\ \bibnamefont {Gornyi}},\ and\
  \bibinfo {author} {\bibfnamefont {P.~M.}\ \bibnamefont {Ostrovsky}},\
  }\bibfield  {title} {\bibinfo {title} {{Anderson Transitions: Criticality,
  Symmetries, and Topologies}},\ }\href
  {https://doi.org/10.1142/9789814299084_0006} {\bibfield  {journal} {\bibinfo
  {journal} {Int. J. Mod. Phys. B}\ }\textbf {\bibinfo {volume} {24}},\
  \bibinfo {pages} {1577} (\bibinfo {year} {2010})}\BibitemShut {NoStop}%
\bibitem [{\citenamefont {K\"onig}\ \emph {et~al.}(2012)\citenamefont
  {K\"onig}, \citenamefont {Ostrovsky}, \citenamefont {Protopopov},\ and\
  \citenamefont {Mirlin}}]{Konig2012}%
  \BibitemOpen
  \bibfield  {author} {\bibinfo {author} {\bibfnamefont {E.~J.}\ \bibnamefont
  {K\"onig}}, \bibinfo {author} {\bibfnamefont {P.~M.}\ \bibnamefont
  {Ostrovsky}}, \bibinfo {author} {\bibfnamefont {I.~V.}\ \bibnamefont
  {Protopopov}},\ and\ \bibinfo {author} {\bibfnamefont {A.~D.}\ \bibnamefont
  {Mirlin}},\ }\bibfield  {title} {\bibinfo {title} {{Metal-insulator
  transition in two-dimensional random fermion systems of chiral symmetry
  classes}},\ }\href {https://doi.org/10.1103/PhysRevB.85.195130} {\bibfield
  {journal} {\bibinfo  {journal} {Phys. Rev. B}\ }\textbf {\bibinfo {volume}
  {85}},\ \bibinfo {pages} {195130} (\bibinfo {year} {2012})}\BibitemShut
  {NoStop}%
\bibitem [{\citenamefont {Ringel}\ \emph {et~al.}(2012)\citenamefont {Ringel},
  \citenamefont {Kraus},\ and\ \citenamefont {Stern}}]{Ringel12}%
  \BibitemOpen
  \bibfield  {author} {\bibinfo {author} {\bibfnamefont {Z.}~\bibnamefont
  {Ringel}}, \bibinfo {author} {\bibfnamefont {Y.~E.}\ \bibnamefont {Kraus}},\
  and\ \bibinfo {author} {\bibfnamefont {A.}~\bibnamefont {Stern}},\ }\bibfield
   {title} {\bibinfo {title} {Strong side of weak topological insulators},\
  }\href {https://doi.org/10.1103/PhysRevB.86.045102} {\bibfield  {journal}
  {\bibinfo  {journal} {Phys. Rev. B}\ }\textbf {\bibinfo {volume} {86}},\
  \bibinfo {pages} {045102} (\bibinfo {year} {2012})}\BibitemShut {NoStop}%
\bibitem [{\citenamefont {Fu}\ and\ \citenamefont {Kane}(2012)}]{Fu12}%
  \BibitemOpen
  \bibfield  {author} {\bibinfo {author} {\bibfnamefont {L.}~\bibnamefont
  {Fu}}\ and\ \bibinfo {author} {\bibfnamefont {C.~L.}\ \bibnamefont {Kane}},\
  }\bibfield  {title} {\bibinfo {title} {{Topology, Delocalization via Average
  Symmetry and the Symplectic Anderson Transition}},\ }\href
  {https://doi.org/10.1103/PhysRevLett.109.246605} {\bibfield  {journal}
  {\bibinfo  {journal} {Phys. Rev. Lett.}\ }\textbf {\bibinfo {volume} {109}},\
  \bibinfo {pages} {246605} (\bibinfo {year} {2012})}\BibitemShut {NoStop}%
\bibitem [{\citenamefont {Slevin}\ and\ \citenamefont
  {Ohtsuki}(2016)}]{Slevin16}%
  \BibitemOpen
  \bibfield  {author} {\bibinfo {author} {\bibfnamefont {K.}~\bibnamefont
  {Slevin}}\ and\ \bibinfo {author} {\bibfnamefont {T.}~\bibnamefont
  {Ohtsuki}},\ }\bibfield  {title} {\bibinfo {title} {{Estimate of the Critical
  Exponent of the Anderson Transition in the Three and Four-Dimensional Unitary
  Universality Classes}},\ }\href {https://doi.org/10.7566/JPSJ.85.104712}
  {\bibfield  {journal} {\bibinfo  {journal} {J. Phys. Soc. Jpn.}\ }\textbf
  {\bibinfo {volume} {85}},\ \bibinfo {pages} {104712} (\bibinfo {year}
  {2016})}\BibitemShut {NoStop}%
\bibitem [{\citenamefont {Roy}\ \emph {et~al.}(2017)\citenamefont {Roy},
  \citenamefont {Alavirad},\ and\ \citenamefont {Sau}}]{Roy17}%
  \BibitemOpen
  \bibfield  {author} {\bibinfo {author} {\bibfnamefont {B.}~\bibnamefont
  {Roy}}, \bibinfo {author} {\bibfnamefont {Y.}~\bibnamefont {Alavirad}},\ and\
  \bibinfo {author} {\bibfnamefont {J.~D.}\ \bibnamefont {Sau}},\ }\bibfield
  {title} {\bibinfo {title} {{Global Phase Diagram of a Three-Dimensional Dirty
  Topological Superconductor}},\ }\href
  {https://doi.org/10.1103/PhysRevLett.118.227002} {\bibfield  {journal}
  {\bibinfo  {journal} {Phys. Rev. Lett.}\ }\textbf {\bibinfo {volume} {118}},\
  \bibinfo {pages} {227002} (\bibinfo {year} {2017})}\BibitemShut {NoStop}%
\bibitem [{\citenamefont {Luo}\ \emph {et~al.}(2018)\citenamefont {Luo},
  \citenamefont {Xu}, \citenamefont {Ohtsuki},\ and\ \citenamefont
  {Shindou}}]{Luo18QMCT}%
  \BibitemOpen
  \bibfield  {author} {\bibinfo {author} {\bibfnamefont {X.}~\bibnamefont
  {Luo}}, \bibinfo {author} {\bibfnamefont {B.}~\bibnamefont {Xu}}, \bibinfo
  {author} {\bibfnamefont {T.}~\bibnamefont {Ohtsuki}},\ and\ \bibinfo {author}
  {\bibfnamefont {R.}~\bibnamefont {Shindou}},\ }\bibfield  {title} {\bibinfo
  {title} {{Quantum multicriticality in disordered Weyl semimetals}},\ }\href
  {https://doi.org/10.1103/PhysRevB.97.045129} {\bibfield  {journal} {\bibinfo
  {journal} {Phys. Rev. B}\ }\textbf {\bibinfo {volume} {97}},\ \bibinfo
  {pages} {045129} (\bibinfo {year} {2018})}\BibitemShut {NoStop}%
\bibitem [{\citenamefont {Yoshioka}\ \emph {et~al.}(2018)\citenamefont
  {Yoshioka}, \citenamefont {Akagi},\ and\ \citenamefont
  {Katsura}}]{Yoshioka18}%
  \BibitemOpen
  \bibfield  {author} {\bibinfo {author} {\bibfnamefont {N.}~\bibnamefont
  {Yoshioka}}, \bibinfo {author} {\bibfnamefont {Y.}~\bibnamefont {Akagi}},\
  and\ \bibinfo {author} {\bibfnamefont {H.}~\bibnamefont {Katsura}},\
  }\bibfield  {title} {\bibinfo {title} {Learning disordered topological phases
  by statistical recovery of symmetry},\ }\href
  {https://doi.org/10.1103/PhysRevB.97.205110} {\bibfield  {journal} {\bibinfo
  {journal} {Phys. Rev. B}\ }\textbf {\bibinfo {volume} {97}},\ \bibinfo
  {pages} {205110} (\bibinfo {year} {2018})}\BibitemShut {NoStop}%
\bibitem [{\citenamefont {Song}\ \emph {et~al.}(2021)\citenamefont {Song},
  \citenamefont {Lian}, \citenamefont {Queiroz}, \citenamefont {Ilan},
  \citenamefont {Bernevig},\ and\ \citenamefont {Stern}}]{song21}%
  \BibitemOpen
  \bibfield  {author} {\bibinfo {author} {\bibfnamefont {Z.-D.}\ \bibnamefont
  {Song}}, \bibinfo {author} {\bibfnamefont {B.}~\bibnamefont {Lian}}, \bibinfo
  {author} {\bibfnamefont {R.}~\bibnamefont {Queiroz}}, \bibinfo {author}
  {\bibfnamefont {R.}~\bibnamefont {Ilan}}, \bibinfo {author} {\bibfnamefont
  {B.~A.}\ \bibnamefont {Bernevig}},\ and\ \bibinfo {author} {\bibfnamefont
  {A.}~\bibnamefont {Stern}},\ }\bibfield  {title} {\bibinfo {title}
  {{Delocalization Transition of a Disordered Axion Insulator}},\ }\href
  {https://doi.org/10.1103/PhysRevLett.127.016602} {\bibfield  {journal}
  {\bibinfo  {journal} {Phys. Rev. Lett.}\ }\textbf {\bibinfo {volume} {127}},\
  \bibinfo {pages} {016602} (\bibinfo {year} {2021})}\BibitemShut {NoStop}%
\bibitem [{\citenamefont {Son}\ and\ \citenamefont {Raghu}(2021)}]{son21}%
  \BibitemOpen
  \bibfield  {author} {\bibinfo {author} {\bibfnamefont {J.~H.}\ \bibnamefont
  {Son}}\ and\ \bibinfo {author} {\bibfnamefont {S.}~\bibnamefont {Raghu}},\
  }\bibfield  {title} {\bibinfo {title} {Three-dimensional network model for
  strong topological insulator transitions},\ }\href
  {https://doi.org/10.1103/PhysRevB.104.125142} {\bibfield  {journal} {\bibinfo
   {journal} {Phys. Rev. B}\ }\textbf {\bibinfo {volume} {104}},\ \bibinfo
  {pages} {125142} (\bibinfo {year} {2021})}\BibitemShut {NoStop}%
\bibitem [{\citenamefont {Pan}\ \emph {et~al.}(2021)\citenamefont {Pan},
  \citenamefont {Wang}, \citenamefont {Ohtsuki},\ and\ \citenamefont
  {Shindou}}]{pan21}%
  \BibitemOpen
  \bibfield  {author} {\bibinfo {author} {\bibfnamefont {Z.}~\bibnamefont
  {Pan}}, \bibinfo {author} {\bibfnamefont {T.}~\bibnamefont {Wang}}, \bibinfo
  {author} {\bibfnamefont {T.}~\bibnamefont {Ohtsuki}},\ and\ \bibinfo {author}
  {\bibfnamefont {R.}~\bibnamefont {Shindou}},\ }\bibfield  {title} {\bibinfo
  {title} {{Renormalization group analysis of Dirac fermions with a random
  mass}},\ }\href {https://doi.org/10.1103/PhysRevB.104.174205} {\bibfield
  {journal} {\bibinfo  {journal} {Phys. Rev. B}\ }\textbf {\bibinfo {volume}
  {104}},\ \bibinfo {pages} {174205} (\bibinfo {year} {2021})}\BibitemShut
  {NoStop}%
\bibitem [{\citenamefont {Wang}\ \emph
  {et~al.}(2021{\natexlab{a}})\citenamefont {Wang}, \citenamefont {Pan},
  \citenamefont {Ohtsuki}, \citenamefont {Gruzberg},\ and\ \citenamefont
  {Shindou}}]{Wang21b}%
  \BibitemOpen
  \bibfield  {author} {\bibinfo {author} {\bibfnamefont {T.}~\bibnamefont
  {Wang}}, \bibinfo {author} {\bibfnamefont {Z.}~\bibnamefont {Pan}}, \bibinfo
  {author} {\bibfnamefont {T.}~\bibnamefont {Ohtsuki}}, \bibinfo {author}
  {\bibfnamefont {I.~A.}\ \bibnamefont {Gruzberg}},\ and\ \bibinfo {author}
  {\bibfnamefont {R.}~\bibnamefont {Shindou}},\ }\bibfield  {title} {\bibinfo
  {title} {{Multicriticality of two-dimensional class-D disordered topological
  superconductors}},\ }\href {https://doi.org/10.1103/PhysRevB.104.184201}
  {\bibfield  {journal} {\bibinfo  {journal} {Phys. Rev. B}\ }\textbf {\bibinfo
  {volume} {104}},\ \bibinfo {pages} {184201} (\bibinfo {year}
  {2021}{\natexlab{a}})}\BibitemShut {NoStop}%
\bibitem [{\citenamefont {Luo}\ \emph {et~al.}(2022)\citenamefont {Luo},
  \citenamefont {Xiao}, \citenamefont {Kawabata}, \citenamefont {Ohtsuki},\
  and\ \citenamefont {Shindou}}]{luo21unifying}%
  \BibitemOpen
  \bibfield  {author} {\bibinfo {author} {\bibfnamefont {X.}~\bibnamefont
  {Luo}}, \bibinfo {author} {\bibfnamefont {Z.}~\bibnamefont {Xiao}}, \bibinfo
  {author} {\bibfnamefont {K.}~\bibnamefont {Kawabata}}, \bibinfo {author}
  {\bibfnamefont {T.}~\bibnamefont {Ohtsuki}},\ and\ \bibinfo {author}
  {\bibfnamefont {R.}~\bibnamefont {Shindou}},\ }\bibfield  {title} {\bibinfo
  {title} {{Unifying the Anderson transitions in Hermitian and non-Hermitian
  systems}},\ }\href {https://doi.org/10.1103/PhysRevResearch.4.L022035}
  {\bibfield  {journal} {\bibinfo  {journal} {Phys. Rev. Research}\ }\textbf
  {\bibinfo {volume} {4}},\ \bibinfo {pages} {L022035} (\bibinfo {year}
  {2022})}\BibitemShut {NoStop}%
\bibitem [{sup()}]{supplemental}%
  \BibitemOpen
  \href@noop {} {}\bibinfo {note} {See Supplemental Material for summary of
  known critical exponents in topological insulators, inverse participation
  ratio, transfer matrix analyses, a relation between weak topological indices
  and Lyapunov exponents, statistical symmetry, details of finite-size scaling
  analyses, quasi-localized phase in other topological models, Anderson
  transition in non-topological models, and detailed numerical results of
  conductance in metal, quasi-localized and Anderson localized phases, which
  includes
  Refs.~\cite{crisanti12,Kawabata19,goldhirsch87,Fulga12,fu07,pendry90,kramer05}}\BibitemShut
  {NoStop}%
\bibitem [{\citenamefont {Burkov}\ \emph {et~al.}(2011)\citenamefont {Burkov},
  \citenamefont {Hook},\ and\ \citenamefont {Balents}}]{Burkov-11}%
  \BibitemOpen
  \bibfield  {author} {\bibinfo {author} {\bibfnamefont {A.~A.}\ \bibnamefont
  {Burkov}}, \bibinfo {author} {\bibfnamefont {M.~D.}\ \bibnamefont {Hook}},\
  and\ \bibinfo {author} {\bibfnamefont {L.}~\bibnamefont {Balents}},\
  }\bibfield  {title} {\bibinfo {title} {{Topological nodal semimetals}},\
  }\href {https://doi.org/10.1103/PhysRevB.84.235126} {\bibfield  {journal}
  {\bibinfo  {journal} {Phys. Rev. B}\ }\textbf {\bibinfo {volume} {84}},\
  \bibinfo {pages} {235126} (\bibinfo {year} {2011})}\BibitemShut {NoStop}%
\bibitem [{\citenamefont {Schnyder}\ and\ \citenamefont
  {Brydon}(2015)}]{Schnyder2015}%
  \BibitemOpen
  \bibfield  {author} {\bibinfo {author} {\bibfnamefont {A.~P.}\ \bibnamefont
  {Schnyder}}\ and\ \bibinfo {author} {\bibfnamefont {P.~M.~R.}\ \bibnamefont
  {Brydon}},\ }\bibfield  {title} {\bibinfo {title} {Topological surface states
  in nodal superconductors},\ }\href
  {https://doi.org/10.1088/0953-8984/27/24/243201} {\bibfield  {journal}
  {\bibinfo  {journal} {J. Phys.: Condens. Matter}\ }\textbf {\bibinfo {volume}
  {27}},\ \bibinfo {pages} {243201} (\bibinfo {year} {2015})}\BibitemShut
  {NoStop}%
\bibitem [{\citenamefont {Fang}\ \emph {et~al.}(2016)\citenamefont {Fang},
  \citenamefont {Weng}, \citenamefont {Dai},\ and\ \citenamefont
  {Fang}}]{fang2016}%
  \BibitemOpen
  \bibfield  {author} {\bibinfo {author} {\bibfnamefont {C.}~\bibnamefont
  {Fang}}, \bibinfo {author} {\bibfnamefont {H.}~\bibnamefont {Weng}}, \bibinfo
  {author} {\bibfnamefont {X.}~\bibnamefont {Dai}},\ and\ \bibinfo {author}
  {\bibfnamefont {Z.}~\bibnamefont {Fang}},\ }\bibfield  {title} {\bibinfo
  {title} {{Topological nodal line semimetals}},\ }\href
  {https://doi.org/10.1088/1674-1056/25/11/117106} {\bibfield  {journal}
  {\bibinfo  {journal} {Chinese Phys. B}\ }\textbf {\bibinfo {volume} {25}},\
  \bibinfo {pages} {117106} (\bibinfo {year} {2016})}\BibitemShut {NoStop}%
\bibitem [{\citenamefont {Armitage}\ \emph {et~al.}(2018)\citenamefont
  {Armitage}, \citenamefont {Mele},\ and\ \citenamefont
  {Vishwanath}}]{Armitage18}%
  \BibitemOpen
  \bibfield  {author} {\bibinfo {author} {\bibfnamefont {N.~P.}\ \bibnamefont
  {Armitage}}, \bibinfo {author} {\bibfnamefont {E.~J.}\ \bibnamefont {Mele}},\
  and\ \bibinfo {author} {\bibfnamefont {A.}~\bibnamefont {Vishwanath}},\
  }\bibfield  {title} {\bibinfo {title} {{Weyl and Dirac semimetals in
  three-dimensional solids}},\ }\href
  {https://doi.org/10.1103/RevModPhys.90.015001} {\bibfield  {journal}
  {\bibinfo  {journal} {Rev. Mod. Phys.}\ }\textbf {\bibinfo {volume} {90}},\
  \bibinfo {pages} {015001} (\bibinfo {year} {2018})}\BibitemShut {NoStop}%
\bibitem [{\citenamefont {Bian}\ \emph {et~al.}(2016)\citenamefont {Bian},
  \citenamefont {Chang}, \citenamefont {Sankar}, \citenamefont {Xu},
  \citenamefont {Zheng}, \citenamefont {Neupert}, \citenamefont {Chiu},
  \citenamefont {Huang}, \citenamefont {Chang}, \citenamefont {Belopolski},
  \citenamefont {Sanchez}, \citenamefont {Neupane}, \citenamefont {Alidoust},
  \citenamefont {Liu}, \citenamefont {Wang}, \citenamefont {Lee}, \citenamefont
  {Jeng}, \citenamefont {Zhang}, \citenamefont {Yuan}, \citenamefont {Jia},
  \citenamefont {Bansil}, \citenamefont {Chou}, \citenamefont {Lin},\ and\
  \citenamefont {Hasan}}]{Bian16}%
  \BibitemOpen
  \bibfield  {author} {\bibinfo {author} {\bibfnamefont {G.}~\bibnamefont
  {Bian}}, \bibinfo {author} {\bibfnamefont {T.-R.}\ \bibnamefont {Chang}},
  \bibinfo {author} {\bibfnamefont {R.}~\bibnamefont {Sankar}}, \bibinfo
  {author} {\bibfnamefont {S.-Y.}\ \bibnamefont {Xu}}, \bibinfo {author}
  {\bibfnamefont {H.}~\bibnamefont {Zheng}}, \bibinfo {author} {\bibfnamefont
  {T.}~\bibnamefont {Neupert}}, \bibinfo {author} {\bibfnamefont {C.-K.}\
  \bibnamefont {Chiu}}, \bibinfo {author} {\bibfnamefont {S.-M.}\ \bibnamefont
  {Huang}}, \bibinfo {author} {\bibfnamefont {G.}~\bibnamefont {Chang}},
  \bibinfo {author} {\bibfnamefont {I.}~\bibnamefont {Belopolski}}, \bibinfo
  {author} {\bibfnamefont {D.~S.}\ \bibnamefont {Sanchez}}, \bibinfo {author}
  {\bibfnamefont {M.}~\bibnamefont {Neupane}}, \bibinfo {author} {\bibfnamefont
  {N.}~\bibnamefont {Alidoust}}, \bibinfo {author} {\bibfnamefont
  {C.}~\bibnamefont {Liu}}, \bibinfo {author} {\bibfnamefont {B.}~\bibnamefont
  {Wang}}, \bibinfo {author} {\bibfnamefont {C.-C.}\ \bibnamefont {Lee}},
  \bibinfo {author} {\bibfnamefont {H.-T.}\ \bibnamefont {Jeng}}, \bibinfo
  {author} {\bibfnamefont {C.}~\bibnamefont {Zhang}}, \bibinfo {author}
  {\bibfnamefont {Z.}~\bibnamefont {Yuan}}, \bibinfo {author} {\bibfnamefont
  {S.}~\bibnamefont {Jia}}, \bibinfo {author} {\bibfnamefont {A.}~\bibnamefont
  {Bansil}}, \bibinfo {author} {\bibfnamefont {F.}~\bibnamefont {Chou}},
  \bibinfo {author} {\bibfnamefont {H.}~\bibnamefont {Lin}},\ and\ \bibinfo
  {author} {\bibfnamefont {M.~Z.}\ \bibnamefont {Hasan}},\ }\bibfield  {title}
  {\bibinfo {title} {{Topological nodal-line fermions in spin-orbit metal
  PbTaSe$_2$}},\ }\href {https://doi.org/10.1038/ncomms10556} {\bibfield
  {journal} {\bibinfo  {journal} {Nat. Commun.}\ }\textbf {\bibinfo {volume}
  {7}},\ \bibinfo {pages} {10556} (\bibinfo {year} {2016})}\BibitemShut
  {NoStop}%
\bibitem [{\citenamefont {Schoop}\ \emph {et~al.}(2016)\citenamefont {Schoop},
  \citenamefont {Ali}, \citenamefont {Straßer}, \citenamefont {Topp},
  \citenamefont {Varykhalov}, \citenamefont {Marchenko}, \citenamefont
  {Duppel}, \citenamefont {Parkin}, \citenamefont {Lotsch},\ and\ \citenamefont
  {Ast}}]{Schoop16}%
  \BibitemOpen
  \bibfield  {author} {\bibinfo {author} {\bibfnamefont {L.~M.}\ \bibnamefont
  {Schoop}}, \bibinfo {author} {\bibfnamefont {M.~N.}\ \bibnamefont {Ali}},
  \bibinfo {author} {\bibfnamefont {C.}~\bibnamefont {Straßer}}, \bibinfo
  {author} {\bibfnamefont {A.}~\bibnamefont {Topp}}, \bibinfo {author}
  {\bibfnamefont {A.}~\bibnamefont {Varykhalov}}, \bibinfo {author}
  {\bibfnamefont {D.}~\bibnamefont {Marchenko}}, \bibinfo {author}
  {\bibfnamefont {V.}~\bibnamefont {Duppel}}, \bibinfo {author} {\bibfnamefont
  {S.~S.~P.}\ \bibnamefont {Parkin}}, \bibinfo {author} {\bibfnamefont {B.~V.}\
  \bibnamefont {Lotsch}},\ and\ \bibinfo {author} {\bibfnamefont {C.~R.}\
  \bibnamefont {Ast}},\ }\bibfield  {title} {\bibinfo {title} {{Dirac cone
  protected by non-symmorphic symmetry and three-dimensional Dirac line node in
  ZrSiS}},\ }\href {https://doi.org/10.1038/ncomms11696} {\bibfield  {journal}
  {\bibinfo  {journal} {Nat. Commun.}\ }\textbf {\bibinfo {volume} {7}},\
  \bibinfo {pages} {11696} (\bibinfo {year} {2016})}\BibitemShut {NoStop}%
\bibitem [{\citenamefont {Chen}\ \emph {et~al.}(2022)\citenamefont {Chen},
  \citenamefont {Zeng}, \citenamefont {Chen}, \citenamefont {Zhao},
  \citenamefont {Sheng},\ and\ \citenamefont {Yang}}]{Chen22}%
  \BibitemOpen
  \bibfield  {author} {\bibinfo {author} {\bibfnamefont {C.}~\bibnamefont
  {Chen}}, \bibinfo {author} {\bibfnamefont {X.-T.}\ \bibnamefont {Zeng}},
  \bibinfo {author} {\bibfnamefont {Z.}~\bibnamefont {Chen}}, \bibinfo {author}
  {\bibfnamefont {Y.~X.}\ \bibnamefont {Zhao}}, \bibinfo {author}
  {\bibfnamefont {X.-L.}\ \bibnamefont {Sheng}},\ and\ \bibinfo {author}
  {\bibfnamefont {S.~A.}\ \bibnamefont {Yang}},\ }\bibfield  {title} {\bibinfo
  {title} {{Second-Order Real Nodal-Line Semimetal in Three-Dimensional
  Graphdiyne}},\ }\href {https://doi.org/10.1103/PhysRevLett.128.026405}
  {\bibfield  {journal} {\bibinfo  {journal} {Phys. Rev. Lett.}\ }\textbf
  {\bibinfo {volume} {128}},\ \bibinfo {pages} {026405} (\bibinfo {year}
  {2022})}\BibitemShut {NoStop}%
\bibitem [{\citenamefont {Song}\ \emph {et~al.}(2019)\citenamefont {Song},
  \citenamefont {He}, \citenamefont {Niu}, \citenamefont {Zhang}, \citenamefont
  {Ren}, \citenamefont {Liu},\ and\ \citenamefont {Jo}}]{Bo2019}%
  \BibitemOpen
  \bibfield  {author} {\bibinfo {author} {\bibfnamefont {B.}~\bibnamefont
  {Song}}, \bibinfo {author} {\bibfnamefont {C.}~\bibnamefont {He}}, \bibinfo
  {author} {\bibfnamefont {S.}~\bibnamefont {Niu}}, \bibinfo {author}
  {\bibfnamefont {L.}~\bibnamefont {Zhang}}, \bibinfo {author} {\bibfnamefont
  {Z.}~\bibnamefont {Ren}}, \bibinfo {author} {\bibfnamefont {X.-J.}\
  \bibnamefont {Liu}},\ and\ \bibinfo {author} {\bibfnamefont {G.-B.}\
  \bibnamefont {Jo}},\ }\bibfield  {title} {\bibinfo {title} {{Observation of
  nodal-line semimetal with ultracold fermions in an optical lattice}},\ }\href
  {https://doi.org/10.1038/s41567-019-0564-y} {\bibfield  {journal} {\bibinfo
  {journal} {Nat. Phys.}\ }\textbf {\bibinfo {volume} {15}},\ \bibinfo {pages}
  {911} (\bibinfo {year} {2019})}\BibitemShut {NoStop}%
\bibitem [{\citenamefont {Gao}\ \emph {et~al.}(2018)\citenamefont {Gao},
  \citenamefont {Yang}, \citenamefont {Liu}, \citenamefont {Guo}, \citenamefont
  {Xia}, \citenamefont {Hibbins},\ and\ \citenamefont {Zhang}}]{Gao2018}%
  \BibitemOpen
  \bibfield  {author} {\bibinfo {author} {\bibfnamefont {W.}~\bibnamefont
  {Gao}}, \bibinfo {author} {\bibfnamefont {B.}~\bibnamefont {Yang},
  \bibfnamefont {Biao~Tremain}}, \bibinfo {author} {\bibfnamefont
  {H.}~\bibnamefont {Liu}}, \bibinfo {author} {\bibfnamefont {Q.}~\bibnamefont
  {Guo}}, \bibinfo {author} {\bibfnamefont {L.}~\bibnamefont {Xia}}, \bibinfo
  {author} {\bibfnamefont {A.~P.}\ \bibnamefont {Hibbins}},\ and\ \bibinfo
  {author} {\bibfnamefont {S.}~\bibnamefont {Zhang}},\ }\bibfield  {title}
  {\bibinfo {title} {{Experimental observation of photonic nodal line
  degeneracies in metacrystals}},\ }\href
  {https://doi.org/10.1038/s41467-018-03407-5} {\bibfield  {journal} {\bibinfo
  {journal} {Nat. Commun.}\ }\textbf {\bibinfo {volume} {9}},\ \bibinfo {pages}
  {950} (\bibinfo {year} {2018})}\BibitemShut {NoStop}%
\bibitem [{\citenamefont {Xia}\ \emph {et~al.}(2019)\citenamefont {Xia},
  \citenamefont {Guo}, \citenamefont {Yang}, \citenamefont {Han}, \citenamefont
  {Liu}, \citenamefont {Zhang},\ and\ \citenamefont {Zhang}}]{xia2019}%
  \BibitemOpen
  \bibfield  {author} {\bibinfo {author} {\bibfnamefont {L.}~\bibnamefont
  {Xia}}, \bibinfo {author} {\bibfnamefont {Q.}~\bibnamefont {Guo}}, \bibinfo
  {author} {\bibfnamefont {B.}~\bibnamefont {Yang}}, \bibinfo {author}
  {\bibfnamefont {J.}~\bibnamefont {Han}}, \bibinfo {author} {\bibfnamefont
  {C.-X.}\ \bibnamefont {Liu}}, \bibinfo {author} {\bibfnamefont
  {W.}~\bibnamefont {Zhang}},\ and\ \bibinfo {author} {\bibfnamefont
  {S.}~\bibnamefont {Zhang}},\ }\bibfield  {title} {\bibinfo {title}
  {{Observation of Hourglass Nodal Lines in Photonics}},\ }\href
  {https://doi.org/10.1103/PhysRevLett.122.103903} {\bibfield  {journal}
  {\bibinfo  {journal} {Phys. Rev. Lett.}\ }\textbf {\bibinfo {volume} {122}},\
  \bibinfo {pages} {103903} (\bibinfo {year} {2019})}\BibitemShut {NoStop}%
\bibitem [{\citenamefont {Deng}\ \emph {et~al.}(2019)\citenamefont {Deng},
  \citenamefont {Lu}, \citenamefont {Li}, \citenamefont {Huang}, \citenamefont
  {Yan}, \citenamefont {Ma},\ and\ \citenamefont {Liu}}]{Deng2019}%
  \BibitemOpen
  \bibfield  {author} {\bibinfo {author} {\bibfnamefont {W.}~\bibnamefont
  {Deng}}, \bibinfo {author} {\bibfnamefont {J.}~\bibnamefont {Lu}}, \bibinfo
  {author} {\bibfnamefont {F.}~\bibnamefont {Li}}, \bibinfo {author}
  {\bibfnamefont {X.}~\bibnamefont {Huang}}, \bibinfo {author} {\bibfnamefont
  {M.}~\bibnamefont {Yan}}, \bibinfo {author} {\bibfnamefont {J.}~\bibnamefont
  {Ma}},\ and\ \bibinfo {author} {\bibfnamefont {Z.}~\bibnamefont {Liu}},\
  }\bibfield  {title} {\bibinfo {title} {{Nodal rings and drumhead surface
  states in phononic crystals}},\ }\href
  {https://doi.org/10.1038/s41467-019-09820-8} {\bibfield  {journal} {\bibinfo
  {journal} {Nat. Commun.}\ }\textbf {\bibinfo {volume} {10}},\ \bibinfo
  {pages} {1769} (\bibinfo {year} {2019})}\BibitemShut {NoStop}%
\bibitem [{\citenamefont {Nandkishore}(2016)}]{nandkishore2016}%
  \BibitemOpen
  \bibfield  {author} {\bibinfo {author} {\bibfnamefont {R.}~\bibnamefont
  {Nandkishore}},\ }\bibfield  {title} {\bibinfo {title} {{Weyl and Dirac loop
  superconductors}},\ }\href {https://doi.org/10.1103/PhysRevB.93.020506}
  {\bibfield  {journal} {\bibinfo  {journal} {Phys. Rev. B}\ }\textbf {\bibinfo
  {volume} {93}},\ \bibinfo {pages} {020506} (\bibinfo {year}
  {2016})}\BibitemShut {NoStop}%
\bibitem [{\citenamefont {Sur}\ and\ \citenamefont
  {Nandkishore}(2016)}]{sur2016}%
  \BibitemOpen
  \bibfield  {author} {\bibinfo {author} {\bibfnamefont {S.}~\bibnamefont
  {Sur}}\ and\ \bibinfo {author} {\bibfnamefont {R.}~\bibnamefont
  {Nandkishore}},\ }\bibfield  {title} {\bibinfo {title} {{Instabilities of
  Weyl loop semimetals}},\ }\href
  {https://doi.org/10.1088/1367-2630/18/11/115006} {\bibfield  {journal}
  {\bibinfo  {journal} {New J. Phys.}\ }\textbf {\bibinfo {volume} {18}},\
  \bibinfo {pages} {115006} (\bibinfo {year} {2016})}\BibitemShut {NoStop}%
\bibitem [{\citenamefont {Syzranov}\ and\ \citenamefont
  {Skinner}(2017)}]{syzranov2017}%
  \BibitemOpen
  \bibfield  {author} {\bibinfo {author} {\bibfnamefont {S.~V.}\ \bibnamefont
  {Syzranov}}\ and\ \bibinfo {author} {\bibfnamefont {B.}~\bibnamefont
  {Skinner}},\ }\bibfield  {title} {\bibinfo {title} {{Electron transport in
  nodal-line semimetals}},\ }\href {https://doi.org/10.1103/PhysRevB.96.161105}
  {\bibfield  {journal} {\bibinfo  {journal} {Phys. Rev. B}\ }\textbf {\bibinfo
  {volume} {96}},\ \bibinfo {pages} {161105} (\bibinfo {year}
  {2017})}\BibitemShut {NoStop}%
\bibitem [{\citenamefont {Gon\ifmmode~\mbox{\c{c}}\else \c{c}\fi{}alves}\ \emph
  {et~al.}(2020)\citenamefont {Gon\ifmmode~\mbox{\c{c}}\else \c{c}\fi{}alves},
  \citenamefont {Ribeiro}, \citenamefont {Castro},\ and\ \citenamefont
  {Ara\'ujo}}]{goncalves2020}%
  \BibitemOpen
  \bibfield  {author} {\bibinfo {author} {\bibfnamefont {M.}~\bibnamefont
  {Gon\ifmmode~\mbox{\c{c}}\else \c{c}\fi{}alves}}, \bibinfo {author}
  {\bibfnamefont {P.}~\bibnamefont {Ribeiro}}, \bibinfo {author} {\bibfnamefont
  {E.~V.}\ \bibnamefont {Castro}},\ and\ \bibinfo {author} {\bibfnamefont
  {M.~A.~N.}\ \bibnamefont {Ara\'ujo}},\ }\bibfield  {title} {\bibinfo {title}
  {{Disorder-Driven Multifractality Transition in Weyl Nodal Loops}},\ }\href
  {https://doi.org/10.1103/PhysRevLett.124.136405} {\bibfield  {journal}
  {\bibinfo  {journal} {Phys. Rev. Lett.}\ }\textbf {\bibinfo {volume} {124}},\
  \bibinfo {pages} {136405} (\bibinfo {year} {2020})}\BibitemShut {NoStop}%
\bibitem [{\citenamefont {Luo}\ \emph {et~al.}(2020)\citenamefont {Luo},
  \citenamefont {Xu}, \citenamefont {Ohtsuki},\ and\ \citenamefont
  {Shindou}}]{Luo20}%
  \BibitemOpen
  \bibfield  {author} {\bibinfo {author} {\bibfnamefont {X.}~\bibnamefont
  {Luo}}, \bibinfo {author} {\bibfnamefont {B.}~\bibnamefont {Xu}}, \bibinfo
  {author} {\bibfnamefont {T.}~\bibnamefont {Ohtsuki}},\ and\ \bibinfo {author}
  {\bibfnamefont {R.}~\bibnamefont {Shindou}},\ }\bibfield  {title} {\bibinfo
  {title} {{Critical behavior of Anderson transitions in three-dimensional
  orthogonal classes with particle-hole symmetries}},\ }\href
  {https://doi.org/10.1103/PhysRevB.101.020202} {\bibfield  {journal} {\bibinfo
   {journal} {Phys. Rev. B}\ }\textbf {\bibinfo {volume} {101}},\ \bibinfo
  {pages} {020202} (\bibinfo {year} {2020})}\BibitemShut {NoStop}%
\bibitem [{\citenamefont {Cohen}\ \emph {et~al.}(1974)\citenamefont {Cohen},
  \citenamefont {Coleman}, \citenamefont {Garito},\ and\ \citenamefont
  {Heeger}}]{cohen1974}%
  \BibitemOpen
  \bibfield  {author} {\bibinfo {author} {\bibfnamefont {M.~J.}\ \bibnamefont
  {Cohen}}, \bibinfo {author} {\bibfnamefont {L.~B.}\ \bibnamefont {Coleman}},
  \bibinfo {author} {\bibfnamefont {A.~F.}\ \bibnamefont {Garito}},\ and\
  \bibinfo {author} {\bibfnamefont {A.~J.}\ \bibnamefont {Heeger}},\ }\bibfield
   {title} {\bibinfo {title} {{Electrical conductivity of tetrathiofulvalinium
  tetracyanoquinodimethan $({\rm TTF})({\rm TCNQ})$}},\ }\href
  {https://doi.org/10.1103/PhysRevB.10.1298} {\bibfield  {journal} {\bibinfo
  {journal} {Phys. Rev. B}\ }\textbf {\bibinfo {volume} {10}},\ \bibinfo
  {pages} {1298} (\bibinfo {year} {1974})}\BibitemShut {NoStop}%
\bibitem [{\citenamefont {Lilly}\ \emph {et~al.}(1999)\citenamefont {Lilly},
  \citenamefont {Cooper}, \citenamefont {Eisenstein}, \citenamefont
  {Pfeiffer},\ and\ \citenamefont {West}}]{lilly1999}%
  \BibitemOpen
  \bibfield  {author} {\bibinfo {author} {\bibfnamefont {M.~P.}\ \bibnamefont
  {Lilly}}, \bibinfo {author} {\bibfnamefont {K.~B.}\ \bibnamefont {Cooper}},
  \bibinfo {author} {\bibfnamefont {J.~P.}\ \bibnamefont {Eisenstein}},
  \bibinfo {author} {\bibfnamefont {L.~N.}\ \bibnamefont {Pfeiffer}},\ and\
  \bibinfo {author} {\bibfnamefont {K.~W.}\ \bibnamefont {West}},\ }\bibfield
  {title} {\bibinfo {title} {{Evidence for an Anisotropic State of
  Two-Dimensional Electrons in High Landau Levels}},\ }\href
  {https://doi.org/10.1103/PhysRevLett.82.394} {\bibfield  {journal} {\bibinfo
  {journal} {Phys. Rev. Lett.}\ }\textbf {\bibinfo {volume} {82}},\ \bibinfo
  {pages} {394} (\bibinfo {year} {1999})}\BibitemShut {NoStop}%
\bibitem [{\citenamefont {Du}\ \emph {et~al.}(1999)\citenamefont {Du},
  \citenamefont {Tsui}, \citenamefont {Stormer}, \citenamefont {Pfeiffer},
  \citenamefont {Baldwin},\ and\ \citenamefont {West}}]{du1999}%
  \BibitemOpen
  \bibfield  {author} {\bibinfo {author} {\bibfnamefont {R.~R.}\ \bibnamefont
  {Du}}, \bibinfo {author} {\bibfnamefont {D.~C.}\ \bibnamefont {Tsui}},
  \bibinfo {author} {\bibfnamefont {H.~L.}\ \bibnamefont {Stormer}}, \bibinfo
  {author} {\bibfnamefont {L.~N.}\ \bibnamefont {Pfeiffer}}, \bibinfo {author}
  {\bibfnamefont {K.~W.}\ \bibnamefont {Baldwin}},\ and\ \bibinfo {author}
  {\bibfnamefont {K.~W.}\ \bibnamefont {West}},\ }\bibfield  {title} {\bibinfo
  {title} {{Strongly anisotropic transport in higher two-dimensional Landau
  levels}},\ }\href {https://doi.org/10.1016/S0038-1098(98)00578-X} {\bibfield
  {journal} {\bibinfo  {journal} {Solid State Commun.}\ }\textbf {\bibinfo
  {volume} {109}},\ \bibinfo {pages} {389} (\bibinfo {year}
  {1999})}\BibitemShut {NoStop}%
\bibitem [{\citenamefont {Ando}\ \emph {et~al.}(2002)\citenamefont {Ando},
  \citenamefont {Segawa}, \citenamefont {Komiya},\ and\ \citenamefont
  {Lavrov}}]{ando2002}%
  \BibitemOpen
  \bibfield  {author} {\bibinfo {author} {\bibfnamefont {Y.}~\bibnamefont
  {Ando}}, \bibinfo {author} {\bibfnamefont {K.}~\bibnamefont {Segawa}},
  \bibinfo {author} {\bibfnamefont {S.}~\bibnamefont {Komiya}},\ and\ \bibinfo
  {author} {\bibfnamefont {A.~N.}\ \bibnamefont {Lavrov}},\ }\bibfield  {title}
  {\bibinfo {title} {{Electrical Resistivity Anisotropy from Self-Organized One
  Dimensionality in High-Temperature Superconductors}},\ }\href
  {https://doi.org/10.1103/PhysRevLett.88.137005} {\bibfield  {journal}
  {\bibinfo  {journal} {Phys. Rev. Lett.}\ }\textbf {\bibinfo {volume} {88}},\
  \bibinfo {pages} {137005} (\bibinfo {year} {2002})}\BibitemShut {NoStop}%
\bibitem [{\citenamefont {Borzi}\ \emph {et~al.}(2007)\citenamefont {Borzi},
  \citenamefont {Grigera}, \citenamefont {Farrell}, \citenamefont {Perry},
  \citenamefont {Lister}, \citenamefont {Lee}, \citenamefont {Tennant},
  \citenamefont {Maeno},\ and\ \citenamefont {Mackenzie}}]{borzi2007}%
  \BibitemOpen
  \bibfield  {author} {\bibinfo {author} {\bibfnamefont {R.~A.}\ \bibnamefont
  {Borzi}}, \bibinfo {author} {\bibfnamefont {S.~A.}\ \bibnamefont {Grigera}},
  \bibinfo {author} {\bibfnamefont {J.}~\bibnamefont {Farrell}}, \bibinfo
  {author} {\bibfnamefont {R.~S.}\ \bibnamefont {Perry}}, \bibinfo {author}
  {\bibfnamefont {S.~J.~S.}\ \bibnamefont {Lister}}, \bibinfo {author}
  {\bibfnamefont {S.~L.}\ \bibnamefont {Lee}}, \bibinfo {author} {\bibfnamefont
  {D.~A.}\ \bibnamefont {Tennant}}, \bibinfo {author} {\bibfnamefont
  {Y.}~\bibnamefont {Maeno}},\ and\ \bibinfo {author} {\bibfnamefont {A.~P.}\
  \bibnamefont {Mackenzie}},\ }\bibfield  {title} {\bibinfo {title} {{Formation
  of a Nematic Fluid at High Fields in Sr$_3$Ru$_2$O$_7$}},\ }\href
  {https://doi.org/10.1126/science.1134796} {\bibfield  {journal} {\bibinfo
  {journal} {Science}\ }\textbf {\bibinfo {volume} {315}},\ \bibinfo {pages}
  {214} (\bibinfo {year} {2007})}\BibitemShut {NoStop}%
\bibitem [{\citenamefont {Hinkov}\ \emph {et~al.}(2008)\citenamefont {Hinkov},
  \citenamefont {Haug}, \citenamefont {Fauqué}, \citenamefont {Bourges},
  \citenamefont {Sidis}, \citenamefont {Ivanov}, \citenamefont {Bernhard},
  \citenamefont {Lin},\ and\ \citenamefont {Keimer}}]{hinkov2008}%
  \BibitemOpen
  \bibfield  {author} {\bibinfo {author} {\bibfnamefont {V.}~\bibnamefont
  {Hinkov}}, \bibinfo {author} {\bibfnamefont {D.}~\bibnamefont {Haug}},
  \bibinfo {author} {\bibfnamefont {B.}~\bibnamefont {Fauqué}}, \bibinfo
  {author} {\bibfnamefont {P.}~\bibnamefont {Bourges}}, \bibinfo {author}
  {\bibfnamefont {Y.}~\bibnamefont {Sidis}}, \bibinfo {author} {\bibfnamefont
  {A.}~\bibnamefont {Ivanov}}, \bibinfo {author} {\bibfnamefont
  {C.}~\bibnamefont {Bernhard}}, \bibinfo {author} {\bibfnamefont {C.~T.}\
  \bibnamefont {Lin}},\ and\ \bibinfo {author} {\bibfnamefont {B.}~\bibnamefont
  {Keimer}},\ }\bibfield  {title} {\bibinfo {title} {{Electronic Liquid Crystal
  State in the High-Temperature Superconductor YBa$_2$Cu$_3$O$_{6.45}$}},\
  }\href {https://doi.org/10.1126/science.1152309} {\bibfield  {journal}
  {\bibinfo  {journal} {Science}\ }\textbf {\bibinfo {volume} {319}},\ \bibinfo
  {pages} {597} (\bibinfo {year} {2008})}\BibitemShut {NoStop}%
\bibitem [{\citenamefont {Chu}\ \emph {et~al.}(2010)\citenamefont {Chu},
  \citenamefont {Analytis}, \citenamefont {Greve}, \citenamefont {McMahon},
  \citenamefont {Islam}, \citenamefont {Yamamoto},\ and\ \citenamefont
  {Fisher}}]{chu2010}%
  \BibitemOpen
  \bibfield  {author} {\bibinfo {author} {\bibfnamefont {J.-H.}\ \bibnamefont
  {Chu}}, \bibinfo {author} {\bibfnamefont {J.~G.}\ \bibnamefont {Analytis}},
  \bibinfo {author} {\bibfnamefont {K.~D.}\ \bibnamefont {Greve}}, \bibinfo
  {author} {\bibfnamefont {P.~L.}\ \bibnamefont {McMahon}}, \bibinfo {author}
  {\bibfnamefont {Z.}~\bibnamefont {Islam}}, \bibinfo {author} {\bibfnamefont
  {Y.}~\bibnamefont {Yamamoto}},\ and\ \bibinfo {author} {\bibfnamefont
  {I.~R.}\ \bibnamefont {Fisher}},\ }\bibfield  {title} {\bibinfo {title}
  {{In-Plane Resistivity Anisotropy in an Underdoped Iron Arsenide
  Superconductor}},\ }\href {https://doi.org/10.1126/science.1190482}
  {\bibfield  {journal} {\bibinfo  {journal} {Science}\ }\textbf {\bibinfo
  {volume} {329}},\ \bibinfo {pages} {824} (\bibinfo {year}
  {2010})}\BibitemShut {NoStop}%
\bibitem [{\citenamefont {Fradkin}\ \emph {et~al.}(2010)\citenamefont
  {Fradkin}, \citenamefont {Kivelson}, \citenamefont {Lawler}, \citenamefont
  {Eisenstein},\ and\ \citenamefont {Mackenzie}}]{fradkin2010}%
  \BibitemOpen
  \bibfield  {author} {\bibinfo {author} {\bibfnamefont {E.}~\bibnamefont
  {Fradkin}}, \bibinfo {author} {\bibfnamefont {S.~A.}\ \bibnamefont
  {Kivelson}}, \bibinfo {author} {\bibfnamefont {M.~J.}\ \bibnamefont
  {Lawler}}, \bibinfo {author} {\bibfnamefont {J.~P.}\ \bibnamefont
  {Eisenstein}},\ and\ \bibinfo {author} {\bibfnamefont {A.~P.}\ \bibnamefont
  {Mackenzie}},\ }\bibfield  {title} {\bibinfo {title} {{Nematic Fermi Fluids
  in Condensed Matter Physics}},\ }\href
  {https://doi.org/10.1146/annurev-conmatphys-070909-103925} {\bibfield
  {journal} {\bibinfo  {journal} {Annu. Rev. Condens. Matter Phys.}\ }\textbf
  {\bibinfo {volume} {1}},\ \bibinfo {pages} {153} (\bibinfo {year}
  {2010})}\BibitemShut {NoStop}%
\bibitem [{\citenamefont {MacKinnon}\ and\ \citenamefont
  {Kramer}(1981)}]{MacKinnon81}%
  \BibitemOpen
  \bibfield  {author} {\bibinfo {author} {\bibfnamefont {A.}~\bibnamefont
  {MacKinnon}}\ and\ \bibinfo {author} {\bibfnamefont {B.}~\bibnamefont
  {Kramer}},\ }\bibfield  {title} {\bibinfo {title} {{One-Parameter Scaling of
  Localization Length and Conductance in Disordered Systems}},\ }\href
  {https://doi.org/10.1103/PhysRevLett.47.1546} {\bibfield  {journal} {\bibinfo
   {journal} {Phys. Rev. Lett.}\ }\textbf {\bibinfo {volume} {47}},\ \bibinfo
  {pages} {1546} (\bibinfo {year} {1981})}\BibitemShut {NoStop}%
\bibitem [{\citenamefont {MacKinnon}\ and\ \citenamefont
  {Kramer}(1983)}]{MacKinnon83}%
  \BibitemOpen
  \bibfield  {author} {\bibinfo {author} {\bibfnamefont {A.}~\bibnamefont
  {MacKinnon}}\ and\ \bibinfo {author} {\bibfnamefont {B.}~\bibnamefont
  {Kramer}},\ }\bibfield  {title} {\bibinfo {title} {{The scaling theory of
  electrons in disordered solids: Additional numerical results}},\ }\href
  {https://doi.org/10.1007/BF01578242} {\bibfield  {journal} {\bibinfo
  {journal} {Z. Physik B}\ }\textbf {\bibinfo {volume} {53}},\ \bibinfo {pages}
  {1} (\bibinfo {year} {1983})}\BibitemShut {NoStop}%
\bibitem [{\citenamefont {Slevin}\ and\ \citenamefont
  {Ohtsuki}(2014)}]{Slevin14}%
  \BibitemOpen
  \bibfield  {author} {\bibinfo {author} {\bibfnamefont {K.}~\bibnamefont
  {Slevin}}\ and\ \bibinfo {author} {\bibfnamefont {T.}~\bibnamefont
  {Ohtsuki}},\ }\bibfield  {title} {\bibinfo {title} {{Critical exponent for
  the Anderson transition in the three-dimensional orthogonal universality
  class}},\ }\href {http://stacks.iop.org/1367-2630/16/i=1/a=015012} {\bibfield
   {journal} {\bibinfo  {journal} {New J. Phys.}\ }\textbf {\bibinfo {volume}
  {16}},\ \bibinfo {pages} {015012} (\bibinfo {year} {2014})}\BibitemShut
  {NoStop}%
\bibitem [{\citenamefont {Markos}(1995)}]{markos1995}%
  \BibitemOpen
  \bibfield  {author} {\bibinfo {author} {\bibfnamefont {P.}~\bibnamefont
  {Markos}},\ }\bibfield  {title} {\bibinfo {title} {Phenomenological theory of
  the metal-insulator transition},\ }\href
  {https://doi.org/10.1088/0953-8984/7/44/006} {\bibfield  {journal} {\bibinfo
  {journal} {J. Phys.: Condens. Matter}\ }\textbf {\bibinfo {volume} {7}},\
  \bibinfo {pages} {8361} (\bibinfo {year} {1995})}\BibitemShut {NoStop}%
\bibitem [{\citenamefont {Mondragon-Shem}\ \emph {et~al.}(2014)\citenamefont
  {Mondragon-Shem}, \citenamefont {Hughes}, \citenamefont {Song},\ and\
  \citenamefont {Prodan}}]{Mondragon-Shem14}%
  \BibitemOpen
  \bibfield  {author} {\bibinfo {author} {\bibfnamefont {I.}~\bibnamefont
  {Mondragon-Shem}}, \bibinfo {author} {\bibfnamefont {T.~L.}\ \bibnamefont
  {Hughes}}, \bibinfo {author} {\bibfnamefont {J.}~\bibnamefont {Song}},\ and\
  \bibinfo {author} {\bibfnamefont {E.}~\bibnamefont {Prodan}},\ }\bibfield
  {title} {\bibinfo {title} {{Topological Criticality in the Chiral-Symmetric
  AIII Class at Strong Disorder}},\ }\href
  {https://doi.org/10.1103/PhysRevLett.113.046802} {\bibfield  {journal}
  {\bibinfo  {journal} {Phys. Rev. Lett.}\ }\textbf {\bibinfo {volume} {113}},\
  \bibinfo {pages} {046802} (\bibinfo {year} {2014})}\BibitemShut {NoStop}%
\bibitem [{\citenamefont {Altland}\ \emph {et~al.}(2014)\citenamefont
  {Altland}, \citenamefont {Bagrets}, \citenamefont {Fritz}, \citenamefont
  {Kamenev},\ and\ \citenamefont {Schmiedt}}]{Altland14}%
  \BibitemOpen
  \bibfield  {author} {\bibinfo {author} {\bibfnamefont {A.}~\bibnamefont
  {Altland}}, \bibinfo {author} {\bibfnamefont {D.}~\bibnamefont {Bagrets}},
  \bibinfo {author} {\bibfnamefont {L.}~\bibnamefont {Fritz}}, \bibinfo
  {author} {\bibfnamefont {A.}~\bibnamefont {Kamenev}},\ and\ \bibinfo {author}
  {\bibfnamefont {H.}~\bibnamefont {Schmiedt}},\ }\bibfield  {title} {\bibinfo
  {title} {{Quantum Criticality of Quasi-One-Dimensional Topological Anderson
  Insulators}},\ }\href {https://doi.org/10.1103/PhysRevLett.112.206602}
  {\bibfield  {journal} {\bibinfo  {journal} {Phys. Rev. Lett.}\ }\textbf
  {\bibinfo {volume} {112}},\ \bibinfo {pages} {206602} (\bibinfo {year}
  {2014})}\BibitemShut {NoStop}%
\bibitem [{\citenamefont {Claes}\ and\ \citenamefont
  {Hughes}(2020)}]{claes21weak}%
  \BibitemOpen
  \bibfield  {author} {\bibinfo {author} {\bibfnamefont {J.}~\bibnamefont
  {Claes}}\ and\ \bibinfo {author} {\bibfnamefont {T.~L.}\ \bibnamefont
  {Hughes}},\ }\bibfield  {title} {\bibinfo {title} {{Disorder driven phase
  transitions in weak AIII topological insulators}},\ }\href
  {https://doi.org/10.1103/PhysRevB.101.224201} {\bibfield  {journal} {\bibinfo
   {journal} {Phys. Rev. B}\ }\textbf {\bibinfo {volume} {101}},\ \bibinfo
  {pages} {224201} (\bibinfo {year} {2020})}\BibitemShut {NoStop}%
\bibitem [{\citenamefont {Molinari}(2003)}]{molinari2003}%
  \BibitemOpen
  \bibfield  {author} {\bibinfo {author} {\bibfnamefont {L.}~\bibnamefont
  {Molinari}},\ }\bibfield  {title} {\bibinfo {title} {{Spectral duality and
  distribution of exponents for transfer matrices of block-tridiagonal
  Hamiltonians}},\ }\href {https://doi.org/10.1088/0305-4470/36/14/311}
  {\bibfield  {journal} {\bibinfo  {journal} {J. Phys. A}\ }\textbf {\bibinfo
  {volume} {36}},\ \bibinfo {pages} {4081} (\bibinfo {year}
  {2003})}\BibitemShut {NoStop}%
\bibitem [{\citenamefont {Altland}\ and\ \citenamefont
  {Zirnbauer}(1997)}]{Altland97}%
  \BibitemOpen
  \bibfield  {author} {\bibinfo {author} {\bibfnamefont {A.}~\bibnamefont
  {Altland}}\ and\ \bibinfo {author} {\bibfnamefont {M.~R.}\ \bibnamefont
  {Zirnbauer}},\ }\bibfield  {title} {\bibinfo {title} {Nonstandard symmetry
  classes in mesoscopic normal-superconducting hybrid structures},\ }\href
  {https://doi.org/10.1103/PhysRevB.55.1142} {\bibfield  {journal} {\bibinfo
  {journal} {Phys. Rev. B}\ }\textbf {\bibinfo {volume} {55}},\ \bibinfo
  {pages} {1142} (\bibinfo {year} {1997})}\BibitemShut {NoStop}%
\bibitem [{\citenamefont {Hatano}\ and\ \citenamefont
  {Nelson}(1996)}]{Hatano96}%
  \BibitemOpen
  \bibfield  {author} {\bibinfo {author} {\bibfnamefont {N.}~\bibnamefont
  {Hatano}}\ and\ \bibinfo {author} {\bibfnamefont {D.~R.}\ \bibnamefont
  {Nelson}},\ }\bibfield  {title} {\bibinfo {title} {{Localization Transitions
  in Non-Hermitian Quantum Mechanics}},\ }\href
  {https://doi.org/https://doi.org/10.1103/PhysRevLett.77.570} {\bibfield
  {journal} {\bibinfo  {journal} {Phys. Rev. Lett.}\ }\textbf {\bibinfo
  {volume} {77}},\ \bibinfo {pages} {570} (\bibinfo {year} {1996})}\BibitemShut
  {NoStop}%
\bibitem [{\citenamefont {Asada}\ \emph {et~al.}(2004)\citenamefont {Asada},
  \citenamefont {Slevin},\ and\ \citenamefont {Ohtsuki}}]{Asada04}%
  \BibitemOpen
  \bibfield  {author} {\bibinfo {author} {\bibfnamefont {Y.}~\bibnamefont
  {Asada}}, \bibinfo {author} {\bibfnamefont {K.}~\bibnamefont {Slevin}},\ and\
  \bibinfo {author} {\bibfnamefont {T.}~\bibnamefont {Ohtsuki}},\ }\bibfield
  {title} {\bibinfo {title} {Numerical estimation of the $\ensuremath{\beta}$
  function in two-dimensional systems with spin-orbit coupling},\ }\href
  {https://doi.org/10.1103/PhysRevB.70.035115} {\bibfield  {journal} {\bibinfo
  {journal} {Phys. Rev. B}\ }\textbf {\bibinfo {volume} {70}},\ \bibinfo
  {pages} {035115} (\bibinfo {year} {2004})}\BibitemShut {NoStop}%
\bibitem [{\citenamefont {Slevin}\ and\ \citenamefont
  {Ohtsuki}(1999)}]{Slevin99}%
  \BibitemOpen
  \bibfield  {author} {\bibinfo {author} {\bibfnamefont {K.}~\bibnamefont
  {Slevin}}\ and\ \bibinfo {author} {\bibfnamefont {T.}~\bibnamefont
  {Ohtsuki}},\ }\bibfield  {title} {\bibinfo {title} {{Corrections to Scaling
  at the Anderson Transition}},\ }\href
  {https://doi.org/10.1103/PhysRevLett.82.382} {\bibfield  {journal} {\bibinfo
  {journal} {Phys. Rev. Lett.}\ }\textbf {\bibinfo {volume} {82}},\ \bibinfo
  {pages} {382} (\bibinfo {year} {1999})}\BibitemShut {NoStop}%
\bibitem [{\citenamefont {Wang}\ \emph
  {et~al.}(2021{\natexlab{b}})\citenamefont {Wang}, \citenamefont {Ohtsuki},\
  and\ \citenamefont {Shindou}}]{Wang21}%
  \BibitemOpen
  \bibfield  {author} {\bibinfo {author} {\bibfnamefont {T.}~\bibnamefont
  {Wang}}, \bibinfo {author} {\bibfnamefont {T.}~\bibnamefont {Ohtsuki}},\ and\
  \bibinfo {author} {\bibfnamefont {R.}~\bibnamefont {Shindou}},\ }\bibfield
  {title} {\bibinfo {title} {{Universality classes of the Anderson transition
  in the three-dimensional symmetry classes AIII, BDI, C, D, and CI}},\ }\href
  {https://doi.org/10.1103/PhysRevB.104.014206} {\bibfield  {journal} {\bibinfo
   {journal} {Phys. Rev. B}\ }\textbf {\bibinfo {volume} {104}},\ \bibinfo
  {pages} {014206} (\bibinfo {year} {2021}{\natexlab{b}})}\BibitemShut
  {NoStop}%
\bibitem [{\citenamefont {Balents}\ and\ \citenamefont
  {Fisher}(1997)}]{Leon97}%
  \BibitemOpen
  \bibfield  {author} {\bibinfo {author} {\bibfnamefont {L.}~\bibnamefont
  {Balents}}\ and\ \bibinfo {author} {\bibfnamefont {M.~P.~A.}\ \bibnamefont
  {Fisher}},\ }\bibfield  {title} {\bibinfo {title} {{Delocalization transition
  via supersymmetry in one dimension}},\ }\href
  {https://doi.org/10.1103/PhysRevB.56.12970} {\bibfield  {journal} {\bibinfo
  {journal} {Phys. Rev. B}\ }\textbf {\bibinfo {volume} {56}},\ \bibinfo
  {pages} {12970} (\bibinfo {year} {1997})}\BibitemShut {NoStop}%
\bibitem [{\citenamefont {Mathur}(1997)}]{Mathur97}%
  \BibitemOpen
  \bibfield  {author} {\bibinfo {author} {\bibfnamefont {H.}~\bibnamefont
  {Mathur}},\ }\bibfield  {title} {\bibinfo {title} {Feynman's propagator
  applied to network models of localization},\ }\href
  {https://doi.org/10.1103/PhysRevB.56.15794} {\bibfield  {journal} {\bibinfo
  {journal} {Phys. Rev. B}\ }\textbf {\bibinfo {volume} {56}},\ \bibinfo
  {pages} {15794} (\bibinfo {year} {1997})}\BibitemShut {NoStop}%
\bibitem [{\citenamefont {Brouwer}\ \emph {et~al.}(1998)\citenamefont
  {Brouwer}, \citenamefont {Mudry}, \citenamefont {Simons},\ and\ \citenamefont
  {Altland}}]{Brouwer98}%
  \BibitemOpen
  \bibfield  {author} {\bibinfo {author} {\bibfnamefont {P.~W.}\ \bibnamefont
  {Brouwer}}, \bibinfo {author} {\bibfnamefont {C.}~\bibnamefont {Mudry}},
  \bibinfo {author} {\bibfnamefont {B.~D.}\ \bibnamefont {Simons}},\ and\
  \bibinfo {author} {\bibfnamefont {A.}~\bibnamefont {Altland}},\ }\bibfield
  {title} {\bibinfo {title} {{Delocalization in Coupled One-Dimensional
  Chains}},\ }\href {https://doi.org/10.1103/PhysRevLett.81.862} {\bibfield
  {journal} {\bibinfo  {journal} {Phys. Rev. Lett.}\ }\textbf {\bibinfo
  {volume} {81}},\ \bibinfo {pages} {862} (\bibinfo {year} {1998})}\BibitemShut
  {NoStop}%
\bibitem [{\citenamefont {Konotop}\ \emph {et~al.}(2016)\citenamefont
  {Konotop}, \citenamefont {Yang},\ and\ \citenamefont {Zezyulin}}]{Konotop16}%
  \BibitemOpen
  \bibfield  {author} {\bibinfo {author} {\bibfnamefont {V.~V.}\ \bibnamefont
  {Konotop}}, \bibinfo {author} {\bibfnamefont {J.}~\bibnamefont {Yang}},\ and\
  \bibinfo {author} {\bibfnamefont {D.~A.}\ \bibnamefont {Zezyulin}},\
  }\bibfield  {title} {\bibinfo {title} {{Nonlinear waves in
  $\mathcal{PT}$-symmetric systems}},\ }\href
  {https://doi.org/10.1103/RevModPhys.88.035002} {\bibfield  {journal}
  {\bibinfo  {journal} {Rev. Mod. Phys.}\ }\textbf {\bibinfo {volume} {88}},\
  \bibinfo {pages} {035002} (\bibinfo {year} {2016})}\BibitemShut {NoStop}%
\bibitem [{\citenamefont {Feng}\ \emph {et~al.}(2017)\citenamefont {Feng},
  \citenamefont {El-Ganainy},\ and\ \citenamefont {Ge}}]{Feng17}%
  \BibitemOpen
  \bibfield  {author} {\bibinfo {author} {\bibfnamefont {L.}~\bibnamefont
  {Feng}}, \bibinfo {author} {\bibfnamefont {R.}~\bibnamefont {El-Ganainy}},\
  and\ \bibinfo {author} {\bibfnamefont {L.}~\bibnamefont {Ge}},\ }\bibfield
  {title} {\bibinfo {title} {{Non-Hermitian photonics based on parity-time
  symmetry}},\ }\href {https://doi.org/10.1038/s41566-017-0031-1} {\bibfield
  {journal} {\bibinfo  {journal} {Nat. Photon.}\ }\textbf {\bibinfo {volume}
  {11}},\ \bibinfo {pages} {752} (\bibinfo {year} {2017})}\BibitemShut
  {NoStop}%
\bibitem [{\citenamefont {El-Ganainy}\ \emph {et~al.}(2018)\citenamefont
  {El-Ganainy}, \citenamefont {Makris}, \citenamefont {Khajavikhan},
  \citenamefont {Musslimani}, \citenamefont {Rotter},\ and\ \citenamefont
  {Christodoulides}}]{El-Ganainy18}%
  \BibitemOpen
  \bibfield  {author} {\bibinfo {author} {\bibfnamefont {R.}~\bibnamefont
  {El-Ganainy}}, \bibinfo {author} {\bibfnamefont {K.~G.}\ \bibnamefont
  {Makris}}, \bibinfo {author} {\bibfnamefont {M.}~\bibnamefont {Khajavikhan}},
  \bibinfo {author} {\bibfnamefont {Z.~H.}\ \bibnamefont {Musslimani}},
  \bibinfo {author} {\bibfnamefont {S.}~\bibnamefont {Rotter}},\ and\ \bibinfo
  {author} {\bibfnamefont {D.~N.}\ \bibnamefont {Christodoulides}},\ }\bibfield
   {title} {\bibinfo {title} {{Non-Hermitian physics and $\mathcal{PT}$
  symmetry}},\ }\href {https://doi.org/10.1038/nphys4323} {\bibfield  {journal}
  {\bibinfo  {journal} {Nat. Phys.}\ }\textbf {\bibinfo {volume} {14}},\
  \bibinfo {pages} {11} (\bibinfo {year} {2018})}\BibitemShut {NoStop}%
\bibitem [{unp()}]{unpublished}%
  \BibitemOpen
  \href@noop {} {}\bibinfo {note} {Z.~Xiao, K.~Kawabata, X.~Luo, T.~Ohtsuki,
  and R.~Shindou, unpublished}\BibitemShut {NoStop}%
\bibitem [{\citenamefont {Crisanti}\ \emph {et~al.}(1993)\citenamefont
  {Crisanti}, \citenamefont {Paladin},\ and\ \citenamefont
  {Vulpiani}}]{crisanti12}%
  \BibitemOpen
  \bibfield  {author} {\bibinfo {author} {\bibfnamefont {A.}~\bibnamefont
  {Crisanti}}, \bibinfo {author} {\bibfnamefont {G.}~\bibnamefont {Paladin}},\
  and\ \bibinfo {author} {\bibfnamefont {A.}~\bibnamefont {Vulpiani}},\ }\href
  {https://doi.org/10.1007/978-3-642-84942-8} {\emph {\bibinfo {title}
  {Products of Random Matrices}}}\ (\bibinfo  {publisher} {Springer},\ \bibinfo
  {address} {Berlin, Heidelberg},\ \bibinfo {year} {1993})\BibitemShut
  {NoStop}%
\bibitem [{\citenamefont {Kawabata}\ \emph {et~al.}(2019)\citenamefont
  {Kawabata}, \citenamefont {Shiozaki}, \citenamefont {Ueda},\ and\
  \citenamefont {Sato}}]{Kawabata19}%
  \BibitemOpen
  \bibfield  {author} {\bibinfo {author} {\bibfnamefont {K.}~\bibnamefont
  {Kawabata}}, \bibinfo {author} {\bibfnamefont {K.}~\bibnamefont {Shiozaki}},
  \bibinfo {author} {\bibfnamefont {M.}~\bibnamefont {Ueda}},\ and\ \bibinfo
  {author} {\bibfnamefont {M.}~\bibnamefont {Sato}},\ }\bibfield  {title}
  {\bibinfo {title} {{Symmetry and Topology in Non-Hermitian Physics}},\ }\href
  {https://doi.org/10.1103/PhysRevX.9.041015} {\bibfield  {journal} {\bibinfo
  {journal} {Phys. Rev. X}\ }\textbf {\bibinfo {volume} {9}},\ \bibinfo {pages}
  {041015} (\bibinfo {year} {2019})}\BibitemShut {NoStop}%
\bibitem [{\citenamefont {Goldhirsch}\ \emph {et~al.}(1987)\citenamefont
  {Goldhirsch}, \citenamefont {Sulem},\ and\ \citenamefont
  {Orszag}}]{goldhirsch87}%
  \BibitemOpen
  \bibfield  {author} {\bibinfo {author} {\bibfnamefont {I.}~\bibnamefont
  {Goldhirsch}}, \bibinfo {author} {\bibfnamefont {P.-L.}\ \bibnamefont
  {Sulem}},\ and\ \bibinfo {author} {\bibfnamefont {S.~A.}\ \bibnamefont
  {Orszag}},\ }\bibfield  {title} {\bibinfo {title} {{Stability and Lyapunov
  stability of dynamical systems: A differential approach and a numerical
  method}},\ }\href {https://doi.org/10.1016/0167-2789(87)90034-0} {\bibfield
  {journal} {\bibinfo  {journal} {Physica D}\ }\textbf {\bibinfo {volume}
  {27}},\ \bibinfo {pages} {311} (\bibinfo {year} {1987})}\BibitemShut
  {NoStop}%
\bibitem [{\citenamefont {Fulga}\ \emph {et~al.}(2012)\citenamefont {Fulga},
  \citenamefont {Akhmerov}, \citenamefont {Tworzyd\l{}o}, \citenamefont
  {B\'eri},\ and\ \citenamefont {Beenakker}}]{Fulga12}%
  \BibitemOpen
  \bibfield  {author} {\bibinfo {author} {\bibfnamefont {I.~C.}\ \bibnamefont
  {Fulga}}, \bibinfo {author} {\bibfnamefont {A.~R.}\ \bibnamefont {Akhmerov}},
  \bibinfo {author} {\bibfnamefont {J.}~\bibnamefont {Tworzyd\l{}o}}, \bibinfo
  {author} {\bibfnamefont {B.}~\bibnamefont {B\'eri}},\ and\ \bibinfo {author}
  {\bibfnamefont {C.~W.~J.}\ \bibnamefont {Beenakker}},\ }\bibfield  {title}
  {\bibinfo {title} {Thermal metal-insulator transition in a helical
  topological superconductor},\ }\href
  {https://doi.org/10.1103/PhysRevB.86.054505} {\bibfield  {journal} {\bibinfo
  {journal} {Phys. Rev. B}\ }\textbf {\bibinfo {volume} {86}},\ \bibinfo
  {pages} {054505} (\bibinfo {year} {2012})}\BibitemShut {NoStop}%
\bibitem [{\citenamefont {Fu}\ \emph {et~al.}(2007)\citenamefont {Fu},
  \citenamefont {Kane},\ and\ \citenamefont {Mele}}]{fu07}%
  \BibitemOpen
  \bibfield  {author} {\bibinfo {author} {\bibfnamefont {L.}~\bibnamefont
  {Fu}}, \bibinfo {author} {\bibfnamefont {C.~L.}\ \bibnamefont {Kane}},\ and\
  \bibinfo {author} {\bibfnamefont {E.~J.}\ \bibnamefont {Mele}},\ }\bibfield
  {title} {\bibinfo {title} {{Topological Insulators in Three Dimensions}},\
  }\href {https://doi.org/10.1103/PhysRevLett.98.106803} {\bibfield  {journal}
  {\bibinfo  {journal} {Phys. Rev. Lett.}\ }\textbf {\bibinfo {volume} {98}},\
  \bibinfo {pages} {106803} (\bibinfo {year} {2007})}\BibitemShut {NoStop}%
\bibitem [{\citenamefont {Pendry}\ \emph {et~al.}(1990)\citenamefont {Pendry},
  \citenamefont {MacKinnon},\ and\ \citenamefont {Pretre}}]{pendry90}%
  \BibitemOpen
  \bibfield  {author} {\bibinfo {author} {\bibfnamefont {J.~B.}\ \bibnamefont
  {Pendry}}, \bibinfo {author} {\bibfnamefont {A.}~\bibnamefont {MacKinnon}},\
  and\ \bibinfo {author} {\bibfnamefont {A.~B.}\ \bibnamefont {Pretre}},\
  }\bibfield  {title} {\bibinfo {title} {{Maximal fluctuations---A new
  phenomenon in disordered systems}},\ }\href
  {https://doi.org/10.1016/0378-4371(90)90391-5} {\bibfield  {journal}
  {\bibinfo  {journal} {Physica A}\ }\textbf {\bibinfo {volume} {168}},\
  \bibinfo {pages} {400} (\bibinfo {year} {1990})}\BibitemShut {NoStop}%
\bibitem [{\citenamefont {Kramer}\ \emph {et~al.}(2005)\citenamefont {Kramer},
  \citenamefont {Ohtsuki},\ and\ \citenamefont {Kettemann}}]{kramer05}%
  \BibitemOpen
  \bibfield  {author} {\bibinfo {author} {\bibfnamefont {B.}~\bibnamefont
  {Kramer}}, \bibinfo {author} {\bibfnamefont {T.}~\bibnamefont {Ohtsuki}},\
  and\ \bibinfo {author} {\bibfnamefont {S.}~\bibnamefont {Kettemann}},\
  }\bibfield  {title} {\bibinfo {title} {Random network models and quantum
  phase transitions in two dimensions},\ }\href
  {https://doi.org/10.1016/j.physrep.2005.07.001} {\bibfield  {journal}
  {\bibinfo  {journal} {Phys. Rep.}\ }\textbf {\bibinfo {volume} {417}},\
  \bibinfo {pages} {211} (\bibinfo {year} {2005})}\BibitemShut {NoStop}%
\end{thebibliography}%


\begin{thebibliography}{25}%
\makeatletter
\providecommand \@ifxundefined [1]{%
 \@ifx{#1\undefined}
}%
\providecommand \@ifnum [1]{%
 \ifnum #1\expandafter \@firstoftwo
 \else \expandafter \@secondoftwo
 \fi
}%
\providecommand \@ifx [1]{%
 \ifx #1\expandafter \@firstoftwo
 \else \expandafter \@secondoftwo
 \fi
}%
\providecommand \natexlab [1]{#1}%
\providecommand \enquote  [1]{``#1''}%
\providecommand \bibnamefont  [1]{#1}%
\providecommand \bibfnamefont [1]{#1}%
\providecommand \citenamefont [1]{#1}%
\providecommand \href@noop [0]{\@secondoftwo}%
\providecommand \href [0]{\begingroup \@sanitize@url \@href}%
\providecommand \@href[1]{\@@startlink{#1}\@@href}%
\providecommand \@@href[1]{\endgroup#1\@@endlink}%
\providecommand \@sanitize@url [0]{\catcode `\\12\catcode `\$12\catcode
  `\&12\catcode `\#12\catcode `\^12\catcode `\_12\catcode `\%12\relax}%
\providecommand \@@startlink[1]{}%
\providecommand \@@endlink[0]{}%
\providecommand \url  [0]{\begingroup\@sanitize@url \@url }%
\providecommand \@url [1]{\endgroup\@href {#1}{\urlprefix }}%
\providecommand \urlprefix  [0]{URL }%
\providecommand \Eprint [0]{\href }%
\providecommand \doibase [0]{https://doi.org/}%
\providecommand \selectlanguage [0]{\@gobble}%
\providecommand \bibinfo  [0]{\@secondoftwo}%
\providecommand \bibfield  [0]{\@secondoftwo}%
\providecommand \translation [1]{[#1]}%
\providecommand \BibitemOpen [0]{}%
\providecommand \bibitemStop [0]{}%
\providecommand \bibitemNoStop [0]{.\EOS\space}%
\providecommand \EOS [0]{\spacefactor3000\relax}%
\providecommand \BibitemShut  [1]{\csname bibitem#1\endcsname}%
\let\auto@bib@innerbib\@empty
\bibitem [{\citenamefont {Obuse}\ \emph {et~al.}(2007)\citenamefont {Obuse},
  \citenamefont {Furusaki}, \citenamefont {Ryu},\ and\ \citenamefont
  {Mudry}}]{Obuse07}%
  \BibitemOpen
  \bibfield  {author} {\bibinfo {author} {\bibfnamefont {H.}~\bibnamefont
  {Obuse}}, \bibinfo {author} {\bibfnamefont {A.}~\bibnamefont {Furusaki}},
  \bibinfo {author} {\bibfnamefont {S.}~\bibnamefont {Ryu}},\ and\ \bibinfo
  {author} {\bibfnamefont {C.}~\bibnamefont {Mudry}},\ }\bibfield  {title}
  {\bibinfo {title} {{Two-dimensional spin-filtered chiral network model for
  the ${\mathbb{Z}}_{2}$ quantum spin-Hall effect}},\ }\href
  {https://doi.org/10.1103/PhysRevB.76.075301} {\bibfield  {journal} {\bibinfo
  {journal} {Phys. Rev. B}\ }\textbf {\bibinfo {volume} {76}},\ \bibinfo
  {pages} {075301} (\bibinfo {year} {2007})}\BibitemShut {NoStop}%
\bibitem [{\citenamefont {Asada}\ \emph {et~al.}(2002)\citenamefont {Asada},
  \citenamefont {Slevin},\ and\ \citenamefont {Ohtsuki}}]{Asada02}%
  \BibitemOpen
  \bibfield  {author} {\bibinfo {author} {\bibfnamefont {Y.}~\bibnamefont
  {Asada}}, \bibinfo {author} {\bibfnamefont {K.}~\bibnamefont {Slevin}},\ and\
  \bibinfo {author} {\bibfnamefont {T.}~\bibnamefont {Ohtsuki}},\ }\bibfield
  {title} {\bibinfo {title} {{Anderson Transition in Two-Dimensional Systems
  with Spin-Orbit Coupling}},\ }\href
  {https://doi.org/10.1103/PhysRevLett.89.256601} {\bibfield  {journal}
  {\bibinfo  {journal} {Phys. Rev. Lett.}\ }\textbf {\bibinfo {volume} {89}},\
  \bibinfo {pages} {256601} (\bibinfo {year} {2002})}\BibitemShut {NoStop}%
\bibitem [{\citenamefont {Wang}\ \emph
  {et~al.}(2021{\natexlab{a}})\citenamefont {Wang}, \citenamefont {Pan},
  \citenamefont {Ohtsuki}, \citenamefont {Gruzberg},\ and\ \citenamefont
  {Shindou}}]{Wang21b}%
  \BibitemOpen
  \bibfield  {author} {\bibinfo {author} {\bibfnamefont {T.}~\bibnamefont
  {Wang}}, \bibinfo {author} {\bibfnamefont {Z.}~\bibnamefont {Pan}}, \bibinfo
  {author} {\bibfnamefont {T.}~\bibnamefont {Ohtsuki}}, \bibinfo {author}
  {\bibfnamefont {I.~A.}\ \bibnamefont {Gruzberg}},\ and\ \bibinfo {author}
  {\bibfnamefont {R.}~\bibnamefont {Shindou}},\ }\bibfield  {title} {\bibinfo
  {title} {{Multicriticality of two-dimensional class-D disordered topological
  superconductors}},\ }\href {https://doi.org/10.1103/PhysRevB.104.184201}
  {\bibfield  {journal} {\bibinfo  {journal} {Phys. Rev. B}\ }\textbf {\bibinfo
  {volume} {104}},\ \bibinfo {pages} {184201} (\bibinfo {year}
  {2021}{\natexlab{a}})}\BibitemShut {NoStop}%
\bibitem [{\citenamefont {Yoshioka}\ \emph {et~al.}(2018)\citenamefont
  {Yoshioka}, \citenamefont {Akagi},\ and\ \citenamefont
  {Katsura}}]{Yoshioka18}%
  \BibitemOpen
  \bibfield  {author} {\bibinfo {author} {\bibfnamefont {N.}~\bibnamefont
  {Yoshioka}}, \bibinfo {author} {\bibfnamefont {Y.}~\bibnamefont {Akagi}},\
  and\ \bibinfo {author} {\bibfnamefont {H.}~\bibnamefont {Katsura}},\
  }\bibfield  {title} {\bibinfo {title} {Learning disordered topological phases
  by statistical recovery of symmetry},\ }\href
  {https://doi.org/10.1103/PhysRevB.97.205110} {\bibfield  {journal} {\bibinfo
  {journal} {Phys. Rev. B}\ }\textbf {\bibinfo {volume} {97}},\ \bibinfo
  {pages} {205110} (\bibinfo {year} {2018})}\BibitemShut {NoStop}%
\bibitem [{\citenamefont {Luo}\ \emph {et~al.}(2018)\citenamefont {Luo},
  \citenamefont {Xu}, \citenamefont {Ohtsuki},\ and\ \citenamefont
  {Shindou}}]{Luo18QMCT}%
  \BibitemOpen
  \bibfield  {author} {\bibinfo {author} {\bibfnamefont {X.}~\bibnamefont
  {Luo}}, \bibinfo {author} {\bibfnamefont {B.}~\bibnamefont {Xu}}, \bibinfo
  {author} {\bibfnamefont {T.}~\bibnamefont {Ohtsuki}},\ and\ \bibinfo {author}
  {\bibfnamefont {R.}~\bibnamefont {Shindou}},\ }\bibfield  {title} {\bibinfo
  {title} {{Quantum multicriticality in disordered Weyl semimetals}},\ }\href
  {https://doi.org/10.1103/PhysRevB.97.045129} {\bibfield  {journal} {\bibinfo
  {journal} {Phys. Rev. B}\ }\textbf {\bibinfo {volume} {97}},\ \bibinfo
  {pages} {045129} (\bibinfo {year} {2018})}\BibitemShut {NoStop}%
\bibitem [{\citenamefont {Slevin}\ and\ \citenamefont
  {Ohtsuki}(2016)}]{Slevin16}%
  \BibitemOpen
  \bibfield  {author} {\bibinfo {author} {\bibfnamefont {K.}~\bibnamefont
  {Slevin}}\ and\ \bibinfo {author} {\bibfnamefont {T.}~\bibnamefont
  {Ohtsuki}},\ }\bibfield  {title} {\bibinfo {title} {{Estimate of the Critical
  Exponent of the Anderson Transition in the Three and Four-Dimensional Unitary
  Universality Classes}},\ }\href {https://doi.org/10.7566/JPSJ.85.104712}
  {\bibfield  {journal} {\bibinfo  {journal} {J. Phys. Soc. Jpn.}\ }\textbf
  {\bibinfo {volume} {85}},\ \bibinfo {pages} {104712} (\bibinfo {year}
  {2016})}\BibitemShut {NoStop}%
\bibitem [{\citenamefont {Song}\ \emph {et~al.}(2021)\citenamefont {Song},
  \citenamefont {Lian}, \citenamefont {Queiroz}, \citenamefont {Ilan},
  \citenamefont {Bernevig},\ and\ \citenamefont {Stern}}]{song21}%
  \BibitemOpen
  \bibfield  {author} {\bibinfo {author} {\bibfnamefont {Z.-D.}\ \bibnamefont
  {Song}}, \bibinfo {author} {\bibfnamefont {B.}~\bibnamefont {Lian}}, \bibinfo
  {author} {\bibfnamefont {R.}~\bibnamefont {Queiroz}}, \bibinfo {author}
  {\bibfnamefont {R.}~\bibnamefont {Ilan}}, \bibinfo {author} {\bibfnamefont
  {B.~A.}\ \bibnamefont {Bernevig}},\ and\ \bibinfo {author} {\bibfnamefont
  {A.}~\bibnamefont {Stern}},\ }\bibfield  {title} {\bibinfo {title}
  {{Delocalization Transition of a Disordered Axion Insulator}},\ }\href
  {https://doi.org/10.1103/PhysRevLett.127.016602} {\bibfield  {journal}
  {\bibinfo  {journal} {Phys. Rev. Lett.}\ }\textbf {\bibinfo {volume} {127}},\
  \bibinfo {pages} {016602} (\bibinfo {year} {2021})}\BibitemShut {NoStop}%
\bibitem [{\citenamefont {Son}\ and\ \citenamefont {Raghu}(2021)}]{son21}%
  \BibitemOpen
  \bibfield  {author} {\bibinfo {author} {\bibfnamefont {J.~H.}\ \bibnamefont
  {Son}}\ and\ \bibinfo {author} {\bibfnamefont {S.}~\bibnamefont {Raghu}},\
  }\bibfield  {title} {\bibinfo {title} {Three-dimensional network model for
  strong topological insulator transitions},\ }\href
  {https://doi.org/10.1103/PhysRevB.104.125142} {\bibfield  {journal} {\bibinfo
   {journal} {Phys. Rev. B}\ }\textbf {\bibinfo {volume} {104}},\ \bibinfo
  {pages} {125142} (\bibinfo {year} {2021})}\BibitemShut {NoStop}%
\bibitem [{\citenamefont {Asada}\ \emph {et~al.}(2005)\citenamefont {Asada},
  \citenamefont {Slevin},\ and\ \citenamefont {Ohtsuki}}]{Asada05}%
  \BibitemOpen
  \bibfield  {author} {\bibinfo {author} {\bibfnamefont {Y.}~\bibnamefont
  {Asada}}, \bibinfo {author} {\bibfnamefont {K.}~\bibnamefont {Slevin}},\ and\
  \bibinfo {author} {\bibfnamefont {T.}~\bibnamefont {Ohtsuki}},\ }\bibfield
  {title} {\bibinfo {title} {{Anderson Transition in the Three Dimensional
  Symplectic Universality Class}},\ }\href
  {https://doi.org/10.1143/JPSJS.74S.238} {\bibfield  {journal} {\bibinfo
  {journal} {J. Phys. Soc. Jpn.}\ }\textbf {\bibinfo {volume} {74}},\ \bibinfo
  {pages} {238} (\bibinfo {year} {2005})}\BibitemShut {NoStop}%
\bibitem [{\citenamefont {Roy}\ \emph {et~al.}(2017)\citenamefont {Roy},
  \citenamefont {Alavirad},\ and\ \citenamefont {Sau}}]{Roy17}%
  \BibitemOpen
  \bibfield  {author} {\bibinfo {author} {\bibfnamefont {B.}~\bibnamefont
  {Roy}}, \bibinfo {author} {\bibfnamefont {Y.}~\bibnamefont {Alavirad}},\ and\
  \bibinfo {author} {\bibfnamefont {J.~D.}\ \bibnamefont {Sau}},\ }\bibfield
  {title} {\bibinfo {title} {{Global Phase Diagram of a Three-Dimensional Dirty
  Topological Superconductor}},\ }\href
  {https://doi.org/10.1103/PhysRevLett.118.227002} {\bibfield  {journal}
  {\bibinfo  {journal} {Phys. Rev. Lett.}\ }\textbf {\bibinfo {volume} {118}},\
  \bibinfo {pages} {227002} (\bibinfo {year} {2017})}\BibitemShut {NoStop}%
\bibitem [{\citenamefont {Luo}\ \emph {et~al.}(2022)\citenamefont {Luo},
  \citenamefont {Xiao}, \citenamefont {Kawabata}, \citenamefont {Ohtsuki},\
  and\ \citenamefont {Shindou}}]{luo21unifying}%
  \BibitemOpen
  \bibfield  {author} {\bibinfo {author} {\bibfnamefont {X.}~\bibnamefont
  {Luo}}, \bibinfo {author} {\bibfnamefont {Z.}~\bibnamefont {Xiao}}, \bibinfo
  {author} {\bibfnamefont {K.}~\bibnamefont {Kawabata}}, \bibinfo {author}
  {\bibfnamefont {T.}~\bibnamefont {Ohtsuki}},\ and\ \bibinfo {author}
  {\bibfnamefont {R.}~\bibnamefont {Shindou}},\ }\bibfield  {title} {\bibinfo
  {title} {{Unifying the Anderson transitions in Hermitian and non-Hermitian
  systems}},\ }\href {https://doi.org/10.1103/PhysRevResearch.4.L022035}
  {\bibfield  {journal} {\bibinfo  {journal} {Phys. Rev. Research}\ }\textbf
  {\bibinfo {volume} {4}},\ \bibinfo {pages} {L022035} (\bibinfo {year}
  {2022})}\BibitemShut {NoStop}%
\bibitem [{\citenamefont {Crisanti}\ \emph {et~al.}(1993)\citenamefont
  {Crisanti}, \citenamefont {Paladin},\ and\ \citenamefont
  {Vulpiani}}]{crisanti12}%
  \BibitemOpen
  \bibfield  {author} {\bibinfo {author} {\bibfnamefont {A.}~\bibnamefont
  {Crisanti}}, \bibinfo {author} {\bibfnamefont {G.}~\bibnamefont {Paladin}},\
  and\ \bibinfo {author} {\bibfnamefont {A.}~\bibnamefont {Vulpiani}},\ }\href
  {https://doi.org/10.1007/978-3-642-84942-8} {\emph {\bibinfo {title}
  {Products of Random Matrices}}}\ (\bibinfo  {publisher} {Springer},\ \bibinfo
  {address} {Berlin, Heidelberg},\ \bibinfo {year} {1993})\BibitemShut
  {NoStop}%
\bibitem [{\citenamefont {Kawabata}\ \emph {et~al.}(2019)\citenamefont
  {Kawabata}, \citenamefont {Shiozaki}, \citenamefont {Ueda},\ and\
  \citenamefont {Sato}}]{Kawabata19}%
  \BibitemOpen
  \bibfield  {author} {\bibinfo {author} {\bibfnamefont {K.}~\bibnamefont
  {Kawabata}}, \bibinfo {author} {\bibfnamefont {K.}~\bibnamefont {Shiozaki}},
  \bibinfo {author} {\bibfnamefont {M.}~\bibnamefont {Ueda}},\ and\ \bibinfo
  {author} {\bibfnamefont {M.}~\bibnamefont {Sato}},\ }\bibfield  {title}
  {\bibinfo {title} {{Symmetry and Topology in Non-Hermitian Physics}},\ }\href
  {https://doi.org/10.1103/PhysRevX.9.041015} {\bibfield  {journal} {\bibinfo
  {journal} {Phys. Rev. X}\ }\textbf {\bibinfo {volume} {9}},\ \bibinfo {pages}
  {041015} (\bibinfo {year} {2019})}\BibitemShut {NoStop}%
\bibitem [{\citenamefont {Molinari}(2003)}]{molinari2003}%
  \BibitemOpen
  \bibfield  {author} {\bibinfo {author} {\bibfnamefont {L.}~\bibnamefont
  {Molinari}},\ }\bibfield  {title} {\bibinfo {title} {{Spectral duality and
  distribution of exponents for transfer matrices of block-tridiagonal
  Hamiltonians}},\ }\href {https://doi.org/10.1088/0305-4470/36/14/311}
  {\bibfield  {journal} {\bibinfo  {journal} {J. Phys. A}\ }\textbf {\bibinfo
  {volume} {36}},\ \bibinfo {pages} {4081} (\bibinfo {year}
  {2003})}\BibitemShut {NoStop}%
\bibitem [{\citenamefont {Goldhirsch}\ \emph {et~al.}(1987)\citenamefont
  {Goldhirsch}, \citenamefont {Sulem},\ and\ \citenamefont
  {Orszag}}]{goldhirsch87}%
  \BibitemOpen
  \bibfield  {author} {\bibinfo {author} {\bibfnamefont {I.}~\bibnamefont
  {Goldhirsch}}, \bibinfo {author} {\bibfnamefont {P.-L.}\ \bibnamefont
  {Sulem}},\ and\ \bibinfo {author} {\bibfnamefont {S.~A.}\ \bibnamefont
  {Orszag}},\ }\bibfield  {title} {\bibinfo {title} {{Stability and Lyapunov
  stability of dynamical systems: A differential approach and a numerical
  method}},\ }\href {https://doi.org/10.1016/0167-2789(87)90034-0} {\bibfield
  {journal} {\bibinfo  {journal} {Physica D}\ }\textbf {\bibinfo {volume}
  {27}},\ \bibinfo {pages} {311} (\bibinfo {year} {1987})}\BibitemShut
  {NoStop}%
\bibitem [{\citenamefont {Fulga}\ \emph {et~al.}(2012)\citenamefont {Fulga},
  \citenamefont {Akhmerov}, \citenamefont {Tworzyd\l{}o}, \citenamefont
  {B\'eri},\ and\ \citenamefont {Beenakker}}]{Fulga12}%
  \BibitemOpen
  \bibfield  {author} {\bibinfo {author} {\bibfnamefont {I.~C.}\ \bibnamefont
  {Fulga}}, \bibinfo {author} {\bibfnamefont {A.~R.}\ \bibnamefont {Akhmerov}},
  \bibinfo {author} {\bibfnamefont {J.}~\bibnamefont {Tworzyd\l{}o}}, \bibinfo
  {author} {\bibfnamefont {B.}~\bibnamefont {B\'eri}},\ and\ \bibinfo {author}
  {\bibfnamefont {C.~W.~J.}\ \bibnamefont {Beenakker}},\ }\bibfield  {title}
  {\bibinfo {title} {Thermal metal-insulator transition in a helical
  topological superconductor},\ }\href
  {https://doi.org/10.1103/PhysRevB.86.054505} {\bibfield  {journal} {\bibinfo
  {journal} {Phys. Rev. B}\ }\textbf {\bibinfo {volume} {86}},\ \bibinfo
  {pages} {054505} (\bibinfo {year} {2012})}\BibitemShut {NoStop}%
\bibitem [{\citenamefont {Fu}\ \emph {et~al.}(2007)\citenamefont {Fu},
  \citenamefont {Kane},\ and\ \citenamefont {Mele}}]{fu07}%
  \BibitemOpen
  \bibfield  {author} {\bibinfo {author} {\bibfnamefont {L.}~\bibnamefont
  {Fu}}, \bibinfo {author} {\bibfnamefont {C.~L.}\ \bibnamefont {Kane}},\ and\
  \bibinfo {author} {\bibfnamefont {E.~J.}\ \bibnamefont {Mele}},\ }\bibfield
  {title} {\bibinfo {title} {{Topological Insulators in Three Dimensions}},\
  }\href {https://doi.org/10.1103/PhysRevLett.98.106803} {\bibfield  {journal}
  {\bibinfo  {journal} {Phys. Rev. Lett.}\ }\textbf {\bibinfo {volume} {98}},\
  \bibinfo {pages} {106803} (\bibinfo {year} {2007})}\BibitemShut {NoStop}%
\bibitem [{\citenamefont {Markos}(1995)}]{markos1995}%
  \BibitemOpen
  \bibfield  {author} {\bibinfo {author} {\bibfnamefont {P.}~\bibnamefont
  {Markos}},\ }\bibfield  {title} {\bibinfo {title} {Phenomenological theory of
  the metal-insulator transition},\ }\href
  {https://doi.org/10.1088/0953-8984/7/44/006} {\bibfield  {journal} {\bibinfo
  {journal} {J. Phys.: Condens. Matter}\ }\textbf {\bibinfo {volume} {7}},\
  \bibinfo {pages} {8361} (\bibinfo {year} {1995})}\BibitemShut {NoStop}%
\bibitem [{\citenamefont {Hatano}\ and\ \citenamefont
  {Nelson}(1996)}]{Hatano96}%
  \BibitemOpen
  \bibfield  {author} {\bibinfo {author} {\bibfnamefont {N.}~\bibnamefont
  {Hatano}}\ and\ \bibinfo {author} {\bibfnamefont {D.~R.}\ \bibnamefont
  {Nelson}},\ }\bibfield  {title} {\bibinfo {title} {{Localization Transitions
  in Non-Hermitian Quantum Mechanics}},\ }\href
  {https://doi.org/https://doi.org/10.1103/PhysRevLett.77.570} {\bibfield
  {journal} {\bibinfo  {journal} {Phys. Rev. Lett.}\ }\textbf {\bibinfo
  {volume} {77}},\ \bibinfo {pages} {570} (\bibinfo {year} {1996})}\BibitemShut
  {NoStop}%
\bibitem [{\citenamefont {Asada}\ \emph {et~al.}(2004)\citenamefont {Asada},
  \citenamefont {Slevin},\ and\ \citenamefont {Ohtsuki}}]{Asada04}%
  \BibitemOpen
  \bibfield  {author} {\bibinfo {author} {\bibfnamefont {Y.}~\bibnamefont
  {Asada}}, \bibinfo {author} {\bibfnamefont {K.}~\bibnamefont {Slevin}},\ and\
  \bibinfo {author} {\bibfnamefont {T.}~\bibnamefont {Ohtsuki}},\ }\bibfield
  {title} {\bibinfo {title} {Numerical estimation of the $\ensuremath{\beta}$
  function in two-dimensional systems with spin-orbit coupling},\ }\href
  {https://doi.org/10.1103/PhysRevB.70.035115} {\bibfield  {journal} {\bibinfo
  {journal} {Phys. Rev. B}\ }\textbf {\bibinfo {volume} {70}},\ \bibinfo
  {pages} {035115} (\bibinfo {year} {2004})}\BibitemShut {NoStop}%
\bibitem [{\citenamefont {Slevin}\ and\ \citenamefont
  {Ohtsuki}(2014)}]{Slevin14}%
  \BibitemOpen
  \bibfield  {author} {\bibinfo {author} {\bibfnamefont {K.}~\bibnamefont
  {Slevin}}\ and\ \bibinfo {author} {\bibfnamefont {T.}~\bibnamefont
  {Ohtsuki}},\ }\bibfield  {title} {\bibinfo {title} {{Critical exponent for
  the Anderson transition in the three-dimensional orthogonal universality
  class}},\ }\href {http://stacks.iop.org/1367-2630/16/i=1/a=015012} {\bibfield
   {journal} {\bibinfo  {journal} {New J. Phys.}\ }\textbf {\bibinfo {volume}
  {16}},\ \bibinfo {pages} {015012} (\bibinfo {year} {2014})}\BibitemShut
  {NoStop}%
\bibitem [{\citenamefont {Luo}\ \emph {et~al.}(2020)\citenamefont {Luo},
  \citenamefont {Xu}, \citenamefont {Ohtsuki},\ and\ \citenamefont
  {Shindou}}]{Luo20}%
  \BibitemOpen
  \bibfield  {author} {\bibinfo {author} {\bibfnamefont {X.}~\bibnamefont
  {Luo}}, \bibinfo {author} {\bibfnamefont {B.}~\bibnamefont {Xu}}, \bibinfo
  {author} {\bibfnamefont {T.}~\bibnamefont {Ohtsuki}},\ and\ \bibinfo {author}
  {\bibfnamefont {R.}~\bibnamefont {Shindou}},\ }\bibfield  {title} {\bibinfo
  {title} {{Critical behavior of Anderson transitions in three-dimensional
  orthogonal classes with particle-hole symmetries}},\ }\href
  {https://doi.org/10.1103/PhysRevB.101.020202} {\bibfield  {journal} {\bibinfo
   {journal} {Phys. Rev. B}\ }\textbf {\bibinfo {volume} {101}},\ \bibinfo
  {pages} {020202} (\bibinfo {year} {2020})}\BibitemShut {NoStop}%
\bibitem [{\citenamefont {Wang}\ \emph
  {et~al.}(2021{\natexlab{b}})\citenamefont {Wang}, \citenamefont {Ohtsuki},\
  and\ \citenamefont {Shindou}}]{Wang21}%
  \BibitemOpen
  \bibfield  {author} {\bibinfo {author} {\bibfnamefont {T.}~\bibnamefont
  {Wang}}, \bibinfo {author} {\bibfnamefont {T.}~\bibnamefont {Ohtsuki}},\ and\
  \bibinfo {author} {\bibfnamefont {R.}~\bibnamefont {Shindou}},\ }\bibfield
  {title} {\bibinfo {title} {{Universality classes of the Anderson transition
  in the three-dimensional symmetry classes AIII, BDI, C, D, and CI}},\ }\href
  {https://doi.org/10.1103/PhysRevB.104.014206} {\bibfield  {journal} {\bibinfo
   {journal} {Phys. Rev. B}\ }\textbf {\bibinfo {volume} {104}},\ \bibinfo
  {pages} {014206} (\bibinfo {year} {2021}{\natexlab{b}})}\BibitemShut
  {NoStop}%
\bibitem [{\citenamefont {Pendry}\ \emph {et~al.}(1990)\citenamefont {Pendry},
  \citenamefont {MacKinnon},\ and\ \citenamefont {Pretre}}]{pendry90}%
  \BibitemOpen
  \bibfield  {author} {\bibinfo {author} {\bibfnamefont {J.~B.}\ \bibnamefont
  {Pendry}}, \bibinfo {author} {\bibfnamefont {A.}~\bibnamefont {MacKinnon}},\
  and\ \bibinfo {author} {\bibfnamefont {A.~B.}\ \bibnamefont {Pretre}},\
  }\bibfield  {title} {\bibinfo {title} {{Maximal fluctuations---A new
  phenomenon in disordered systems}},\ }\href
  {https://doi.org/10.1016/0378-4371(90)90391-5} {\bibfield  {journal}
  {\bibinfo  {journal} {Physica A}\ }\textbf {\bibinfo {volume} {168}},\
  \bibinfo {pages} {400} (\bibinfo {year} {1990})}\BibitemShut {NoStop}%
\bibitem [{\citenamefont {Kramer}\ \emph {et~al.}(2005)\citenamefont {Kramer},
  \citenamefont {Ohtsuki},\ and\ \citenamefont {Kettemann}}]{kramer05}%
  \BibitemOpen
  \bibfield  {author} {\bibinfo {author} {\bibfnamefont {B.}~\bibnamefont
  {Kramer}}, \bibinfo {author} {\bibfnamefont {T.}~\bibnamefont {Ohtsuki}},\
  and\ \bibinfo {author} {\bibfnamefont {S.}~\bibnamefont {Kettemann}},\
  }\bibfield  {title} {\bibinfo {title} {Random network models and quantum
  phase transitions in two dimensions},\ }\href
  {https://doi.org/10.1016/j.physrep.2005.07.001} {\bibfield  {journal}
  {\bibinfo  {journal} {Phys. Rep.}\ }\textbf {\bibinfo {volume} {417}},\
  \bibinfo {pages} {211} (\bibinfo {year} {2005})}\BibitemShut {NoStop}%
\end{thebibliography}%
	
	 \clearpage
	 \begin{widetext}
	 	\section{Supplemental Material for \\
``Topological Anderson Transitions in Chiral Symmetry Classes"}

This Supplemental Material is organized as follows. 
In Sec.~\ref{supplement-sec-CE}, we summarize known critical exponents between metal and topological-insulator phases and between metal and ordinary-insulator phases.
In Sec.~\ref{supplement-sec-IPR}, we introduce the inverse participation ratio along different directions and prove that coupling among low-dimensional systems makes wave functions more extended even in the small coupling limit. 
In Sec.~\ref{supplement-sec-fitting}, we review the polynomial fitting of the finite-size scaling function and details of 
Table~I 
in the main text. 
In Sec.~\ref{supplement-sec-transfer}, we introduce the transfer matrix method and explain properties of transfer matrices for chiral-symmetric Hamiltonians. 
In Sec.~\ref{supplement-sec-Lyapunov}, we summarize a relation between weak topological indices and distributions of Lyapunov exponents (LEs). 
In Sec.~\ref{sec_stat}, we show how statistical symmetries require the weak topological indices to be zero 
and 
LEs of the non-Hermitian matrix (right upper part of a chiral-symmetric Hamiltonian) to 
come in opposite-sign pairs.
In Sec.~\ref{supplement-sec-Lyapunov-scaling}, we summarize a scaling form for the maximal and minimal LEs within a continuum spectrum 
and show numerical fittings based on this scaling form. 
In Secs.~\ref{supplement-sec-Q1D-topo} and \ref{supplement-sec-Q1D-triv}, we provide detailed numerical studies of the criticality in chiral-symmetric 
models with and without non-trivial 
topological indices, respectively. 
In Sec.~\ref{conductance}, we provide detailed numerical results of the two-terminal conductance of the model with non-trivial topological indices. 
The conductance shows the anisotropic transport behavior in the quasi-localized phase.

\tableofcontents
\clearpage

\subsection{Summary of known critical exponents }
 \label{supplement-sec-CE}

Table~\ref{CE_all} summarizes known results of critical exponents of the Anderson transitions between metal and topological-insulator phases and between metal and trivial-insulator phases in the same spatial dimensions and symmetry classes. 
For each symmetry class and spatial dimensions, the evaluated critical exponents for the two types of 
the Anderson
transitions are consistent with each other. 
However, the critical exponents between 
topological-insulator and trivial-insulator 
phases, as well as those between topological-semimetal and diffusive-metal phases, can be different from the ones in Table~\ref{CE_all}.
In our work, we focus on the Anderson transitions
in the 3D chiral classes and demonstrate the different critical exponents due to the weak topological indices.

\begin{table}[h]
    \centering
    \caption{Correlation-length critical exponents of the Anderson transitions between metal and topological-insulator phases 
    and those between metal and trivial-insulator phases 
    in the same symmetry classes and spatial dimensions.
    }
    \begin{tabular}{cc|cc}
    \hline \hline
     & 
     ~Class~
     & 
     ~~Metal-topological-insulator~~
     & 
     ~~Metal-trivial-insulator~~
     \\
     \hline
     2D & AII & $2.74 \pm 0.12$~\cite{Obuse07} & $2.73\pm 0.02$\footnote[1]{95\% confidence interval.}~\cite{Asada02} \\
      & D & $ 1.371[1.311, 1.437]$\footnotemark[1]~\cite{Wang21b} & $ 1.348[1.279, 1.402]$\footnotemark[1]~\cite{Wang21b} \\
     & DIII & $1.5 \pm 0.1$~\cite{Yoshioka18} & $1.5\pm 0.1$~\cite{Yoshioka18} \\
     \hline
     3D & A & $1.34[1.23,1.53]$\footnotemark[1]~\cite{Luo18QMCT}\footnote[2]{Layered Chern insulator.} & $1.443 [1.437, 1.449]$\footnotemark[1]~\cite{Slevin16} \\
      &  &  $1.42\pm0.12$~\cite{song21}\footnote[3]{Axion insulator.} &  \\
      & AII & $1.311 \pm 0.033$\footnotemark[1]~\cite{son21} \footnote[4]{Reference~\cite{son21} obtained the critical exponent between the metal and topological-insulator phases in a three-dimensional network model belonging to symmetry class AII. 
The critical exponent is different from the one between the metal and trivial-insulator phases in the same symmetry class and spatial dimensions obtained by the SU(2) model~\cite{Asada05}.  
However, a more careful error analysis is needed, because system sizes in Ref.~\cite{son21} 
may not be large enough ($\leq 10$), and the difference between the two exponents is small.
}
    & $1.375 \pm 0.016$\footnotemark[1]~\cite{Asada05}\\
      & DIII  & $0.85 \pm 0.05$~\cite{Roy17} & $0.903[0.896, 0.908]$\footnotemark[1]~\cite{luo21unifying} \\
    \hline
    \end{tabular}
    \label{CE_all}
\end{table}

\subsection{Inverse participation ratio}
    \label{supplement-sec-IPR}

\subsubsection{Inverse participation ratio along different directions}

The inverse participation ratio $P_2$ measures localization properties of a wave function $\Phi( \bm r)$ in $d$ dimension, defined by 
\begin{equation}
    P_2 \equiv \sum_{\bm r} |\Phi( \bm r)|^4
\end{equation}
with the normalization condition $\sum_{\bm r} |\Phi( \bm r)|^2  = 1$.
We have $P_2 \leq 1$, where the equality holds only when $\Phi( \bm r)$ is fully localized at one lattice site. 
$P_2$ in extended and localized phases show the different scaling relations with 
the system size $L$, 
\begin{equation}
    P_2 \sim \begin{cases}
        L^{-d} & [\Phi( \bm r) \text{ is extended}] \, , \\
         {\rm constant}<1 & [\Phi( \bm r) \text{ is localized}] \, 
    \end{cases}
\end{equation}
for $L \to \infty$.
The one-dimensional inverse participation ratio $P^\mu_2$ ($\mu=x,y,z$) measures 
localization properties of $\Phi(x,y,z)$ along the $\mu$ direction. 
The integrated weight of the wave function at $z$,  
$|\phi(z)|^2 \equiv \sum_{x,y} |\Phi( x,y,z)|^2$, is regarded as 
the squared one-dimensional normalized wave function along the $z$ direction and describes 
how the three-dimensional wave function $\Phi(x,y,z)$ is localized along the $z$ direction. 
Thus, the inverse participation ratio $P_2^{z}$ along the $z$ direction 
is introduced as
\begin{equation}
    P_2^z = \sum_z |\phi(z)|^4 \, \quad, \quad
     |\phi(z)|^2 = \sum_{x,y} |\Phi( x,y,z)|^2 \, .
\end{equation}
Notably, $P_2^{z}$ provides an upper bound of $P_2$,
\begin{equation}
     P_2   = \sum_z |\phi(z) |^4 \left[ \sum_{x,y} \left \lvert  \frac{\Phi({\bm r}) }{\phi(z)}  \right \rvert^4 \right]
    \leq P_2^{z},
\end{equation}
where the equality holds only when we have $ \sum_{x,y} \left \lvert  \Phi({\bm r})/\phi(z)  \right \rvert^4 = 1$ for all $z$. 
The one-dimensional inverse participation ratio $P_2^{\mu}$ along the other two directions 
($\mu=x,y$) is defined in the same manner. 
In a similar manner, the $d$-dimensional inverse participation ratio is defined for a normalized wave function in $d^{\prime}$ dimension ($d<d^{\prime}$). 
For example, the following two-dimensional inverse participation ratio $P^{(x,y)}_2$ measures the localization properties of $\Phi(x,y,z)$ within 
the $xy$ plane, 
\begin{equation}
    P_2^{(x,y)} \equiv \sum_{x,y} | \phi(x,y)|^4 \,,  
    \quad | \phi(x,y)|^2 = \sum_z  |\Phi( x,y,z)|^2 \, ,
\end{equation}
which satisfies 
\begin{equation}
    P_2 \leq P_2^{(x,y)} \,.
\end{equation}

\subsubsection{Wave-function hybridization and inverse participation ratio} 
Suppose that a $d'$-dimensional
disordered non-interacting Hamiltonian ${\cal H}^{\bm R}$ in ${\bm R} \equiv ({\bm r},{\bm s})$ with ${\bm r}=(r_1,\cdots,r_d)$ and ${\bm s}=(s_1,\cdots, s_{d'-d})$ ($d<d'$) consists of $d$-dimensional 
Hamiltonians ${\cal H}^{\bm r}_{\bm s}$ at different ${\bm s}$ and coupling ${\cal H}^{\prime}$ among the $d$-dimensional systems. 
On-site disorder potential $V({\bm R})$
is chosen to
distribute 
uniformly in the range $[-W/2,W/2]$ for all the lattice sites ${\bm R}$. 
Then, ${\bm s}$ can be regarded as different disorder realizations from 
the same ensemble for the $d$-dimensional system with the disorder strength $W$.
In this section, we show that even in the small coupling limit, an eigenstate $\Phi({\bm R})$ of ${\cal H}^{\bm R}$ with eigenenergy $E$ is more extended along 
the ${\bm r}$ direction than an eigenstate $\psi({\bm r})$ of ${\cal H}^{\bm r}_{\bm s}$ with the same eigenenergy.
Here, the small coupling limit means that the maximal eigenvalue of ${\cal H}^{\prime}$ is much smaller than the mean level spacing of ${\cal H}^{\bm r}_{\bm s}$ around $E$. 

For ${\cal H}^{\prime} = 0$, eigenstates of ${\cal H}^{\bm R}$ are 
given by eigenstates of ${\cal H}_{\bm s}^{\bm r}$.
In the small coupling limit, we can treat ${\cal H}^{\prime}$ perturbatively. 
We introduce an energy window $[E - \Delta E, E + \Delta E]$ and choose $\Delta E$ to be small enough 
that each ${\cal H}^{\bm r}_{\bm s}$ has at most one eigenstate $\psi_{\bm s}({\bm r})$ with eigenenergy $E_{\bm s}$ in the energy window and that we have
$|E_{\bm s}-E|\ll \Delta E$.  
In the small coupling limit, the maximal eigenvalue of ${\cal H}^{\prime}$ 
can be much smaller than $\Delta E$.
Thus, in the lowest order of degenerate perturbation theory, $ {\cal H}^{\prime}$ does not mix unperturbed eigenstates inside the energy window with those outside the energy window, and an eigenstate $\Phi({\bm R})$ of ${\cal H}^{\bm R}$  
is given by a linear superposition of $\psi_{\bm s}({\bm r})$ over different ${\bm s}$,
\begin{equation}
    \Phi({\bm r},{\bm s}) = \sum_{{\bm s}^{\prime}} a_{{\bm s}^{\prime}} \psi_{\bm s^{\prime}}({\bm r}) \delta_{{\bm s},{\bm s}^{\prime}} 
    = a_{{\bm s}} \psi_{\bm s}({\bm r})
    \, ,
\end{equation}
where $ \delta_{\bm s, \bm s^{\prime}} $ is the Kronecker delta. 
Here, we 
impose
the normalization conditions $\sum_{\bm s} |a_{\bm s}|^2 = 1$ and $\sum_{\bm r} |\psi_{\bm s}({\bm r}) |^2 = 1$, where we sum only over such ${\bm s}$ that
${\cal H}^{\bm r}_{\bm s}$ has an eigenenergy inside the window $[E - \Delta E, E + \Delta E]$. 
$a_{\bm s}$ is the ${\bm s}$-component of an eigenstate of an 
effective Hamiltonian $H^{\bm R}$ given as
\begin{align}
({H}^{\bm R})_{{\bm s},{\bm s}^{\prime}} = 
\sum_{\bm r}\sum_{{\bm r}^{\prime}} \!\  
\psi^{*}_{\bm s}({\bm r}) \!\ ({\cal H}^{\prime})_{{\bm r},{\bm r}^{\prime}} \!\  
\psi_{{\bm s}^{\prime}}({\bm r}^{\prime}) + E_s \delta_{\bm s, \bm s^{\prime}} . 
\end{align}

In the following, we show that the inverse participation ratio $P_2^{\bm r}$ of $\Phi({\bm R})$ along the ${\bm r}$ direction is always smaller than the $d$-dimensional inverse participation 
ratio of $\psi_{\bm s}({\bm r})$. 
The weight $|\phi({\bm r}_0)|^2$ of 
the wave function $\Phi({\bm R})$ on a hyperplane ${\bm r} = {\bm r}_0$ is given as  
\begin{equation}
    |\phi({\bm r})|^2 = \sum_{\bm s} \left \lvert  \Phi({\bm r},{\bm s}) \right \rvert^2 \, .
\end{equation}
The inverse participation ratio $P_2^{\bm r}$ of $\Phi({\bm R})$ along the ${\bm r}$ direction measures the localization properties of $\Phi({\bm R})$ within the ${\bm r}$ direction and is 
given by the sum of the square of the weight over ${\bm r}$,
\begin{equation}
\begin{aligned}
    P_2^{\bm r} &= \sum_{\bm r} |\phi({\bm r})|^4 
    = \sum_{\bm r} \sum_{\bm s_1} \sum_{\bm s_2} \left \lvert  \Phi({\bm r},{\bm s_1}) \right \rvert^2   \left \lvert  \Phi({\bm r},{\bm s_2}) \right \rvert^2  
      = \sum_{\bm s_1} \sum_{\bm s_2}  |a_{{\bm s_1}}|^2 |a_{{\bm s_2}}|^2 \left[ \sum_{\bm r} |\psi_{\bm s_1}({\bm r})|^2 |\psi_{\bm s_2}({\bm r})|^2 \right] \\ 
     &  \leq \frac{1}{2} \sum_{\bm s_1} \sum_{\bm s_2}  |a_{{\bm s_1}}|^2 |a_{{\bm s_2}}|^2 \sum_{\bm r}  \left[   |\psi_{\bm s_1}({\bm r})|^4 +  |\psi_{\bm s_2}({\bm r})|^4 
    \right] =  \sum_{\bm s} |a_{{\bm s}}|^2 \sum_{\bm r}|\psi_{\bm s}({\bm r})|^4\,  .
\end{aligned}
\end{equation}
Here, the equality holds true only when we have $\psi_{\bm s_1}({\bm r})= \psi_{\bm s_2}({\bm r})$ for all $\bm r$, ${\bm s}_1$, and ${\bm s}_2$. 
Notably, $\sum_{\bm r} |\psi_{\bm s}({\bm r})|^4$ is the $d$-dimensional inverse participation ratio $P_2^{\psi_{\bm s}(\bm r)}$ of $\psi_{\bm s}(\bm r)$, and 
${\cal H}^{\bm r}_{\bm s}$ at different ${\bm s}$ belongs to the same ensemble with the same disorder strength $W$.
In the thermodynamic limit 
($N_{\bm r} \equiv \sum_{\bm r} \rightarrow \infty$),
$\Delta E$ goes to zero as the mean level spacing goes to zero, and $P_2^{\psi_{\bm s}(\bm r)}$ at different ${\bm s}$ takes the same value $P_2^{\psi_(\bm r)}$.
Then, we have
\begin{align}
    P^{\bm r}_{2} \leq P^{\psi({\bm r})}_2 \sum_{\bm s}  |a_{{\bm s}}|^2 
    = P^{\psi({\bm r})}_2,
\end{align}
which proves that within the 
lowest order in 
${\cal H}^{\prime}$, the small coupling ${\cal H}^{\prime}$ 
among the $d$-dimensional systems always makes $d$-dimensional 
wave functions spatially more extended.

\subsection{Polynomial fitting 
}
    \label{supplement-sec-fitting}

In this section, we present more details about the polynomial fitting [Eq.~(7) in the main text] of the normalized localization length $\Lambda_x(W,L) = \xi_x(W,L)/L$ and show details of Table~I in the main text (see Table~\ref{fitting_table_SM_0}). 
The scaling function for 
$\Lambda_x(W,L)$ is Taylor-expanded with respect to the relevant 
scaling variable $\phi(w)$ and the least irrelevant scaling 
variable $\psi(w)$ up to the $n$th order and first order, respectively, 
\begin{align}
\Lambda_x(W,L) = \sum^n_{i=0} \sum^1_{j=0} a_{i,j} 
\big(\phi(w)L^{1/\nu}\big)^i \big(\psi(w) L^{-y}\big)^j, \label{fit1}
\end{align}
with $w \equiv (W-W^{(x)}_c)/W^{(x)}_c$, and the scaling dimension $-y$ $(<0)$ of the least irrelevant scaling variable around a saddle-point fixed point. 
The relevant scaling variable is further expanded around $w=0$ up to the 
$m$th order, while only the 
zeroth-order
in $w$ is kept for the irrelevant scaling variable $\psi(w)$,
\begin{align}
    \phi(w) = \sum^m_{k=1} b_k w^k,\quad
    \psi(w) = c. \label{fit2}
\end{align}
Here, $\{W^{(x)}_c,\nu,y,a_{i,j},b_k,c\}$ are the fitting parameters. 
To avoid the ambiguity in the Taylor expansion of the scaling function, we should set $a_{0,1} = a_{1,0} = 1$. Thus, the number $N_f$ of the free parameters in the fitting is $N_f = 2(n + 1) + m + 2$. 
We minimize $\chi^2$ statistics 
\begin{equation}
    \chi^2 = \sum_{j = 1}^{N_D} \left( \frac{F_j - \Lambda_{j}}{ \sigma_j}\right)^2 \, , 
\end{equation}
where $\Lambda_j$ and $\sigma_j$ are the normalized localization length and its standard deviation for $(W,L)$ evaluated by the transfer matrix method, respectively, 
$F_j$ is the value of the polynomial fitting function for $(W,L)$,
and $N_D$ is the number of data points. 
The confidence error bars for the optimal parameters are determined by the fittings for 1000 sets of synthetic data for $\Lambda_x(W,L)$. 
The synthetic data are generated according to a standard deviation from the transfer matrix calculation. 

\begin{table*}[bt]
    \centering
    \caption{
    Polynomial fitting results of the normalized localization length 
    $\Lambda_{\mu} \equiv \xi_{\mu}/L$ along the $\mu$ direction ($\mu = x, y, z$) around critical points of different models with the quasi-one-dimensional geometry $L\times L\times L_{\mu}$. 
    ``$\surd$'' in the column ``topology'' shows that the weak topological index $\nu_z$ is non-zero around the critical point, and ``$\times$'' shows that all the weak topological indices always vanish around the 
    critical point. 
    We show the critical disorder strength $W^{(\mu)}_c$, critical exponent $\nu$, scaling dimension $-y$ of the least irrelevant scaling variable, critical localization length $\Lambda_c$, the goodness of fitting (GOF), and Taylor-expansion order of $(m,n)$ in Eqs.~(\ref{fit1}) and (\ref{fit2}). The square brackets denote the 95\% confidence interval.
    Note that $\Lambda_c$'s here are critical values in the presence of anisotropic spatial geometry and take non-universal values.}
    \begin{tabular}{c|c|c|cc|c|cccccc}
    \hline \hline
    symmetry class& topology & direction &$m$ & $n$ & GOF  &$W^{(\mu)}_c$& $\nu$ & $y$ & $\Lambda_c$ \\ \hline
    BDI & $\surd$   & $\mu=x$  & 2 & 3 & 0.15 & 27.241[27.194,27.303] & 0.820[0.783,0.846] &  2.584[2.175,2.955] & 0.134[0.130,0.137] \\
    BDI & $\surd$
     & $\mu=x$  & 3 & 3 & 0.14 &27.243[27.192,27.301] & 0.820[0.787,0.848] &  2.574[2.212,2.947] & 0.134[0.130,0.138] \\
    AIII  & $\surd$  & $\mu=x$ & 3 & 3 & 0.47 & 9.143[9.125,9.168] & 0.824[0.776,0.862] &  2.157[1.727,2.519] & 0.225[0.213,0.232] \\ \hline
    BDI & $\times$ & $\mu=z$ &  2 & 3 & 0.19 & 23.220[23.167,23.293] & 1.089[1.005,1.128] &  1.926[1.074,3.034] & 0.374[0.352,0.385] \\
    BDI & $\times$ & $\mu=z$ & 3 & 3 & 0.18 & 23.223[23.138,23.409] & 1.088[0.991,1.141] &  1.906[0.604,3.677] & 0.373[0.302,0.389] \\
   BDI & $\times$ & $\mu=x$ & 2 & 3 & 0.23 & 23.170[23.098,23.279] & 1.042[0.943,1.099] &  1.591[0.889,2.543] & 0.281[0.254,0.293] \\
   BDI & $\times$  & $\mu=x$  & 3 & 3  & 0.31 & 23.167[23.101,23.310] & 1.039[0.937,1.100] &  1.607[0.753,2.425] & 0.281[0.239,0.292] \\
    AIII& $\times$ & $\mu=z$ & 2 & 3 & 0.20 & 8.091[8.074,8.096] & 1.024[0.973,1.070] 
    &  0.470[0.450,1.481] &   0.650[0.639,0.706]    \\  
    \hline
    \end{tabular}
    \label{fitting_table_SM_0}
\end{table*}

\subsection{Transfer matrix, Lyapunov exponents, and localization length}
    \label{supplement-sec-transfer}

The transfer matrix method solves an eigenvalue problem of a non-interacting disordered Hamiltonian $\cal{H}$ recursively. 
This method is efficient for obtaining 
the localization length along one spatial direction, which we call the $\mu$ direction in the following.
In this formulation, the Hamiltonian 
is decomposed into a layer structure along the $\mu$ direction,  
\begin{equation}
    \mathcal{H}_{i,j} = H_{i} \delta_{i,j} + V_{i,i+1} \delta_{i,j-1} +  V_{i,i-1} \delta_{i,j+1} \, , 
\end{equation}
where $i,j = 1,2,\cdots,L_{\mu}$ 
are indices of the layers, 
$H_i$ is a block of matrix elements within the $i$th layer, and 
$V_{i,i\pm 1}$ is a block of matrix elements between the 
$i$th layer and the $(i\pm 1)$th layer. 
The decomposition assumes that matrix elements appear 
only between the nearest neighboring layers or within each layer. 
In the presence of next-nearest hopping,
one can redefine two neighboring layers as one layer. 
Let $H_i$, $V_{i,i\pm 1}$ be $m$ by $m$ matrices and 
$(\cdots, A_{i-1}, A_{i}, A_{i+1},\cdots)^{T}$
be an eigenvector of ${\cal H}$ 
for an eigenenergy $E$:   
\begin{equation}
    H_i A_{i} + V_{i,i-1} A_{i-1} + V_{i,i+1} A_{i+1} = E A_{i}.
    \label{TM}
\end{equation}
For simplicity, suppose that the disorder terms are present only in 
the diagonal matrix elements and that $V_{i,i-1} = V_+$ 
and $V_{i,i+1} = V_-$ are free from disorder.
The eigenvectors are solved layer by layer recursively 
by a transfer matrix $M_i$,  
\begin{equation}
    \left( \begin{matrix}
    A_{i+1} \\ A_{i}
    \end{matrix} \right) = 
 M_i
    \left( \begin{matrix}
    A_{i} \\ A_{i-1}
    \end{matrix} \right) , \!\ \!\  \, 
    M_i \equiv  
     \left( 
    \begin{matrix}
    -V_-^{ -1}( H_{i}-E) &  -V_-^{ -1} V_+ \\
    1_{m \times m} & 0_{m \times m} \\
    \end{matrix}
    \right) \, .
    \label{TMv0}
\end{equation}
The product of the transfer matrices,
$M = M_{L_{\mu}} M_{L_{\mu} - 1} \cdots M_1$, relates the components of the eigenvector at 
the $(L_{\mu}+1)$th and $L_{\mu}$th layers with the components 
at the first and zeroth layer. 
According to Oseledec's theorem~\cite{crisanti12},  
the matrix
\begin{equation}
P(E) = \lim_{L_{\mu} \rightarrow \infty} \ln{(M^{\dagger}M)^{\frac{1}{2L_{\mu}}}} 
= - \lim_{L_{\mu} \rightarrow \infty} \ln{(M^{-1}M^{-1 \dagger})^{\frac{1}{2L_{\mu}}}}    
\end{equation}
well converges in the limit $L_{\mu} \to \infty$. 
Eigenvalues of $P(E)$ 
are known as Lyapunov exponents (LEs). 
If $\mathcal{H}$ is Hermitian, LEs 
come in opposite-sign pairs~\cite{crisanti12}. 
The inverse of the smallest positive or the largest negative LE corresponds to the localization length $\xi_{\mu}$ along the $\mu$ direction.

\subsubsection{Transfer matrix of a chiral-symmetric Hamiltonian}

Suppose that a $2n\times 2n$ Hermitian Hamiltonian ${\cal H}$ satisfies chiral symmetry $\mathcal{C} {\cal H} \mathcal{C}^{-1} = -{\cal H}$ 
with a chiral operator $\mathcal{C}$ satisfying $\mathcal{C}^2 = 1$. 
Eigenvalues of $\mathcal{C}$ are $\pm 1$, the numbers of which are assumed to be the same.
Then, the unitary matrix ${\cal C}$ is diagonalized as $\mathcal{C} = \sum_{i=1}^{n} \ket{v_i} \bra{v_i} - \sum_{i =  1}^{n} \ket{u_i} \bra{u_i}$. 
Here, $\ket{v_1},\cdots , \ket{v_n}$ and $\ket{u_{1}},\cdots , \ket{u_{n}}$ are eigenvectors of $\mathcal{C}$ with eigenvalues $+1$ and $-1$, respectively.
Because of chiral symmetry, 
we have $\braket{v_i|{\cal H}|v_{j}}=\braket{u_i|{\cal H}|u_{j}}=0$. 
Thus, the $2n \times 2n$ matrix ${\cal H}$ is decomposed into 
two $n \times n$ matrices $h$ and $h^{\prime}$ in the off-diagonal parts, 
\begin{align}
    {\cal H} = \left(\begin{array}{cc} 
    0 & h \\
    h^{\prime} & 0 \\
    \end{array}\right),  \label{chiral0}
\end{align}
with
\begin{align}
(h)_{i,j} = \braket{v_i|{\cal H}|u_{j}},\quad
(h^{\prime})_{i,j} = \braket{u_{i}|{H}|v_{j}}, \label{relation}
\end{align}
satisfying ${h}^{\prime} = {h}^{\dagger}$. 

Equation~(\ref{relation}) does not determine 
${h}$ uniquely up to $n \times n$ unitary transformations,   
$h \rightarrow {\cal V}^{\dagger}{h}{\cal U}$, where  
the unitary transformations ${\cal V}$ and ${\cal U}$ 
change bases among the $n$-fold degenerate eigenstates 
of $\mathcal{C}$.  Nonetheless, with a certain choice of the bases 
for $\ket{v_1},\cdots , \ket{v_n}$ and $\ket{u_{1}},\cdots , \ket{u_{n}}$, 
any Hermitian Hamiltonian with chiral symmetry can be decomposed into the 
off-diagonal form as Eq.~(\ref{chiral0}).
The off-diagonal parts 
thus introduced are non-Hermitian matrices, in general.

If the chiral operator ${\cal C}$ is diagonal with respect to the 
layer index and its matrix elements do not depend on 
the layer index, the Hamiltonian $H_i$ within the $i$th layer
and the hopping matrix $V_{\pm}$ between the $i$th layer and the 
$(i\pm 1)$th layers also take the block off-diagonal structure,   
\begin{equation}
    H_i = \begin{pmatrix}
        0 & \tilde{h}_i   \\
        \tilde{h}_i^{\dagger} & 0
    \end{pmatrix} ,\quad 
    V_{+} =  \begin{pmatrix}
        0 & v_{+} \\
        v_-^{\dagger} & 0
    \end{pmatrix} ,\quad
    V_{-} =  \begin{pmatrix}
        0 & v_{-} \\
        v_+^{\dagger} & 0 
    \end{pmatrix},
    \label{layer-chiral}
\end{equation}
where $v_{\pm}$ are free from the disorder and independent of the layer index. 
The transfer matrix $M^{(\cal H)}_i$ of ${\cal H}$ for zero energy $E = 0$ reads
\begin{equation}
    \begin{aligned}
    M^{(\cal H)}_i& = \begin{pmatrix}
        -\begin{pmatrix}
        0 & v_-\\
        v_+^{\dagger} & 0
        \end{pmatrix}^{ -1}
        \begin{pmatrix}
        0 & \tilde{h}_{i} \\
       \tilde{h}_{i}^{\dagger} & 0
        \end{pmatrix}
        &  -  \begin{pmatrix}
        0 & v_-\\
        v_+^{\dagger} & 0
        \end{pmatrix}^{ -1}   \begin{pmatrix}
        0 & v_+\\
        v_-^{\dagger} & 0
        \end{pmatrix} \\
         \begin{matrix}
         1 & 0 \\
         0 & 1 \\
         \end{matrix}
         & \begin{matrix}
         0 & 0 \\
         0 & 0 \\
         \end{matrix}
        \end{pmatrix} \\
    & = \begin{pmatrix}
        -v_+^{\dagger -1} \tilde{ h}_{i}^{\dagger} & 0 & -v_+^{\dagger -1} v_-^{\dagger} & 0  \\
        0 & -v_-^{ -1} \tilde{h}_i & 0 & -v_-^{ -1} v_+  \\
        1 & 0 & 0 & 0 \\
        0 & 1 & 0 & 0 \\
        \end{pmatrix} \, .
    \end{aligned}
\end{equation}
With a proper unitary transformation ${\cal U}$, the transfer matrix 
$M_i^{({\cal H})}$ is block-diagonalized into $M_i$ and $M^{\prime}_i$, 
\begin{align}
    {\cal U}^{\dagger} M_i^{(\cal H)} {\cal U} = \begin{pmatrix}
    M_i^{\prime} & 0 \\
    0 & M_i \\
    \end{pmatrix},  \!\ \!\ M_i^{\prime} \equiv  
    \begin{pmatrix}
        -v_+^{\dagger -1} \tilde{ h}_{i}^{\dagger} & -v_+^{\dagger -1} v_-^{\dagger}   \\
    1 & 0 \\
    \end{pmatrix}, \!\ \!\
     M_i \equiv  
    \begin{pmatrix}
        -v_-^{ -1} \tilde{h}_i &  -v_-^{ -1} v_+ \\
    1 & 0 \\
    \end{pmatrix}. \label{TMv1}
\end{align}
Notably, $M_i$ and $M^{\prime}_i$ are the transfer matrices of the 
right-upper part $h$ and left-lower part $h^{\dagger}$ of the Hamiltonian ${\cal H}$ in Eq.~(\ref{chiral0}), respectively. 
In the canonical basis of Eqs.~(\ref{chiral0}) and (\ref{layer-chiral}), 
$h$ and $h' = h^{\dag}$ are
decomposed into the 
layer structure along the $\mu$ direction,  
\begin{align}
    h_{i,j} &= \tilde{ h}_{i} \delta_{i,j} + v_{+} \delta_{i,j-1} +  v_{-} \delta_{i,j+1}, \nonumber \\
    h^{\prime}_{i,j} &= \tilde{h}^{\dagger}_{i} \delta_{i,j} + v^{\dagger}_{-} \delta_{i,j-1} +  
    v^{\dagger}_{+} \delta_{i,j+1}, \nonumber
\end{align} 
where $i,j=1,\cdots,L_{\mu}$ are the indices of layers. 
From Eqs.~(\ref{TM}) and (\ref{TMv0}), 
we obtain $M_i$ of $h$ and $M^{\prime}_i$ of $h^{\prime}$ as in Eq.~(\ref{TMv1}). 
Note that $M_i^{\prime}$ is equivalent to $( M_i^{\dagger} )^{-1}$ under a certain transformation,
\begin{equation}
    S \equiv \begin{pmatrix}
    0 & -v_-^{\dagger} \\
    v_+^{\dagger}& 0 \\
    \end{pmatrix}
    ,\quad 
    S M_i^{\prime} S^{ -1} = (M_i^{\dagger})^{-1} = 
    \begin{pmatrix}
    0 & -v_-^{\dagger} v_+^{\dagger -1} \\
    1 & - \tilde{h}_{i}^{\dagger} v_+^{\dagger -1}
    \end{pmatrix} \, . \label{MtoM^prime}
\end{equation} 
Thus, the LEs obtained by the product of $S M_i^{\prime} S^{ -1}$ 
have signs opposite to
the LEs obtained by the product of $M_i$. 
The non-singular similarity transformation $S$ 
does not
change LEs. 
Thereby,
the LEs of $h$
and the LEs of $h^{\prime}=h^{\dagger}$ come in opposite-sign pairs. 
The LEs of ${\cal H}$ are the sum of the LEs of 
$h$ and the LEs of $h^{\prime}$.

\subsubsection{Transfer matrix of the nodal-line semimetal model}
\label{TM-nl}
The Hamiltonian of the nodal-line semimetal model reads
\begin{equation}
    {\cal H} = \sum_{\bm r=(r_x,r_y,r_z)} \left\{ (\Delta + \epsilon_{\bm{r}}) c^{\dagger}_{\bm{r}} \sigma_z c_{\bm{r}} + \left[  \sum_{\mu = x,y}\left(  t_{\perp}c^{\dagger}_{\bm r+\bm{e_{\mu}}} \sigma_z c_{\bm{r}}  \right)  -{\rm i} t_{\|}   c^{\dagger}_{\bm r+\bm{ e_z}}  \sigma_y c_{\bm{r}} + t_{\|}^{\prime}   c^{\dagger}_{\bm r+\bm{ e_z}}  \sigma_z c_{\bm{r}} +\text{H.c.} \right]  \right\} \, ,
    \label{nodal-line-h}
\end{equation}
where $c_{\bm{r}}$ is a two-component annihilation operator on 
the cubic lattice site ${\bm r}$, $\sigma_{\mu} (\mu = x, y,z)$ are the Pauli matrices, 
$\Delta$, $t_{\perp}$, $t_{\|}$, and $t_{\|}^{\prime}$ are real-valued parameters,  
$\epsilon_{\bm{r}}$ is a random potential that distributes uniformly in $[-W/2,W/2]$, 
and ${\bm e}_x=(1,0,0)$, ${\bm e}_y=(0,1,0)$, and ${\bm e}_z=(0,0,1)$ are the unit vectors. 
Note that the Hamiltonian in Eq.~(\ref{nodal-line-h}) reduces to Eq.~(5) in the main text for $\Delta = 0$. 
Depending on $\Delta$ and the other parameters, Eq.~(\ref{nodal-line-h}) describes an ordinary insulator, topological insulator, and nodal-line semimetal [see Eq.~(\ref{hk})].
${\cal H}$ satisfies time-reversal symmetry ${\cal H} = {\cal H}^*$ 
and chiral symmetry ${\cal H} =-\mathcal{C}^{\dagger} {\cal H}^{\dagger} \mathcal{C}$ with a unitary operator $\mathcal{C}_{ {\bm r , \bm r^{\prime}}} = \delta_{{\bm r , \bm r^{\prime}}} \sigma_x$
with $\mathcal{C}^{*} = \mathcal{C}$, and thus belongs to the chiral orthogonal class (class BDI).

The chiral operator ${\cal C}$ has eigenvalues $+1$ and $-1$. Since ${\cal C}$ is diagonal with 
respect to the lattice site, eigenvectors of ${\cal C}$ can be labelled by the cubic-lattice 
site ${\bm s}\equiv (s_x,s_y,s_z)$:
 \begin{align}
    \langle {\bm r}| v_{\bm s}\rangle = \delta_{{\bm r},{\bm s}}  \frac{1}{\sqrt{2}}\left(\begin{array}{c}
    1 \\
    1  \\
    \end{array}\right),\quad
    \langle {\bm r}| u_{\bm s}\rangle = \delta_{{\bm r},{\bm s}} \frac{1}{\sqrt{2}}\left(\begin{array}{c}
    1 \\
    -1  \\
    \end{array}\right),
    \textcolor{blue}{}\end{align}
satisfying ${\cal C}|v_{\bm s}\rangle = |v_{\bm s}\rangle$ and 
${\cal C}|u_{\bm s}\rangle = - |u_{\bm s}\rangle$. Following Eq.~(\ref{relation}), 
we construct the right-upper part $h$ of $\cal H$ on the same cubic lattice as, 
\begin{align}
    (h)_{{\bm s},{\bm s}} &= \braket{v_{\bm s}|{\cal H}|u_{\bm s}}  = \Delta + \epsilon_{\bm s}, \\
    (h)_{{\bm s}+{\bm e}_{\mu},{\bm s}} 
    = (h)_{{\bm s},{\bm s}+{\bm e}_{\mu}} &= 
    \braket{v_{{\bm s} + {\bm e}_{\mu}}|{\cal H}|u_{\bm s}}  =
     \braket{v_{\bm s}|{\cal H}|u_{{\bm s} + {\bm e}_{\mu}}} = t_{\perp}, \\
    (h)_{{\bm s}+{\bm e}_{z},{\bm s}} &= 
    \braket{v_{{\bm s} + {\bm e}_{z}}|{\cal H}|u_{\bm s}}  =  
     t_{\|}  + t_{\|}^{\prime},  \\
    (h)_{{\bm s},{\bm s}+{\bm e}_{z}} &= 
    \braket{v_{\bm s}|{\cal H}|u_{{\bm s}+ {\bm e}_{z}}}  =  
     - t_{\|}  + t_{\|}^{\prime},   
\end{align}
for $\mu=x,y$. All the other matrix elements of $h$ are zero. 
Notably, $h$ can be regarded as a single-orbital tight-binding model, 
\begin{align}
h&= \sum_{\bm r=(r_x,r_y,r_z)} \left[ (\Delta + \epsilon_{\bm{r}}) f^{\dagger}_{\bm{r}} f_{\bm{r}} + \sum_{\mu = x,y}  \left(  t_{\perp}f^{\dagger}_{{\bm r}+{\bm e}_{\mu}}  f_{\bm r} + \text{H.c.}  \right) + (t^{\prime}_{\|} + t_{\|})f^{\dagger}_{{\bm r}+{\bm e}_{z}}  f_{\bm r} + (t^{\prime}_{\|} - t_{\|})f^{\dagger}_{\bm r}  f_{{\bm r}+{\bm e}_{z}} \right] \, , \label{NDSM-NH}
\end{align}
where $f_{\bm{r}}$ and $f^{\dagger}_{\bm r}$ are annihilation and creation operators 
at site ${\bm r}$. 
While $h$ respects $h = h^*$, we have $h \neq h^{\dagger}$ for $t_{\|} \ne 0$.
Hence, $h$ generally belongs to the non-Hermitian symmetry class AI~\cite{Kawabata19, luo21unifying}. 

The transfer matrix of $h$ along 
the $z$ direction is given by 
\begin{equation}
     M_i = \left( 
    \begin{matrix}
    - \frac{1}{t^{\prime}_{\|} - t_{\|}}  \tilde{h}_i & - \frac{t^{\prime}_{\|} + t_{\|}}{t^{\prime}_{\|} - t_{\|}}1_{m \times m}\\
    1_{m \times m} & 0_{m \times m} \\
    \end{matrix}
    \right)  \, , \label{Mi-h-z}
\end{equation} 
where $m= L^2$ is the degrees of freedom in each layer and the quasi-1D geometry 
($L\times L\times L_z$, $L_z\gg L$) is considered. $\tilde{h}_i$ is the 
Hamiltonian within the $i$th layer, which has $\Delta + \epsilon_{\bm s}$ in its 
diagonal elements and $t_{\perp}$ 
in its nearest-neighbor hopping.

For $t^{\prime}_{\|} - t_{\|} = 0$, $M_i$ is singular, and $m$ eigenvalues of 
$ \frac{1}{L_{z}}\ln M \equiv\frac{1}{L_z}\ln (M_{L_z}M_{L_z-1}\cdots M_1)$ diverge to $\infty$. 
In fact, $M^{-1}_i$ has zero eigenvalues with multiplicity $m$ for $t^{\prime}_{\|} = t_{\|}$, 
\begin{equation}
    M_i^{-1} = \begin{pmatrix}
        0_{m \times m} & 1_{m \times m} \\
        -\frac{t^{\prime}_{\|} - t_{\|}}{t^{\prime}_{\|} + t_{\|}}1_{m \times m} & -\frac{1}{t^{\prime}_{\|} + t_{\|}} \tilde{h}_i \\ 
    \end{pmatrix} \, \rightarrow \begin{pmatrix}
        0_{m \times m} & 1_{m \times m} \\
        0_{m \times m} & -\frac{1}{t^{\prime}_{\|} + t_{\|}} \tilde{h}_i \\ 
    \end{pmatrix}\quad ( t^{\prime}_{\|} - t_{\|}  \to 0),
    \label{TM-nl-1}
\end{equation}
where $M^{-1} = M^{-1}_{1}\cdots M^{-1}_{L_z-1} M^{-1}_{L_z}$ has zero eigenvalue with multiplicity at least $m$. 
Therefore, $m$ eigenvalues of $\frac{1}{2L_{z}}\ln (M^{-1})^{\dagger} M^{-1}$ always diverge to $- \infty$, while $m$ eigenvalues of $\frac{1}{2L_{z}}\ln M^{\dagger} M$ always diverge to $+\infty$. 
The other $m$ finite-valued LEs of $h$ are determined from the following product: 
\begin{align}
    p & \equiv - \lim_{L_z \rightarrow \infty} \frac{1}{2L_z}
    \ln{\left( \frac{1}{t^{\prime}_{\|} + t_{\|}} \tilde{h}_1 \cdots 
    \frac{1}{t^{\prime}_{\|} + t_{\|}} \tilde{h}_{L_z}
    \frac{1}{t^{\prime}_{\|} + t_{\|}} \tilde{h}_{L_z}^{\dagger}
    \cdots
    \frac{1}{t^{\prime}_{\|} + t_{\|}} \tilde{h}_1^{\dagger} \right)}, \nonumber \\ 
      &= \lim_{L_z \rightarrow \infty} \frac{1}{2L_z} 
      \ln{\left( (t^{\prime}_{\|} + t_{\|}) \tilde{h}_1^{-1\dagger} \cdots 
      (t^{\prime}_{\|} + t_{\|}) \tilde{h}_{L_z}^{-1 \dagger}
      (t^{\prime}_{\|} + t_{\|}) \tilde{h}_{L_z}^{-1}
      \cdots
      (t^{\prime}_{\|} + t_{\|}) \tilde{h}_1^{-1} \right)}.
\end{align}

\subsection{Weak topological indices and Lyapunov exponents}
    \label{supplement-sec-Lyapunov}

We summarize a relationship between the weak topological indices of chiral-symmetric Hamiltonians and the numbers of positive and negative LEs of 
its right-upper part $h$ in the canonical basis in Eq.~(\ref{chiral0})~\cite{molinari2003}. Consider a chiral-symmetric 
Hamiltonian ${\cal H}(\phi)$, in which a magnetic flux $\phi$ is inserted through 
a closed loop along the $\mu$ direction. 
Similarly to ${\cal H}$ in Eq.~(\ref{chiral0}),
${\cal H}(\phi)$ takes a 
block off-diagonal structure in a basis where the chiral operator 
is diagonal,
\begin{align}
{\cal H} (\phi)=
\begin{pmatrix}
0& h(\phi)\\
h(\phi)^{\dagger}&0\\
\end{pmatrix}.\label{chiral}
\end{align}
The right-upper block $h(\phi)$ is decomposed into a layer structure along the 
$\mu$ direction, 
\begin{equation}
    h(\phi) = 
    \begin{pmatrix}
    \tilde{h}_1 & v_- & 0 & \cdots & 0  & \frac{1}{z} v_+  \\
    v_+ & \tilde{h}_2 & v_- & \cdots & \cdots  & 0 \\
    0 & v_+ & \tilde{h}_3 & v_-  & \cdots  & 0 \\
    \vdots & \vdots & \ddots & \ddots & \ddots & \vdots \\
    z v_- & 0 & \cdots  & 0 & v_+ & \tilde{h}_{L_{\mu}} \\
    \end{pmatrix} \, ,
\end{equation}
with $z = e^{{\rm i} \phi}$. 
We assume that the hopping appears only between the 
nearest neighboring layers or within each layer. 
In the presence of 
next-nearest neighbor hopping, we can redefine two neighboring 
layers as one layer. 
$v_-$, $v_+$, and $\tilde{h}_{i} (i=1,2,\cdots,L_{\mu})$ are 
$m \times m$ matrices, where $m$ is the degrees of freedom of $h$ in each layer. 
The 1D winding number $w_{\mu}$ along the $\mu$ direction is defined in terms 
of $h(\phi)$,  
\begin{align}
    w_{\mu} \equiv {\rm i} \int^{2\pi}_{0}  \frac{d\phi}{2\pi } 
    \partial_{\phi} {\rm Tr}\big[\log\big[h(\phi)]\big] \, . 
\label{topological} 
\end{align}
The winding number $w_{\mu}$ is given by the contour integral 
\begin{equation}
\begin{aligned}
    w_{\mu} & = {\rm i} \oint_{|z| = 1}   \frac{dz }{2\pi } \partial_z{\rm Tr}\big[\log\big[h(z)]\big] \\
    & = {\rm i} \oint_{|z| = 1}  \frac{dz }{2\pi } \partial_z \log \left[ \det \left[ h(z)\right] \right]. \\
\end{aligned}
\end{equation}
Here, $\det \left[ h(z)\right]$ is a polynomial function in terms of $z$ with the lowest order $z^{-m}$ and the highest order $z^m$ if we have $\det{v_-} \neq 0$ and $\det{v_+} \neq 0$. 
For simplicity, we assume $\det{v_-} \neq 0$ and $\det{v_+} \neq 0$ while the following argument can be generalized to other cases. 
Then, $z^m \det \left[ h(z)\right]$ is an analytic function of $z$, and $w_{\mu}$ is related to the number of zeros of $z^m \det \left[ h(z)\right]$ within the circle $|z|=1$, 
\begin{equation}
\begin{aligned}
    w_{\mu} & = {\rm i} \oint_{|z| = 1}  \frac{dz }{2\pi } \partial_z \log \left[z^{-m} z^m \det \left[ h(z)\right] \right] \\
    & =m + {\rm i}\oint_{|z| = 1}  \frac{dz }{2\pi } \partial_z \log \left[ z^m \det \left[ h(z)\right] \right] \\
    & = m + {\rm i} \oint_{|z| = 1}  \frac{dz }{2\pi } \frac{\partial_z \left[ z^m \det \left[ h(z)\right]\right]} {z^m \det \left[ h(z)\right] } \\
    & = m - Z, 
    \label{winding}
\end{aligned}
\end{equation}
where $Z$ is the weighted number of the zeros, and the residue theorem is used in the last equality. $z = 0$ should not be a zero of $z^m \det \left[ h(z)\right]$, since 
the lowest order of $\det \left[ h(z)\right]$ is $z^{-m}$.
Thus, $Z$ is equal to the weighted number of the zeros of $\det \left[ h(z)\right]$ in the disk $|z| < 1$. 

The number of the zeros of $\det \left[ h(z)\right]$ is determined by 
the LEs of $h$ along the $\mu$ direction. 
For $\det \left[h(z)\right]=0$, 
$h(z)$ has a zero mode. 
The presence of the zero modes is given by the
transfer matrices of $h(z)$ along the $\mu$ direction.
The transfer matrices for each layer are given by,  
\begin{equation}
    M_i(z) = 
    \begin{cases}
        \begin{pmatrix} 
            -v_-^{ -1} \tilde{h}_1 &  -\frac{1}{z }v_-^{ -1} v_+ \\
        1 & 0 \\
        \end{pmatrix} &  (i = 1), \\
        \begin{pmatrix}
            -v_-^{ -1} \tilde{h}_i &  -v_-^{ -1} v_+ \\
        1 & 0 \\
        \end{pmatrix} = M_i &  (i = 2,3,\cdots,L_{\mu}-1), \\
        \begin{pmatrix}
            -\frac{1}{z}v_-^{ -1} \tilde{h}_{L_{\mu}} &  -\frac{1}{z }v_-^{ -1} v_+ \\
            1 & 0 \\
            \end{pmatrix} &  (i = L_{\mu}). \\
    \end{cases}
\end{equation}
Here, $M_1(z)$ and $M_{L_{\mu}}(z)$ satisfy 
\begin{equation}
    M_1(z) M_{L_{\mu}}(z) = 
    \begin{pmatrix}
    \frac{1}{z} v_-^{-1} \tilde{h}_1 v_-^{-1}  \tilde{h}_{L_{\mu}} -\frac{1}{z} v_-^{-1}v_+ &  \frac{1}{z} v_-^{-1} H_1 v_-^{-1}v_+ \\
     -\frac{1}{z}v_-^{-1} \tilde{h}_{L_{\mu}} & -\frac{1}{z}v_-^{-1}v_+ \\
    \end{pmatrix} = \frac{1}{z} M_1(z = 1) M_{L_{\mu}}(z=1)\, .
\end{equation}
Suppose that $(A_1,A_2,\cdots,A_{L_{\mu}})^{T}$ is a zero mode of $h(z)$ under the periodic boundary conditions.
Then, we have
\begin{equation}
    \begin{pmatrix}
    A_{2} \\ A_1 
    \end{pmatrix} = M_1(z)  \begin{pmatrix} A_{1} \\ A_{L_{\mu}} \end{pmatrix} , \quad
     \begin{pmatrix}
    A_{3} \\ A_2 
    \end{pmatrix} = M_2  \begin{pmatrix} A_{2} \\ A_{1} \end{pmatrix} , 
    \quad \cdots , \quad
    \begin{pmatrix}
    A_{1} \\ A_{L_{\mu}} 
    \end{pmatrix} = M_{L_{\mu}}(z) \begin{pmatrix} A_{L_{\mu}} \\ A_{L_{\mu}-1} \end{pmatrix},
\end{equation}
and hence  
\begin{equation}
    \begin{aligned}
    \begin{pmatrix}
    A_{L_{\mu}} \\ A_{L_{\mu}-1} 
    \end{pmatrix}  & =  M_{L_{\mu}-1} \cdots M_{2} M_{1 }(z) M_{L_{\mu}}(z) \begin{pmatrix}
    A_{L_{\mu}} \\ A_{L_{\mu}-1}
    \end{pmatrix}  \equiv  \frac{1}{z} M \begin{pmatrix}
    A_{L_{\mu}} \\ A_{L_{\mu}-1}
    \end{pmatrix},
    \end{aligned}
    \label{condition1}
\end{equation}
where $M \equiv M_{L_{\mu}-1} \cdots M_{2} M_{1 }(z = 1) M_{L_{\mu}}(z = 1)$  
is the product of the transfer matrices without the magnetic flux ($\phi=0$, $z=1$). 
Since 
$M_{1}(z=1)$ and $M_{L_{\mu}}(z=1)$ are statistically equivalent to 
$M_{L_{\mu}-1}$, $M_{L_{\mu}-2}$, $\cdots$, $M_3$, and $M_2$, 
the eigenvalues of $\frac{1}{L_{\mu}}\ln M$ in the limit $L_{\mu}\rightarrow \infty$ are characterized by the LEs of $h$. 

If $h(z)$ has a zero mode for a complex value $z$, 
$M$ has an eigenvalue of $z$ from Eq.~(\ref{condition1}), 
and vice versa. 
Thus, the number $Z$ of 
the zeros of $\det \left[h(z)\right]$ within the 
disk
$|z|<1$ is equivalent to the number of eigenvalues of $M$ whose absolute values are smaller than $1$. 
The product $M$ of $L_{\mu}$ random matrices
has eigenvalues $e^{\alpha_j  + {\rm i} \beta_j} (j= 1,2,\cdots,2m; \alpha_j, \beta_j \in \mathbb{R})$,  
and generally, $\alpha_j$ grows linearly with $L_{\mu}$, satisfying 
\begin{equation}
    \gamma_j = \lim_{L_{\mu} \rightarrow \infty}  \frac{\alpha_j}{L_{\mu}} \, ,
\end{equation}
with the LE $\gamma_j$ of 
$M$~\cite{molinari2003,crisanti12,goldhirsch87}.
Thus, $Z$
is also the same as 
the number $N_{-}$ of the negative LEs of $h$:
\begin{equation}
    Z = N_- . \,
\end{equation}
In terms of Eq.~(\ref{winding}), the 1D winding number is given by 
\begin{equation}
    \begin{aligned}
        w_{\mu} & = m -  N_-  = \frac{1}{2} ( N_+ - N_-),
    \end{aligned}
    \label{winding_LE}
\end{equation}
where $N_+$ and $N_-$ are the numbers of positive and negative LEs, 
satisfying $N_+ + N_- = 2m$. 
The weak topological index along the $\mu$ direction is  
the 1D winding number normalized by the degrees of freedom of $h$ in each layer,   
\begin{align}
    \nu_{\mu} = \frac{1}{m} w_{\mu} = \frac{1}{2m} (N_{+}-N_{-}). \label{weakTI_LE}
\end{align}
Notably, if a LE is exactly zero, the localization length $\xi_{\mu}$ 
along the $\mu$ direction diverges and the winding number $w_{\mu}$ is 
ill defined. 
In the quasi-1D geometry of a 3D disordered Hamiltonian $(L\times L\times L_{\mu}, L_{\mu} \gg L)$, $L^2$ LEs are distributed within a finite range and form a continuous spectrum in the thermodynamic limit $L\rightarrow \infty$. 
When the spectrum crosses zero with changing $W$, $w_{\mu}$ also changes 
from an integer to another integer,
$\nu_{\mu}$ continuously changes with $W$, 
and the localization length always diverges.

\subsubsection{Winding number in the clean limit}

When a $d$-dimensional system has
translation invariance
in all the $d$-dimensional 
coordinates, Eq.~(\ref{topological}) reduces to 
\begin{align}
w_{\mu} = {\rm i} \int_0^{2 \pi}  \frac{d k_{\mu}}{2\pi } 
\!\  \partial_{k_{\mu}} 
{\rm Tr} \big[\log \left[h({\bm k})\right]\big] \, , 
\end{align}
with the momentum ${\bm k} \equiv (k_1,k_2,\cdots,k_{d-1},k_{d})$ and the Bloch Hamiltonian $h({\bm k})$.
The trace includes the sum over momenta 
along the directions complementary to the $\mu$ direction.  
For example, 
the 
Bloch 
Hamiltonian for the 3D nodal-line semimetal model is given 
by the two-by-two matrix 
\begin{equation}
    {\cal H}(\bm{k}) =  [\Delta + 2t_{\perp} \left( \cos k_x + \cos k_y \right) + 2t_{\|}^{\prime} \cos k_z ] \sigma_z - 2t_{\|}\sin k_z \sigma_y .
\end{equation}
In the canonical basis 
where the chiral operator is diagonal, the matrix takes the block off-diagonal 
structure with
\begin{equation}
    h({\bm k})= 2{\rm i} \!\ t_{\|} \sin k_z  -
[\Delta + 2t_{\perp} \left( \cos k_x + \cos k_y \right) + 2t_{\|}^{\prime} \cos k_z ].
\label{hk}
\end{equation}
The complex number
$h({\bm k})$
winds around zero when $k_z$ 
changes from $0$ to $2\pi$. 
For $|\Delta + 2t_{\perp} \left( \cos k_x + \cos k_y \right)| > 2 |t_{\|}^{\prime}|$, 
${\cal H}$ has an energy gap around $E=0$ and $h({\bm k})$ winds around zero  
clockwise for all $k_x$ and $k_y$ in the first Brillouin zone, leading to $w_z=L^2$ and 
$\nu_z=1$. 
Here, $L^2$ is the system size within the 
$xy$ plane.
For $|\Delta + 2t_{\perp} \left( \cos k_x + \cos k_y \right)| < 2 |t_{\|}^{\prime}|$,
${\cal H}$ has a gap at $E=0$, but $h({\bm k})$ does not wind around zero  
for any $k_x$ and $k_y$, 
leading to $w_z=\nu_z=0$. 
For $2|t_{\|}^{\prime}| - 4t_{\perp} < \Delta < 2|t_{\|}^{\prime}| +4t_{\perp}$, zero modes of ${\cal H}$ 
form a nodal ring 
in momentum space, 
and the winding number is 
$+1$ ($0$) for the wave numbers $k_x,k_y$ inside (outside) the nodal 
ring, leading to $0 < \nu_z < 1$.

\subsection{Statistical symmetry}
    \label{sec_stat}
    
An ensemble of disordered Hamiltonians, as a whole, can be invariant under a symmetry operation even if each disorder realization in the ensemble breaks the symmetry. 
Such symmetry of the ensemble is dubbed statistical symmetry~\cite{Fulga12}.
Statistical symmetry does not influence the symmetry class of the 
Hamiltonians since it is not a symmetry of each disordered Hamiltonian. 
An example of statistical 
symmetry is translation symmetry in 3D weak topological 
insulators with disorder~\cite{fu07}.

Statistical symmetry can make the ensemble averages of physical 
observables or topological indices be zero. A prime example is zero 
Hall conductance due to statistical time-reversal symmetry. 
Suppose that ${\cal H}_{\alpha}$ is a 
time-reversal-breaking
Hamiltonian in an 
ensemble with statistical 
time-reversal symmetry
and 
has a finite Hall conductance 
$\sigma_{xy}^{\alpha}$. 
A time-reversed counterpart 
${\cal H}_{\bar{\alpha}}$ of ${\cal H}_{\alpha}$ exists in 
the same ensemble and has the opposite value $-\sigma_{xy}^{\alpha}$ of the Hall 
conductance. 
Such an ensemble has zero Hall conductance on average,  
\begin{equation}
    \langle  \sigma_{xy} \rangle =  \frac{1}{2N_{\rm sample}}  \sum_{\alpha} \left( \sigma_{xy}^{\alpha} -\sigma_{xy}^{\alpha} \right) = 0 \,.
\end{equation}

\subsubsection{Statistical symmetry, Lyapunov exponents, and one-dimensional winding number}

In Sec.~\ref{supplement-sec-Lyapunov},
the 1D winding numbers and weak topological 
indices are defined in terms of the right-upper part $h$ of the chiral-symmetric 
Hamiltonian ${\cal H}$ in the canonical basis 
[see Eq.~(\ref{chiral0})].
Now, we 
introduce statistical symmetry of $h$ and show that it
requires LEs of $h$ to come in opposite-sign pairs. Statistical symmetry also makes the 1D winding numbers and weak topological indices 
be zero as a whole. 
Suppose that an ensemble of $h$ with different disorder realizations, 
$\big\{ \!\ h \!\ \big| \!\ \epsilon_{\bm r} \in [-W/2,W/2] \big\}$,  
is symmetric under transposition of $h$ together with a certain 
unitary transformation ${\cal U}$:
\begin{align}
    \Big\{ \!\ h \!\ \Big| \!\ \epsilon_{\bm r} \in \big[-W/2,W/2\big] \Big\} = \Big\{ \!\ h^{\prime} \!\ \Big| \!\ \epsilon_{\bm r} \in \big[-W/2,W/2\big] \Big\}, \quad 
    {\rm with} \quad h^{\prime}\equiv {\cal U} h^T {\cal U}^{\dagger}. 
    \label{statistical-symmetry}
\end{align}
Here, we assume that the unitary transformation  ${\cal U}$ is diagonal in a spatial coordinate 
$r_{\mu}$ and is independent of $r_{\mu}$ while it can be non-diagonal in the other 
coordinates ${\bm s}$, 
\begin{align}
    ({\cal U})_{{\bm r},{\bm r}^{\prime}} = 
    \delta_{r_{\mu},r^{\prime}_{\mu}}   (u)_{{\bm s},{\bm s}^{\prime}},  \label{unitary}
\end{align}
with ${\bm r} \equiv ({\bm s},r_{\mu})$ and 
${\bm r}^{\prime} \equiv ({\bm s}^{\prime},r^{\prime}_{\mu})$. 
Then, LEs of $h$ along the $\mu$ direction 
come in opposite-sign pairs.
The 1D winding number and weak topological index of $h$ along the $\mu$ direction vanish from Eqs.~(\ref{winding_LE}) and (\ref{weakTI_LE}). 

To see this, we decompose $h$ into a 
layer structure along the 
$\mu$ direction,  
\begin{equation}
    (h)_{i,j} = \tilde{h}_{i} \delta_{i,j} + v_{+} \delta_{i,j-1} +  v_{-} \delta_{i,j+1} \, ,
\end{equation}
with $i,j=1,\cdots,L_{\mu}$. $\tilde{h}_i$ is a block of matrix elements of $h$ 
within the $i$th layer, and $v_{\pm}$ is a block of matrix elements of $h$ between 
the $i$th layer and $(i\mp 1)$th layer. $v_{\pm}$ are free from disorder and 
independent of the layer index. If the degree of freedom in each layer of 
$h$ is $m$, $\tilde{h}_i$ and $v_{\pm}$ are $m \times m$ matrices. 
Similarly, $h^{\prime} \equiv {\cal U}h^{T} {\cal U}^{\dagger}$ is 
also decomposed into a 
layer structure along the $\mu$ direction, 
\begin{equation}
    (h^{\prime})_{i,j} = u \tilde{h}_{i}^{T} u^{\dagger} \delta_{i,j} +u v^T_{-}u^{\dagger} \delta_{i,j-1} + u v^T_{+} u^{\dagger} \delta_{i,j+1} \, .
\end{equation}
From Eqs.~(\ref{TM}) and (\ref{TMv0}), 
the transfer matrices of $h$ and $h^{\prime}$ are obtained as
\begin{align}
    M_i = \left(\begin{array}{cc}
    -(v_{-})^{-1} \tilde{h}_i & - (v_{-})^{-1} v_{+} \\
    1 & 0 \\
    \end{array}\right), \ \ \ M^{\prime}_i = \left(\begin{array}{cc}
    -u(v^T_{+})^{-1} \tilde{h}^T_i u^{\dagger} & - u(v^T_{+})^{-1} v^T_{-} u^{\dagger} \\
    1 & 0 \\
    \end{array}\right). 
\end{align}
The two matrices are related to each other by the following symmetry, 
\begin{align}
S {M^{\prime}_i}^T S^{-1} = M^{-1}_i,  \label{M-Mprime}
\end{align}
with
\begin{align}
    S \equiv \left(\begin{array}{cc}
     0 & - (v_{-})^{-1} u^T \\
     (v_{+})^{-1} u^T & 0 \\
    \end{array}\right). 
\end{align}
Since $u$ and $v_{\pm}$ in $S$ are independent of the layer index, 
the same symmetry holds between the products of the 
transfer matrices, $M \equiv M_{L_{\mu}}\cdots M_1$ and 
$M^{\prime}\equiv M^{\prime}_{L_{\mu}}\cdots M^{\prime}_1$,   
\begin{align}
S {M^{\prime}}^T S^{-1} = M^{-1}.  \label{M-Mprime-2}
\end{align}
Since a non-singular similarity transformation $S$ does not change LEs, 
eigenvalues of 
$P^{\prime}(0) \equiv \lim_{L_{\mu \rightarrow \infty}} \frac{1}{2L_{\mu}}\ln ({M^{\prime}}^{\dagger} M^{\prime})$ are 
opposite to eigenvalues of $P(0) \equiv  \lim_{L_{\mu \rightarrow \infty}}
\frac{1}{2L_{\mu}} \ln (M^{\dagger} M)$. 
Since $h$ and $h^{\prime}$ 
are in the same ensemble, $M_i$ and $M^{\prime}_{j}$ ($i,j=1,\cdots, L_{\mu}$) 
are random matrices with the same possibility distribution. 
Thus, according to Oseledec's   
theorem~\cite{crisanti12}, the eigenvalues of $P(0)$ and $P^{\prime}(0)$ converge to the same 
values
in the 
limit $L_{\mu} \rightarrow \infty$. 
Then, the eigenvalues of $P(0)$, 
as well as the LEs of $h$,
must come in opposite-sign pairs
in the limit $L_{\mu}\rightarrow \infty$. 
Because of $N_{+}=N_{-}$ for the LEs 
along the $\mu$ direction, $\nu_{\mu}$ and $w_{\mu}$ of $h$ vanish
from Eqs.~(\ref{winding_LE}) and (\ref{weakTI_LE}). 

As discussed in Sec.~\ref{supplement-sec-transfer}, the LEs of $h$ and the LEs of $h^{\dagger}$ generally come in opposite-sign pairs. Thus, 
statistical Hermitian-conjugation symmetry of $h$ also requires 
the LEs of $h$ 
to come in the opposite-sign pairs. 
Suppose that an ensemble of $h$ with different disorder realizations, 
$\big\{ \!\ h \!\ \big| \!\ \epsilon_{\bm r} \in [-W/2,W/2] \big\}$,  
is symmetric under Hermitian conjugation together with a certain 
unitary transformation ${\cal U}$ defined in Eq.~(\ref{unitary}):
\begin{align}
    \Big\{ \!\ h \!\ \Big| \!\ \epsilon_{\bm r} \in \big[-W/2,W/2\big] \Big\} = \Big\{ \!\ h^{\prime} \!\ \Big| \!\ \epsilon_{\bm r} \in \big[-W/2,W/2\big] \Big\}, \quad 
    {\rm with} \quad h^{\prime}\equiv {\cal U} h^{\dagger} {\cal U}^{\dagger}. 
    \label{statistical-symmetry-2}
\end{align}
Then, the LEs of $h$ along the $\mu$ direction 
should come in opposite-sign pairs, 
leading to $\nu_{\mu} = w_{\mu} = 0$. 

It is also notable that statistical symmetry of $h$ leads to statistical symmetry of ${\cal H}$. 
If an ensemble of $h$ is invariant under the combination of transposition and a unitary operation in Eq.~(\ref{statistical-symmetry}), the corresponding ensemble of ${\cal H}$ is invariant under the combination of 
time reversal 
and a unitary operation:
\begin{equation}
    \Big\{ \!\  {\cal H} = \begin{pmatrix}
       0 & h \\
       h^{\dagger} & 0
   \end{pmatrix} \!\ \Big| \!\ \epsilon_{\bm r} \in \big[-W/2,W/2\big] \Big\} = \Big\{ \!\ {\cal H}^{\prime} = \begin{pmatrix}
       0 & h^{\prime} \\
       h^{\prime \dagger} & 0 \end{pmatrix} \!\ \Big| \!\ \epsilon_{\bm r} \in \big[-W/2,W/2\big] \Big\}, \quad 
    {\cal H}^{\prime}\equiv \begin{pmatrix}
       0 & {\cal U} \\
       {\cal U} & 0
   \end{pmatrix} {\cal H}^{*} \begin{pmatrix}
       0 & {\cal U}^{\dagger} \\
       {\cal U}^{\dagger} & 0
   \end{pmatrix} . 
\end{equation}
If an ensemble of $h$ is invariant under the combination of Hermitian conjugation and a unitary transformation in Eq.~(\ref{statistical-symmetry-2}), the corresponding ensemble of ${\cal H}$ is invariant under the following unitary operation:
\begin{equation}
    \Big\{ \!\  {\cal H} = \begin{pmatrix}
       0 & h \\
       h^{\dagger} & 0
   \end{pmatrix} \!\ \Big| \!\ \epsilon_{\bm r} \in \big[-W/2,W/2\big] \Big\} = \Big\{ \!\ {\cal H}^{\prime} = \begin{pmatrix}
       0 & h^{\prime} \\
       h^{\prime \dagger} & 0 \end{pmatrix} \!\ \Big| \!\ \epsilon_{\bm r} \in \big[-W/2,W/2\big] \Big\}, \quad 
    {\cal H}^{\prime}\equiv \begin{pmatrix}
       0 & {\cal U} \\
       {\cal U} & 0
   \end{pmatrix} {\cal H} \begin{pmatrix}
       0 & {\cal U}^{\dagger} \\
       {\cal U}^{\dagger} & 0
   \end{pmatrix} . 
\end{equation}


\subsubsection{Statistical symmetry in the nodal-line semimetal model}

We show that the 3D nodal-line semimetal model has
statistical transposition symmetries for 
the $\mu=x,y$ directions. 
The nodal-line semimetal model takes a block off-diagonal structure 
in the canonical basis, where the upper-right part $h$ is given by Eq.~(\ref{NDSM-NH}). 
Transposition of $h$ exchanges $t^{\prime}_{\|}+t_{\|}$ and $t^{\prime}_{\|}-t_{\|}$ 
in Eq.~(\ref{NDSM-NH}). 
Since the disorder potential $\epsilon_{\bm r}$ is 
statistically equivalent for different lattice points ${\bm r}$, we can introduce  
a mirror operation with respect to the $xy$ plane as a unitary transformation of 
Eqs.~(\ref{statistical-symmetry}) and (\ref{unitary}), 
\begin{align}
    {\cal U}_{(r_x,r_y,r_z|r^{\prime}_x,r^{\prime}_y,r^{\prime}_z)} = 
    \delta_{r_x,r^{\prime}_x} \delta_{r_y,r^{\prime}_y} \delta_{r_z,-r^{\prime}_z}.  \label{mirror}
\end{align}
This mirror operation exchanges $t^{\prime}_{\|}+t_{\|}$ and $t^{\prime}_{\|}-t_{\|}$ 
as well as $\epsilon_{r_x,r_y,r_z}$ and $\epsilon_{r_x,r_y,-r_z}$
while $\epsilon_{{\bm r}}$ is statistically 
equivalent for different ${\bm r}$. Thus, an ensemble for $h$ defined by 
Eq.~(\ref{NDSM-NH}) is statistically 
invariant under transposition with the unitary transformation. The 
symmetry of Eqs.~(\ref{statistical-symmetry}) and (\ref{mirror}) 
requires the LEs of $h$ along the $x$ ($y$) direction 
to come in opposite-sign pairs,
leading to $w_{x(y)} = \nu_{x(y)} = 0$.

\subsection{Finite-size scaling of Lyapunov exponents}
    \label{supplement-sec-Lyapunov-scaling}

The LEs of a chiral-symmetric Hamiltonian ${\cal H}$ are the sum of LEs of the right-upper part $h$ 
of ${\cal H}$ in the canonical basis and 
their opposite-sign exponents 
[see Eq.~(\ref{chiral0})].
In the quasi-one-dimensional (quasi-1D) 
geometry, the LEs of ${\cal H}$, as well as the LEs of $h$, comprise continuum spectra for the limit $L \to \infty$~\cite{markos1995}. 
In the nodal-line semimetal model ${\cal H}$ with $t_{\|} = t_{\|}^{\prime}$, 
$m=L^2$ LEs of
diverge to $+\infty$,
and the other $m$ 
LEs of $h$ 
form a finite spectrum around $\gamma=0$ (see also Fig.~2 in the main text). 
For $t_{\|} \neq t_{\|}^{\prime}$, on the other hand, all the $2 m=2 L^2$ LEs of 
$h$ comprise either one or two continuum spectra around $\gamma=0$, depending on 
the disorder strength (
see Figs.~\ref{LE_M1} and \ref{LE_M2}).

In the following discussion, we focus on the case with 
the two continuum LE spectra. 
Generalization to the other cases 
is straightforward. 
For finite $L$, 
the transfer matrix study of the non-Hermitian Hamiltonian 
$h$ in the quasi-1D geometry gives a discrete set of $2m$ LEs,
\begin{align}
    \big\{\gamma^{(1)}_{\rm min}(W,L), \cdots, \gamma^{(1)}_{\rm max}(W,L), \gamma^{(2)}_{\rm min}(W,L), \cdots, \gamma^{(2)}_{\rm max}(W,L)\big\},
    \label{before-gauge}
\end{align}
with $\gamma^{(1)}_{\rm min}(W,L)<\cdots<\gamma^{(1)}_{\rm max}(W,L)<\gamma^{(2)}_{\rm min}(W,L)<\cdots<\gamma^{(2)}_{\rm max}(W,L)$. 
In the limit $L \to \infty$, all the $m$ LEs from $\gamma^{(j)}_{\rm min}(W,L)$ 
to $\gamma^{(j)}_{\rm max}(W,L)$ ($j=1,2$) form a continuum 
spectrum that ranges from $\gamma^{(j)}_{\rm min}(W)$ to $\gamma^{(j)}_{\rm max}(W)$, satisfying
\begin{align}
\lim_{L\rightarrow \infty} \gamma^{(j)}_{\rm min}(W,L) \equiv 
\gamma^{(j)}_{\rm min}(W), \ \ \lim_{L\rightarrow \infty} \gamma^{(j)}_{\rm max}(W,L) \equiv 
\gamma^{(j)}_{\rm max}(W), 
\end{align}
with $j=1,2$. 
For some disorder strength, a finite gap $2\Delta \equiv \gamma^{(2)}_{\rm min}(W)-\gamma^{(1)}_{\rm max}(W)$ exists 
between the two continuum LEs spectra (see Figs.~\ref{LE_SM}, \ref{LE_M1}, and \ref{LE_M2}).

When the gap $2\Delta$ is much larger 
than $L^{-1}$, $\gamma^{(1)}_{\rm max}(W,L)$ and $\gamma^{(2)}_{\rm min}(W,L)$ 
can be fitted well by the following scaling functions: 
\begin{align}
    \gamma^{(1)}_{\rm max}(W,L) = -\frac{a}{L} + \gamma^{(1)}_{\rm max}(W). 
        \label{3-0a}
\end{align} 
and 
\begin{align}
    \gamma^{(2)}_{\rm min}(W,L) = \frac{a}{L} + \gamma^{(2)}_{\rm min}(W), 
        \label{3-0b}
\end{align}
Notably, this scaling holds irrespective of whether their limits  
$\gamma^{(2)}_{\rm min}(W)$ and $\gamma^{(1)}_{\rm max}(W)$ are 
close to zero. 
To see this, we first note that LEs of $h$ with different $t_{\|}$ and 
$t^{\prime}_{\|}$ are related by an imaginary 
gauge transformation along the $z$ direction~\cite{Hatano96}. 
Let the imaginary gauge transformation with an imaginary gauge ${\rm i}g$ act 
on $h$ by \begin{equation}
    h \rightarrow h_g \equiv V_{g} h V_{g}^{-1},
\end{equation}
where $V_{g}$ is a diagonal matrix whose diagonal element takes $e^{j g}$ in the 
$j$th layer along the $z$ direction:
\begin{equation}
    V_{g} = \begin{pmatrix}
        e^{g} 1_{m \times m} & 0 & 0 & \cdots & 0\\
        0 & e^{2g} 1_{m \times m} & 0 & \cdots & 0 \\
        \vdots &  \vdots & \ddots & \vdots &  \vdots \\
        0  & \cdots   & \cdots  & \cdots  &  e^{L_z g} 1_{m \times m} \\
    \end{pmatrix} \, .
    \label{gauge_matrix}
\end{equation}
The transfer matrix $M_i(g)$ of $h_g \equiv V_{g} h V_{g}^{-1}$ along the $z$ direction is obtained 
from Eq.~(\ref{Mi-h-z}) as 
\begin{equation}
     M_i(g) = \left( 
    \begin{matrix}
    - e^{g} \frac{1}{t^{\prime}_{\|} - t_{\|}} \tilde{h}_i & - e^{2g} 
    \frac{t^{\prime}_{\|} + t_{\|}}{t^{\prime}_{\|} - t_{\|}}1_{m \times m}\\
    1_{m \times m} & 0_{m \times m} \\
    \end{matrix}
    \right) \, ,
\end{equation} 
and satisfies 
\begin{equation}
    M_i(g) =   e^{g} S\left( 
        \begin{matrix}
            -  \frac{1}{t^{\prime}_{\|} - t_{\|}} \tilde{h}_i 
            & - \frac{t^{\prime}_{\|} + t_{\|}}{t^{\prime}_{\|} - t_{\|}}1_{m \times m}\\
            1_{m \times m} & 0_{m \times m} \\
        \end{matrix}
        \right)  S^{-1} = e^g S M_i S^{-1},\quad
        S =  \left( 
        \begin{matrix}
        1_{m \times m} &  0_{m\times m}\\
        0_{m \times m} &e^{-g}1_{m \times m} \\
        \end{matrix}
        \right) \, .
\end{equation}
Thus, the LEs of $h_g$ are 
obtained from
the LEs of $h$ in Eq.~(\ref{before-gauge}), 
\begin{equation}
    \big\{ \gamma^{(1)}_{\rm min}(W,L)+g,\cdots,\gamma^{(1)}_{\rm max}(W,L)+g, 
\gamma^{(2)}_{\rm min}(W,L)+g,\cdots,\gamma^{(2)}_{\rm max}(W,L)+g \big\}. 
\end{equation}
Similarly, the LEs of $h^{\dagger}_g$ differ from the LEs of $h^{\dagger}$ by $-g$. 
Suppose that the gap between $\gamma^{(1)}_{\rm max}(W)$ and 
$\gamma^{(2)}_{\rm min}(W)$ is much larger than $L^{-1}$, and  
choose the imaginary gauge $g$ in such a way 
that a midpoint of the gap comes around zero, 
\begin{align}
 \gamma^{(1)}_{\rm max}(W)+g < 0 < \gamma^{(2)}_{\rm min}(W)+g. \ \
\end{align}
Then, a zero mode of a chiral-symmetric Hamiltonian 
${\cal H}_g$ that has $h_g$ and $h^{\dagger}_g$ in the off-diagonal blocks 
is in the Anderson insulator phase, and its localization length $\xi_z$ 
is much shorter than $L$. 
Depending on $g$, the localization length is 
given by either  
\begin{align}
    \frac{1}{\xi_z(W,L)} = -(\gamma^{(1)}_{\rm max}(W,L)+g), \label{relation2-a}
\end{align}
or  
\begin{align}
    \frac{1}{\xi_z(W,L)} = \gamma^{(2)}_{\rm min}(W,L)+g. \label{relation2-b}
\end{align}
In the Anderson insulator phase, a finite-size scaling of the normalized 
localization length $\Lambda(L) \equiv \xi(L)/L$ is 
described by a function of the single parameter 
$\Lambda$ (i.e., single-parameter scaling)~\cite{Asada04}: 
\begin{equation}
    \frac{d\ln \Lambda}{d\ln L} = \beta({\Lambda}).
\end{equation}
For small $\Lambda$, we have $\beta(\Lambda) \to -1$ and 
$\ln \Lambda \simeq \ln \xi(L=\infty) - \ln L$. 
When the localization length $\xi$ is much shorter than $L$, 
one may expand the $\beta$ function in small $\Lambda$, $\beta(\Lambda) = -1 + a\Lambda 
+ {\cal O}(\Lambda^2)$, and retain the zeroth and first order in $\Lambda$, 
\begin{equation}
    \frac{d\ln \Lambda}{d\ln L} = -1 + a \Lambda. 
\end{equation}
This differential equation may be solved by an 
integration in a domain of $[L,L_0]$ with $L\ll L_0$, 
\begin{align}
    \frac{L}{\xi} \equiv \frac{1}{\Lambda} = a + \frac{L}{L_0} \Big(\frac{1}{\Lambda_0}-a\Big)
    \equiv a + \frac{L}{L_0} \Big(\frac{L_0}{\xi_0}-a\Big), 
\end{align}
with $\xi_0 \equiv \xi(L=L_0)$. 
When $L_0$ goes to infinity, $\xi_0$ converges to finite 
$\xi(L=\infty) \equiv \xi(\infty)$. 
Thus, we obtain 
the lowest-order finite-size scaling form of the quasi-1D localization length,
\begin{align}
 \frac{1}{\xi(L)} = \frac{a}{L} + \frac{1}{\xi(\infty)}.    \label{beta_lambda} 
\end{align}
Now that the localization length 
along the $z$ direction is much shorter than $L$ in Eqs.~(\ref{relation2-a}) and (\ref{relation2-b}), we may use the same scaling function 
not only for $\xi_z(W,L)$ but also for $\gamma^{(2)}_{\rm min}(W,L)$ and 
$\gamma^{(1)}_{\rm max}(W,L)$. In fact, the scaling forms of Eqs.~(\ref{3-0a}) and (\ref{3-0b}) 
work well for the numerical fittings. Note that the coefficient 
$a$ in Eq.~(\ref{beta_lambda}) takes a non-universal value in general 
(see the fitting values in Table~\ref{gamma_m_fit2}). 
To obtain the scaling form for $\gamma^{(1)}_{\rm min}(W,L)$, 
let us choose large positive $g$ and make all the LEs of $h_g$ be positive, 
\begin{align}
 0 < \gamma^{(1)}_{\rm min}(W,L)+g.  
\end{align}
In the chiral-symmetric Hamiltonian ${\cal H}_g$ that has such $h_g$ 
and its Hermitian conjugate $h_g^{\dag}$, $E=0$ is in the weak topological 
insulator phase $(\nu_z=1)$ and the localization length $\xi_z$ is given by 
$\gamma^{(1)}_{\rm min}(W,L)+g$. If we assume that the finite-size 
scaling of $\xi_z$ in the weak topological insulator phase is also described by 
the single parameter scaling function of $\Lambda_z \equiv \xi_z/L$, we  
also obtain the scaling function for $\gamma^{(1)}_{\rm min}(W,L)$ as
\begin{align}
\gamma^{(1)}_{\rm min}(W,L) = \frac{a}{L} + \gamma^{(1)}_{\rm min}(W). \label{3-0c}
\end{align}
This scaling form also works well for the numerical 
data of $\gamma^{(1)}_{\rm min}(W,L)$. 

\begin{figure}[t]
    \centering
    \subfigure[]{
		\begin{minipage}[t]{0.45\linewidth}
			\centering
			\includegraphics[width=1\linewidth]{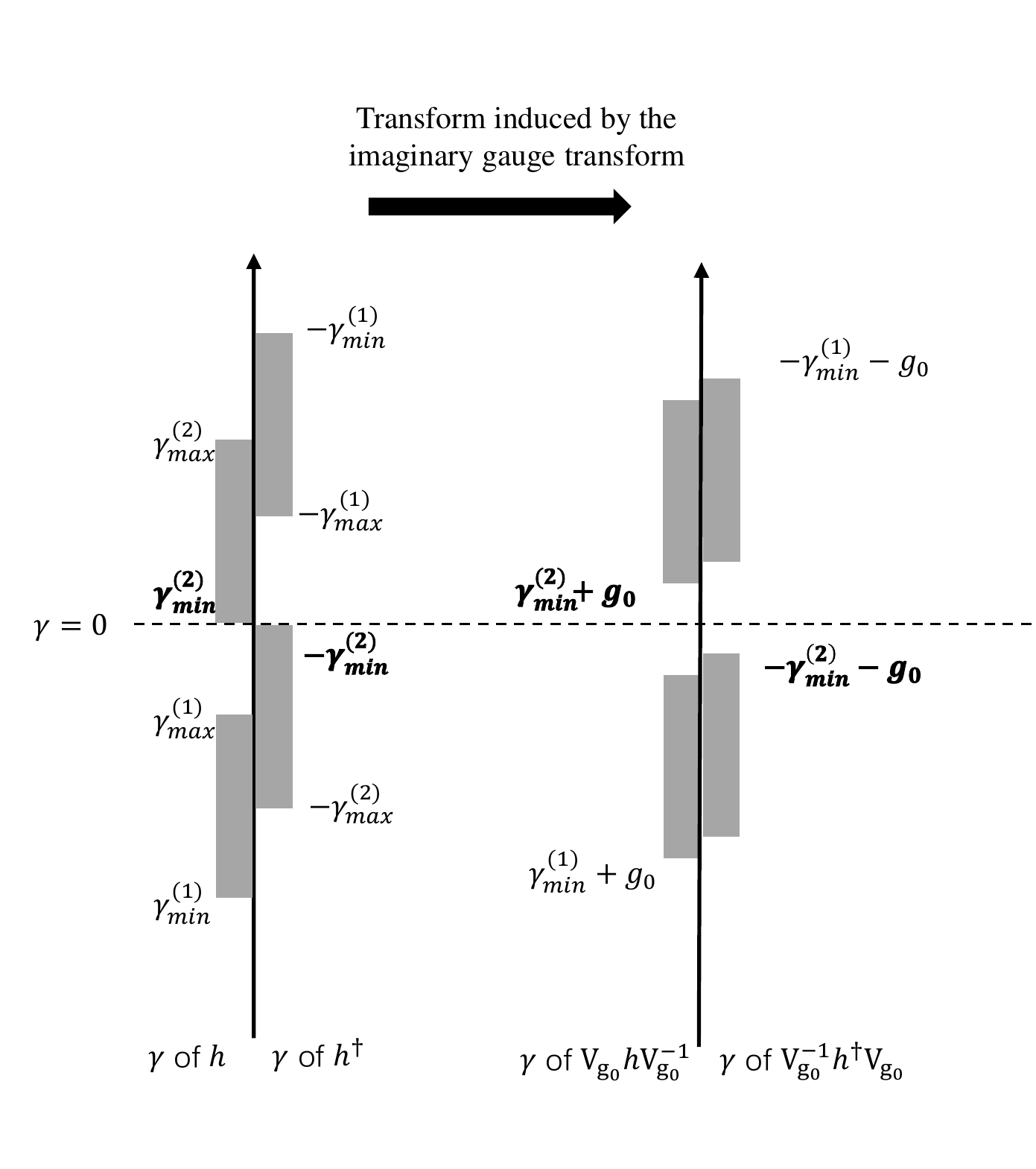}
		\end{minipage}
	}
	\subfigure[]{
		\begin{minipage}[t]{0.45\linewidth}
			\centering
			\includegraphics[width=1\linewidth]{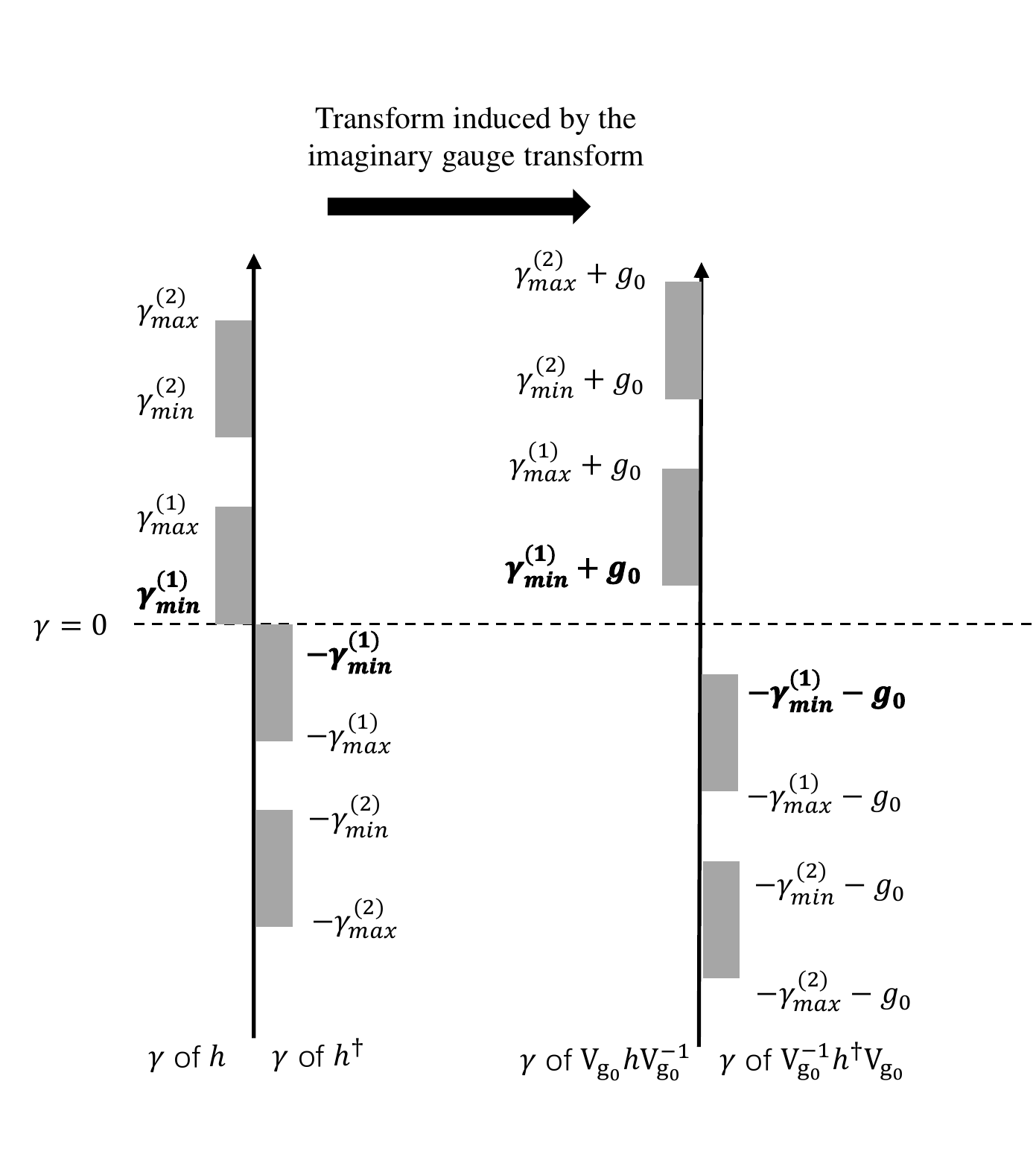}
		\end{minipage}%
	}%
	\caption{Schematic pictures of Lyapunov exponents (LEs) of the nodal-line semimetal model ${\cal H}$ and LEs of ${\cal H}_{g}$ generated by an imaginary gauge transformation. The LEs of 
    ${\cal H}$ and ${\cal H}_{g}$
	are the sum of LEs of its right-upper non-Hermitian Hamiltonian $h$ 
	and their Hermitian conjugate
    $h^{\dag}$ in Eq.~(\ref{chiral0}). 
	The LEs calculated in the quasi-1D geometry ($L\times L \times L_z$, $L_z\gg L$) comprise two continuum LEs spectra in the limit $L\rightarrow \infty$. 
	The grey-shaded regions in the left (right) sides of the vertical axis denote the continuous spectra formed by LEs of $h$ ($h^{\dagger}$). 
	The dotted horizontal line denotes zero $\gamma = 0$.
    The LEs of 
    ${\cal H}$ and ${\cal H}_{g}$,
    as a whole, are symmetric around zero. 
    The smallest positive or the largest negative LE (marked in bold)
    corresponds to the inverse of the quasi-1D localization length. 
    In (a), $g$ is chosen such that finite $\gamma^{(2)}_{\rm min}(W)+g$ corresponds to the inverse of the localization length of ${\cal H}_g$. 
    In (b), $g$ is chosen such that $\gamma^{(1)}_{\rm min}(W)+g$ corresponds to the inverse of the localization length.}
	\label{imag_gauge_herm}
\end{figure}

\subsubsection{Numerical fitting}

To show the validity of the scaling forms in Eqs.~(\ref{3-0a}) and (\ref{3-0b}),
we use the standard $\chi ^2$ fitting method to fit the data of 
$\gamma^{(i)}_{\rm min/max}(W,L)$ ($i=1,2$) with larger $L$. 
For fixed $W$, we minimize the following $\chi^2$ function with respect 
to $a$ and $\gamma^{(i)}_{\rm min/max}(W)$ in Eqs.~(\ref{3-0a}) and (\ref{3-0b}), 
\begin{equation}
\chi^2 =  \sum^D_{j=1} \left( \frac{F_j - \gamma_{\rm min/max}^{(i)}(W,L)}{\sigma_j }\right)^2,
\end{equation}
where $j$ specifies the data point of $\gamma^{(i)}_{\rm min/max}(W,L)$ with different $L$. 
The number $D$ of the data points is 
typically
$7$ ($14 \leq L \leq 28$) and $9$ ($18 \leq L \leq 34$).
$F_j$ is the fitted value from Eqs.~(\ref{3-0a}) and (\ref{3-0b})
for different $L$ specified by $j$. $\sigma_j$ is the standard deviation of 
$\gamma_{\rm min/max}^{(i)}(W,L)$ estimated from the transfer matrix 
calculation~\cite{Slevin14}. The finite-size scaling fit works well in the 
nodal-line semimetal models with or without time-reversal symmetry that are studied in this work. 
Some fitting results for the nodal-line semimetal models 
($\Delta = 0$, $t_{\|} = t_{\|}^{\prime} = 1/2$, $ t_{\perp} = 1$) 
that are studied in the main text are shown in Fig.~\ref{LE_W_L} (a,c) and 
Table~\ref{gamma_m_fit}. 
In Table~\ref{gamma_m_fit}, 
the Monte Carlo method is used to generate pseudo-data sets and evaluate 
the 95\% confidence interval of the fitted values of $``a"$ and 
$``\gamma_{\rm min/max}^{(1)}(W)"$.

Using $\gamma^{(1)}_{\rm max}(W)$ with the confidence interval, 
we determine the critical disorder strength $W_c^{(z)}$ between 
the non-localized region and the localized phase. 
For example, the fitting results in Table~\ref{gamma_m_fit} show 
$\gamma_{\text{max}}=0.0020 \!\ \!\ [0.0017,0.0022]$ 
for $W=29.5$ and $\gamma_{\text{max}}=-0.0042 \!\ \!\ [-0.0044,0.0040]$ 
for $W=29.4$ in the nodal-line semimetal model in symmetry class BDI. 
The results suggest that $W_c^{(z)}$ must be between $29.4$ and $29.5$ 
with the 95\% confidence: $W_c^{(z)}=29.45 \!\ \!\ [29.4,29.5]$. 
$W^{(z)}_{c}$ for the nodal-line semimetal models in symmetry 
classes BDI and AIII are summarized in Table~\ref{W_x_z_2}. 
Table~\ref{W_x_z_2} also shows $W^{(x)}_{c}$ in the same nodal-line 
semimetal models for the same sets of parameters. 
Comparisons between $W_c^{(z)}$ and $W_c^{(x)}$
illustrate that $W^{(x)}_c < W^{(z)}_c$ and $|W^{(z)}_c-W^{(x)}_c|$ is around 10\% of 
$W^{(x)}_c$. This concludes the presence of the quasi-localized phase 
in these nodal-line semimetal models.  
Similarly, comparisons of $W_c^{(z)}$ and $W_c^{(x)}$ in other types of nodal-line semimetal models (Tables~\ref{W_x_z_1} and \ref{W_x_z_3}) also suggest 
the presence of the quasi-localized phases (see below).

\begin{table}[t]
\centering
\caption{
Finite-size scaling analysis of $\gamma^{(1)}_{\rm max}(W,L)$ for several disorder strength $W$ around $W = W_c^{(z)}$ 
in the nodal-line semimetal models ($\Delta = 0$, $t_{\|} = t_{\|}^{\prime} = 1/2$, $ t_{\perp} = 1$) with time-reversal symmetry (symmetry class BDI) and without time-reversal symmetry (symmetry class AIII).  
The square 
brackets are the 95\% confidence 
intervals
determined by the Monte Carlo analyses.}
\centering
\begin{tabular}{cccccccccc}
\hline \hline
\makecell[c]{symmetry \\ class} & $W$ & $L$ &$\gamma_{\rm max}(W,L = \infty)$  & $a$ & GOF \\ \hline
BDI &~29.4~ & ~18 - 34~ & ~0.0020 [0.0017,0.0022]~ & ~-3.890 [-3.893,-3.888]~ & ~0.58~ \\
BDI & 29.5 & 18 - 34 & -0.0042 [-0.0044,-0.0040] & -3.689 [3.691,-3.687] & 0.14\\
AIII & 9.70 & 14 - 28 & 0.0143 [0.0128,0.0156] & -1.921 [-1.948,-1.892] & 0.02\\
AIII & 9.80 & 14 - 28 & 0.0015 [0.0000,0.0029] & -1.929 [-1.957,-1.899] & 0.50 \\
AIII & 9.90 & 14 - 28 & -0.0114 [-0.0129,-0.0099] & -1.931 [-1.961,-1.904] & 0.22\\
\hline
\hline
\quad \quad
\end{tabular}
\label{gamma_m_fit}
\end{table}

\begin{table}[t]
    \centering
    \caption{Comparison between the critical disorder strength 
    $W_c^{(z)}$
    in the $z$ direction (weak topological index $\nu_z \neq 0$)
    and 
    the critical disorder strength $W_c^{(x)}$ 
    in the $x$ direction (weak topological index $\nu_x = 0$)
    for 
    the nodal-line semimetal models 
    in symmetry classes BDI and AIII
    ($\Delta = 0$, $t_{\|} = t_{\|}^{\prime} = 1/2$, $ t_{\perp} = 1$). 
    The square brackets denote 
    the 95\%
    intervals.  
    The confidence intervals of $W_c^{(x)}$ are 
    determined by the 95\% confidence 
    intervals of $\gamma_{\rm max}(W)$.
    }
\begin{tabular}{cccc}
    \hline
    \hline
    \makecell[c]{symmetry \\ class}  & $W_c^{(x)}$ & $W_c^{(z)}$ \\
    \hline
    BDI  &~~27.24[27.19,27.30]~~ & ~~29.45[29.4,29.5]~~ \\
    AIII &   9.14[9.12,9.17]&9.8[9.7,9.9]\\ 
     \hline
      \quad \quad
\end{tabular}
\label{W_x_z_2}
\end{table}

\begin{figure}
    \centering
	\subfigure[class BDI]{
		\begin{minipage}[t]{0.45\linewidth}
			\centering
			\includegraphics[width=1\linewidth]{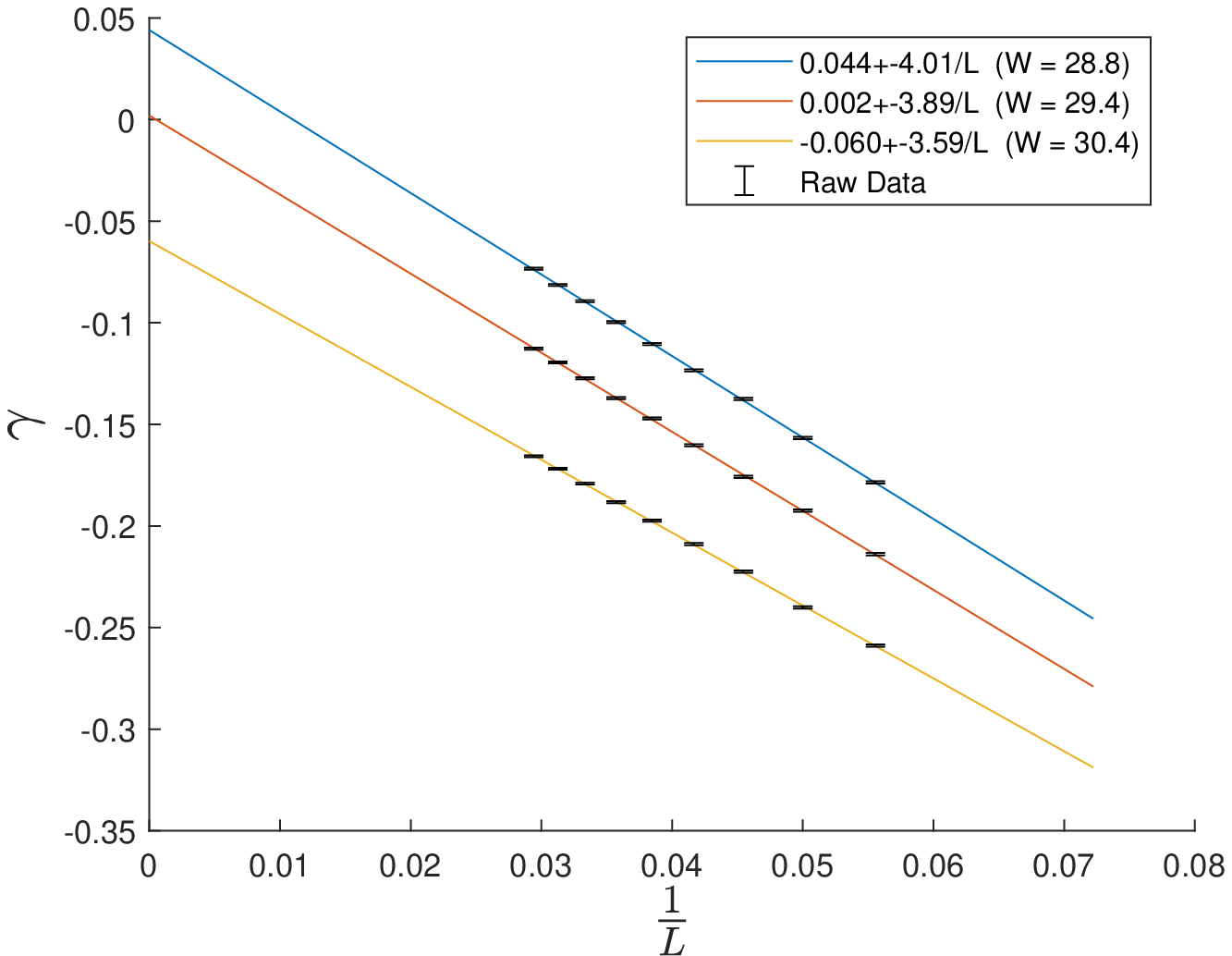}
			\label{LE_W_L_BDI}
		\end{minipage}%
	}%
	\subfigure[class BDI]{
		\begin{minipage}[t]{0.45\linewidth}
			\centering
			\includegraphics[width=1\linewidth]{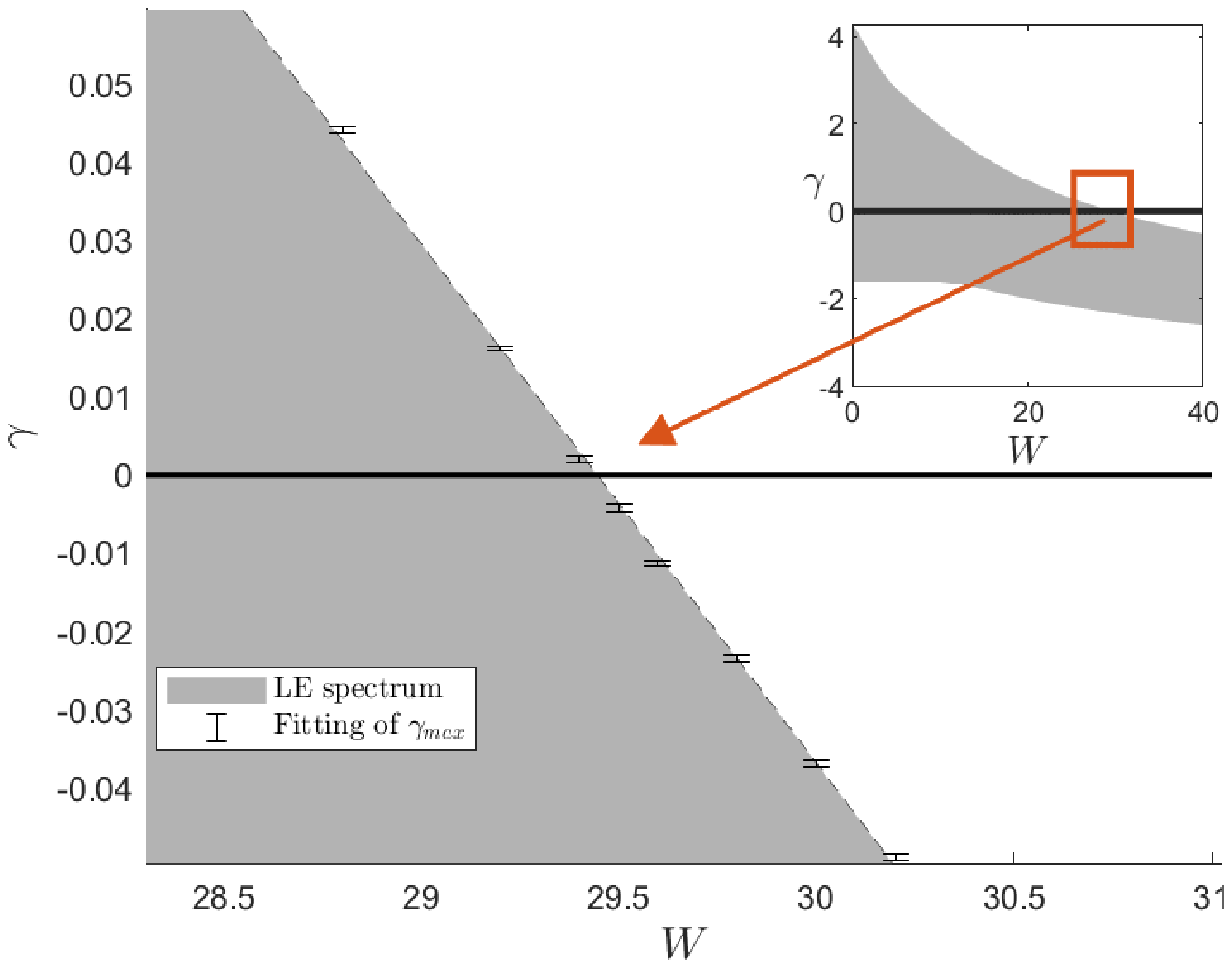}
		\end{minipage}%
	}%

	\subfigure[class AIII]{
		\begin{minipage}[t]{0.45\linewidth}
			\centering
			\includegraphics[width=1\linewidth]{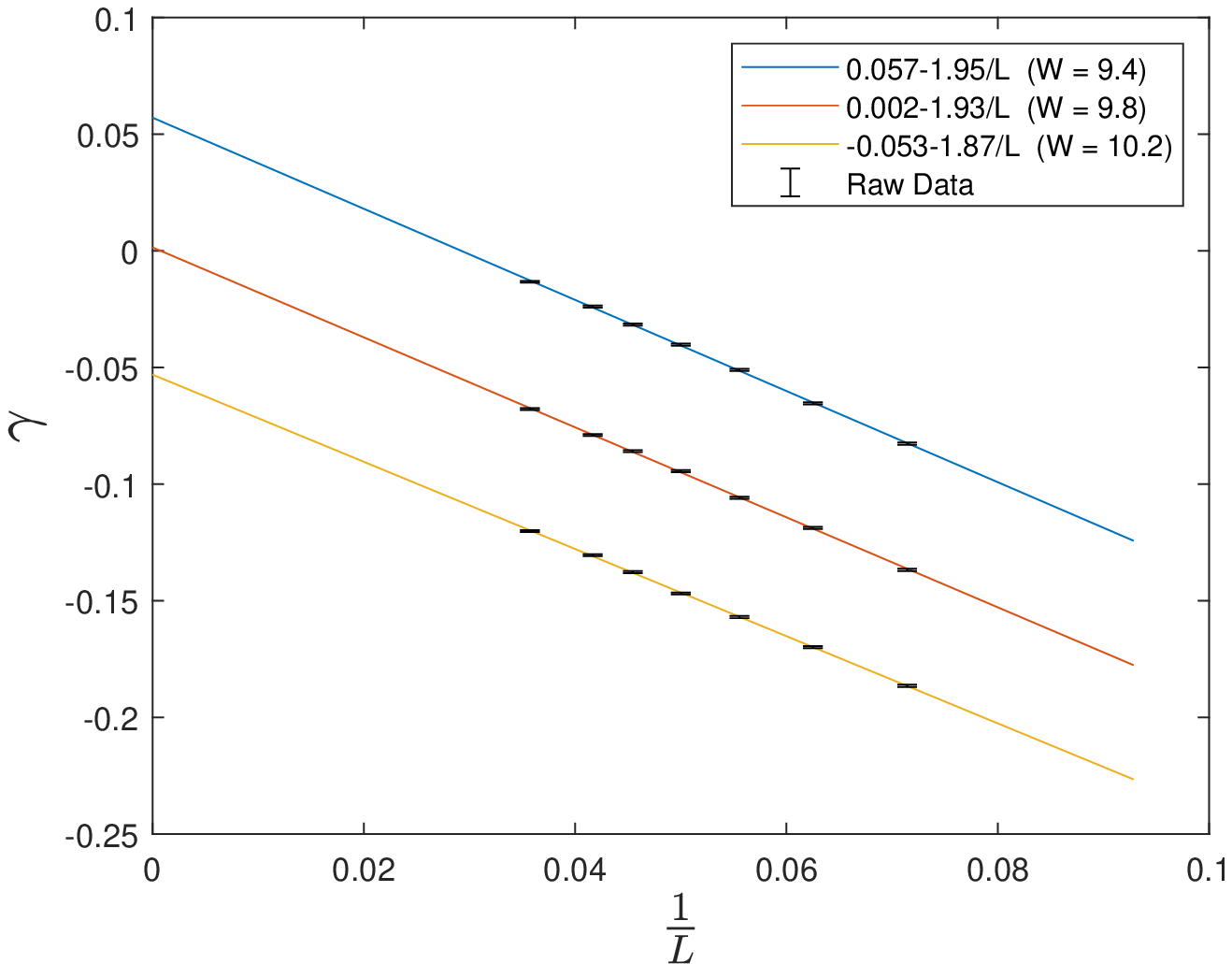}
			\label{LE_W_L_AIII}
		\end{minipage}%
	}%
	\subfigure[class AIII]{
		\begin{minipage}[t]{0.45\linewidth}
			\centering
			\includegraphics[width=1\linewidth]{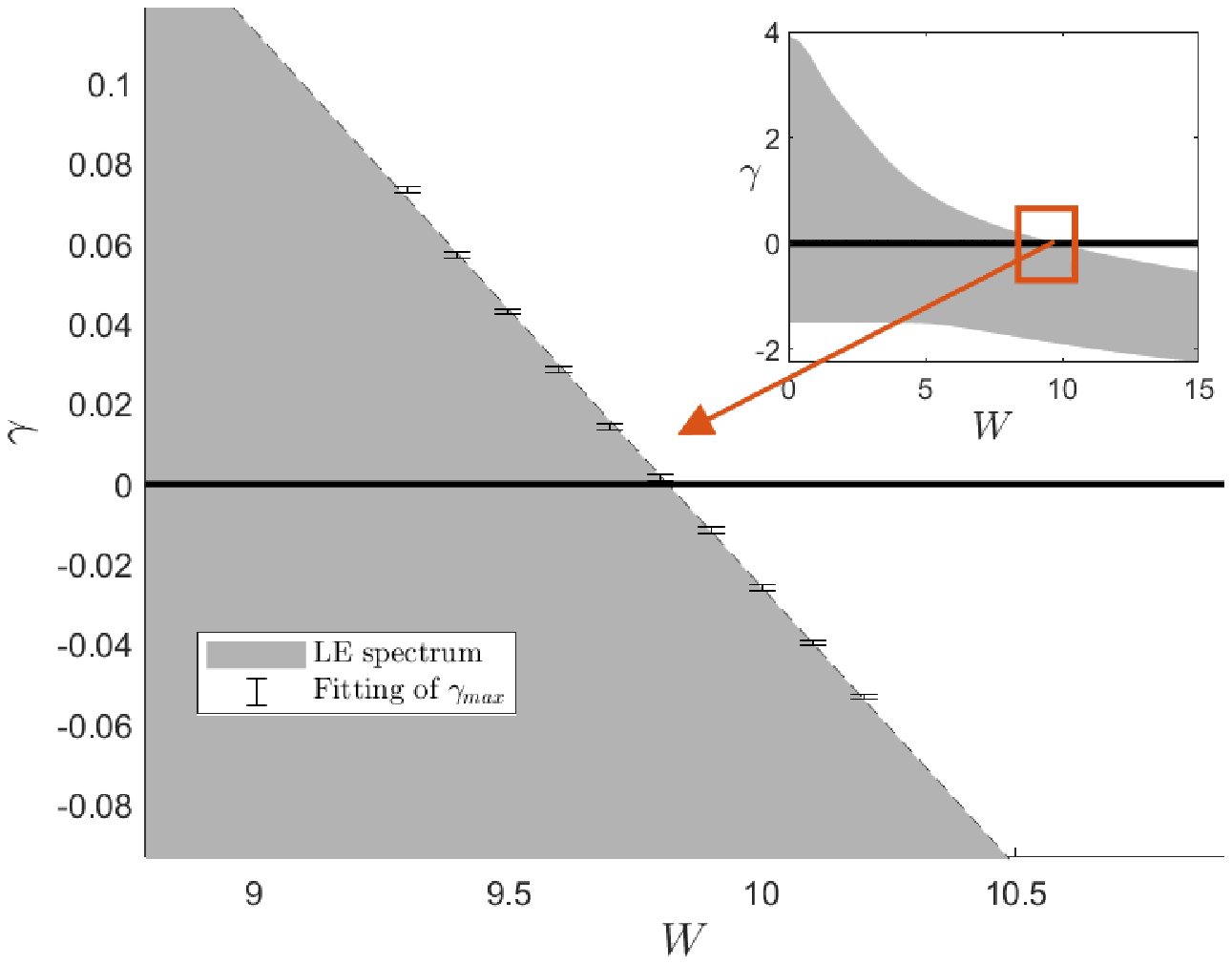}
		\end{minipage}%
	}%
	\caption{(a,c) $\gamma^{(1)}_{\text{max}}(W,L)$ 
	as a function of the system size $L$ for 
	the different disorder strength $W$ in the nodal-line semimetal models ($\Delta = 0$, $t_{\|} = t_{\|}^{\prime} = 1/2$, $ t_{\perp} = 1$) in (a) symmetry class BDI 
	and (c) symmetry class AIII. The solid lines are the fitting curves from Eq.~(\ref{3-0b}). A cross-section of the fitting curve at $1/L=0$ determines $\gamma^{(1)}_{\text{max}}(W) \equiv \lim_{L \rightarrow \infty}\gamma^{(1)}_{\text{max}}(W,L)$. (b,d) $\gamma^{(1)}_{\rm max}(W)$ as a 
	function of $W$ around $W = W_c^{(z)}$ in the nodal-line semimetal models 
	in (b) symmetry class BDI and (d) symmetry class AIII. 
	Insets of (b,d):~distributions of the Lyapunov exponents (LEs)
    of the right-upper part $h$ of the nodal-line semimetal model ${\cal H}$ as a function of $W$ in the larger range of $\gamma$ and $W$.
    The LEs of the nodal-line semimetal models are the sum of the LEs of $h$ and their opposite-sign exponents.
    }
	\label{LE_W_L}
\end{figure}

\subsection{Quasi-localized phases in chiral-symmetric models with weak topological indices}
    \label{supplement-sec-Q1D-topo}

In Sec.~\ref{supplement-sec-Lyapunov-scaling}, we describe the finite-size scaling analysis of the LEs along  
the $z$ direction in the nodal-line models with the weak topological index $\nu_z \neq 0$. 
The analysis enables determinations of the phase boundary $W^{(z)}_{c}$ of the 
non-localized region. 
The non-localized region comprises the metal and 
quasi-localized phases. 
In the nodal-line models with $\nu_x=\nu_y=0$, 
the phase transition between the metal and quasi-localized phases 
is characterized by  
the localization properties along the $x$ or $y$ direction.
In this section, we discuss the localization properties along 
the $x$ direction in the chiral-symmetric models with $\nu_z \neq 0$ and 
$\nu_x=0$. 
We demonstrate the presence of the quasi-localized phases inside the non-localized region for all the models.

\subsubsection{Nodal-line semimetal in class AIII}

We discuss a nodal-line semimetal in Eq.~(\ref{nodal-line-h}) with the time-reversal-breaking disorder.
The Hamiltonian ${\cal H}_1$ has the two types of random potentials, 
\begin{equation}
    {\cal H}_1 = \sum_{\bm r=(r_x,r_y,r_z)} \left\{ (\Delta + \epsilon_{\bm{r}}) c^{\dagger}_{\bm{r}} \sigma_y c_{\bm{r}} + \epsilon_{\bm{r}}^{\prime} c^{\dagger}_{\bm{r}} \sigma_z c_{\bm{r}} + \left[  \sum_{\mu = x,y}\left(  t_{\perp}c^{\dagger}_{\bm r+\bm{e_{\mu}}} \sigma_z c_{\bm{r}}  \right)  -{\rm i} t_{\|}   c^{\dagger}_{\bm r+\bm{ e_z}}  \sigma_y c_{\bm{r}} + t_{\|}^{\prime}   c^{\dagger}_{\bm r+\bm{ e_z}}  \sigma_z c_{\bm{r}} +\text{H.c.} \right]  \right\},
    \label{nodal-line-h-aiii}
\end{equation}
where the random potential $\epsilon_{\bm{r}}$ ($\epsilon_{\bm{r}}^{\prime}$) respects (breaks) time-reversal symmetry and distributes uniformly in 
$\epsilon_{\bm{r}}^2 +  \epsilon_{\bm{r}}^{\prime 2} \leq W^2$. 
The parameters are chosen to be $\Delta = 0$, $t_{\|} = t_{\|}^{\prime} = 1/2$, and $ t_{\perp} = 1$. The Hamiltonian ${\cal H}_1$ only 
satisfies chiral symmetry ${\cal H}_1 = -\sigma_x {\cal H}_1 \sigma_x$, and 
hence belongs to class AIII.  

According to Eq.~(\ref{relation}), the chiral-symmetric Hamiltonian is decomposed into 
the block-off diagonal structure in the basis that diagonalizes the chiral operator 
${\cal C} \equiv \sigma_x$. The right-upper part $h_1$ of ${\cal H}_1$ in this basis 
is given by
\begin{equation}
    h_1= \sum_{\bm r=(r_x,r_y,r_z)} \left[ (\Delta + \epsilon_{\bm{r}} + {\rm i} \epsilon_{\bm{r}}^{\prime}) f^{\dagger}_{\bm{r}} f_{\bm{r}} + \sum_{\mu = x,y}  \left(  t_{\perp}f^{\dagger}_{{\bm r}+{\bm e}_{\mu}}  f_{\bm r} + \text{H.c.}  \right) + (t^{\prime}_{\|} + t_{\|})f^{\dagger}_{{\bm r}+{\bm e}_{z}}  f_{\bm r} + (t^{\prime}_{\|} - t_{\|})f^{\dagger}_{\bm r}  f_{{\bm r}+{\bm e}_{z}} \right] \, .
    \label{NDSM-NH1}
\end{equation}
Transposition exchanges $t^{\prime}_{\|}+t_{\|}$ and $t^{\prime}_{\|}-t_{\|}$. 
Thus, as a unitary transformation in Eqs.~(\ref{statistical-symmetry}) and (\ref{unitary}), 
we can consider the mirror operation with respect to the $xy$ plane 
as in Eq.~(\ref{mirror}). 
Since both $\epsilon_{\bm r}$ and 
$\epsilon_{\bm{r}}^{\prime}$ are statistically equivalent 
for different lattice points ${\bm r}$, an ensemble of $h_1$ 
defined in Eq.~(\ref{NDSM-NH1}) is statistically invariant 
under 
the combination of transposition and 
the mirror operation. 
The symmetry in Eqs.~(\ref{statistical-symmetry}) 
and (\ref{mirror}) requires the LEs of $h_1$ along the $x$ and $y$ directions to come in opposite-sign pairs, leading to $\nu_x=\nu_y=0$. 

We calculate the localization length $\xi_x$ of $h_1$ along the $x$ direction in the quasi-1D geometry ($L_{x}\times L \times L$, $L_x\gg L$). The normalized 
localization length $\Lambda_x \equiv \xi_x/L$ shows scale-invariant behavior around 
$W_c^{(x)} = 9.14 \pm 0.01$ (Fig.~\ref{non_R_lambda_x_A}).  
Fitting by the polynomial expansion of the finite-size scaling functions 
[see Eqs.~(\ref{fit1}) and (\ref{fit2})], we determine the critical disorder strength 
$W^{(x)}_{c}$ and the critical exponent (see Table~\ref{fitting_table_SM_0}). 
Figure~\ref{non_R_lambda_x_A} shows the normalized localization length 
for different $W$ and $L$ together with the fitting curves.  

In Sec.~\ref{supplement-sec-Lyapunov-scaling}, we use the finite-size scaling of LEs to obtain the critical 
disorder strength $W^{(z)}_{c} = 9.8 \!\ [9.7,9.9]$. For $W<W_c^{(z)}$, the 
localization length along the $z$ direction diverges. $W^{(x)}_{c}$ is well 
within the non-localized region, $W^{(x)}_c<W^{(z)}_{c}=9.8 \!\ [9.7,9.9]$, 
demonstrating the presence of the quasi-localized phase in the 
nodal-line semimetal model without time-reversal symmetry.  
 
\begin{figure}[t]
    \centering
	\begin{minipage}[t]{0.8\linewidth}
			\centering
			\includegraphics[width=1\linewidth]{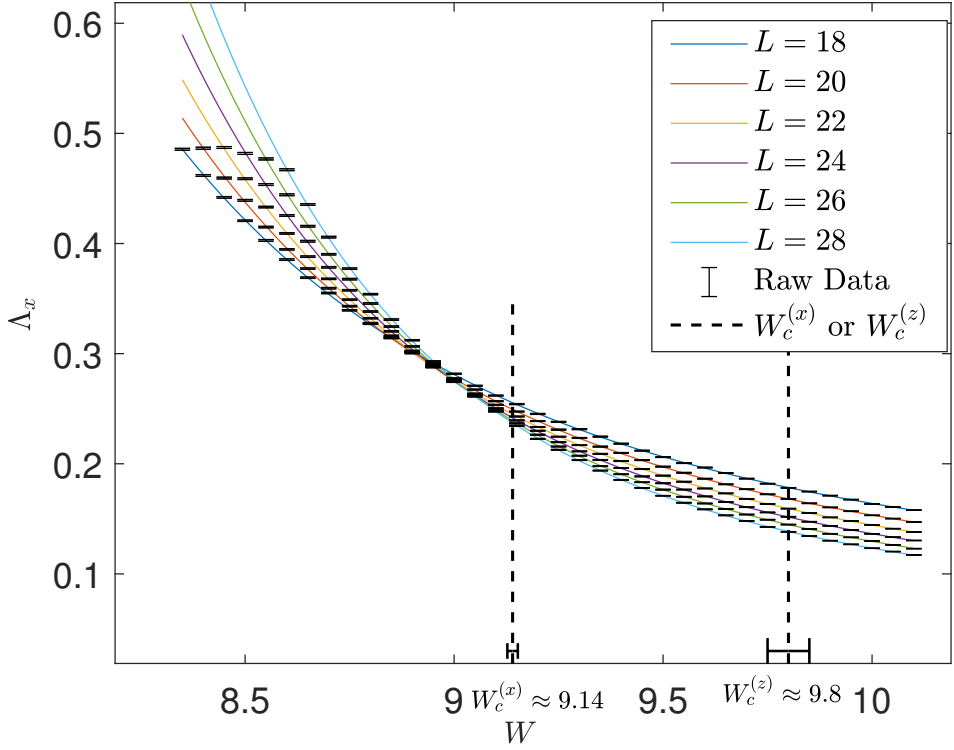}
		\end{minipage}%
			\caption{
            Normalized localization length $\Lambda_x \equiv \xi_x/L$ along 
            the $x$ direction as a function of the disorder strength $W$ in 
            the nodal-line semimetal model in class AIII [Eq.~(\ref{nodal-line-h}) with $\Delta = 0$, $t_{\|} = t_{\|}^{\prime} = 1/2$, $ t_{\perp} = 1$]. 
            $\xi_x$ is calculated in 
            the quasi-1D geometry ($L\times L\times L_x$). The black points are the raw data with the error bars. 
            The solid lines for different $L$ and the dashed vertical line $W_c^{(x)} \approx 9.14$ with the error bars are the results of the fitting according to Eqs.~(\ref{fit1}) and (\ref{fit2}) with 
            $(m,n) = (2,3)$. 
            The dashed line $W_c^{(z)} \approx 9.8$ is evaluated by the fitting of the Lyapunov exponents along the $z$ 
            direction by Eq.~(\ref{3-0a}).}
			\label{non_R_lambda_x_A}
\end{figure}

\subsubsection{Nodal-line semimetals in class BDI}
We also study the localization properties along the $x$ direction in other nodal-line semimetal models in class BDI: 
i) nodal-line semimetal model in Eq.~(\ref{nodal-line-h}) with the different parameters $\Delta = 0$, $t_{\|} = \sinh g, t_{\|}^{\prime} = \cosh g$ ($g=0.22,1$), $ t_{\perp} = 1$ 
and ii) another nodal-line semimetal model ${\cal H}_2$ with 
an extended Fermi line running across the Brillouin zone,  
\begin{equation}
    {\cal H}_2 = \sum_{\bm r=(r_x,r_y,r_z)} \left\{ (\Delta + \epsilon_{\bm{r}}) c^{\dagger}_{\bm{r}} \sigma_z c_{\bm{r}} +  \left[   t_{\perp}c^{\dagger}_{\bm r+\bm{e_{x}}} \sigma_z c_{\bm{r}} + \sum_{\mu = y,z} \left(  -{\rm i} t_{\|}   c^{\dagger}_{\bm r+\bm{ e_{\mu}}}  \sigma_y c_{\bm{r}} + t_{\|}^{\prime}   c^{\dagger}_{\bm r+\bm{ e_{\mu}}}  \sigma_z c_{\bm{r}} \right) +\text{H.c.} \right]  \right\},
    \label{nodal-line-h2}
\end{equation}
where $\epsilon_{\bm r}$ 
takes real values and distributes uniformly in $[-W/2,W/2]$,
and the parameters are chosen to be
$\Delta = 0$, $t_{\|} = \cosh g, t_{\|}^{\prime} = \sinh g$ ($g=0.2$), $ t_{\perp} = 1$. 
The Hamiltonian ${\cal H}_2$ satisfies time-reversal symmetry ${\cal H}_2 = {\cal H}^*_2$ 
and chiral symmetry ${\cal H}_2 = -\sigma_x {\cal H}_2 \sigma_x$, and hence belongs to class BDI.  
In terms of Eq.~(\ref{relation}), the chiral-symmetric Hamiltonian ${\cal H}_2$ is decomposed into 
the block-off diagonal structure in the canonical basis where 
$\sigma_x$ is diagonalized. The right-upper part $h_2$ of 
${\cal H}_2$ in this basis is given by
\begin{equation}
    h_2 = \sum_{\bm r=(r_x,r_y,r_z)} \left\{ (\Delta + \epsilon_{\bm{r}}) f^{\dagger}_{\bm{r}} f_{\bm{r}} +  \left(  t_{\perp}f^{\dagger}_{{\bm r}+{\bm e}_{x}}  f_{\bm r} + \text{H.c.}  \right) + \sum_{\mu=y,z} \left[ (t^{\prime}_{\|} + t_{\|})f^{\dagger}_{{\bm r}+{\bm e}_{\mu}}  f_{\bm r} + (t^{\prime}_{\|} - t_{\|})f^{\dagger}_{\bm r}  f_{{\bm r}+{\bm e}_{\mu}} \right] \right\} \, . \label{NDSM-NH2}
\end{equation}
Transposition exchanges $t^{\prime}_{\|}+t_{\|}$ and 
$t^{\prime}_{\|}-t_{\|}$. 
Thus, as a unitary transformation in Eqs.~(\ref{statistical-symmetry}) and (\ref{unitary}), 
we can apply a $\pi$-rotation around the $x$ axis,
\begin{align}
    {\cal U}_{(r_x,r_y,r_z|r^{\prime}_x,r^{\prime}_y,r^{\prime}_z)} 
    = \delta_{r_x,r^{\prime}_x} \delta_{r_y,-r^{\prime}_y} \delta_{r_z,-r^{\prime}_z}. \label{pi-rot}
\end{align}
Since $\epsilon_{\bm r}$ is statistically equivalent 
for different lattice points ${\bm r}$, an ensemble of $h_2$ 
defined in Eq.~(\ref{NDSM-NH2}) is statistically invariant 
under 
the combination of transposition and the $\pi$-rotation.
The symmetry in Eqs.~(\ref{statistical-symmetry}) and (\ref{pi-rot}) 
requires the LEs of $h_2$ along the $x$ direction to come in opposite-sign pairs, giving rise to $\nu_x=0$. 
The hopping along the $y$ direction and the hopping along the $z$ direction 
are symmetric in Eq.~(\ref{NDSM-NH2}). 
Thus, after transposition, 
we can also apply a mirror operation with respect to the plane with fixed $y+z$:
\begin{align}
    {\cal U}_{(r_x,r_y,r_z|r^{\prime}_x,r^{\prime}_y,r^{\prime}_z)} 
    = \delta_{r_x,r^{\prime}_x} \delta_{r_y,-r^{\prime}_z} \delta_{r_z,-r^{\prime}_y}. \label{mirror-2}
\end{align}
Transposition exchanges $t^{\prime}_{\|}+ t_{\|}$ and $t^{\prime}_{\|}- t_{\|}$ 
and the mirror operation puts them back. 
Thus, an ensemble of $h_2$ is statistically invariant
under the combination of transposition and the mirror operation.
The symmetry in Eqs.~(\ref{statistical-symmetry}) 
and (\ref{mirror-2}) requires the LEs of $h_2$ along the $r_{(0,1,-1)}\equiv r_y-r_z$ direction to come in opposite-sign pairs, leading to $\nu_y=\nu_z$. 

\begin{figure}[bt]
    \centering
    \includegraphics[width=0.5\linewidth]{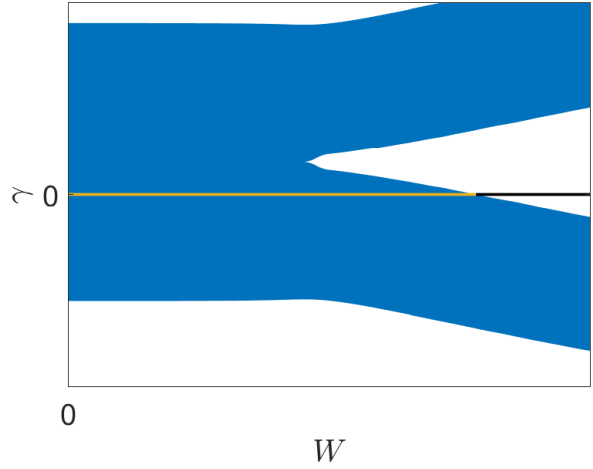}
    \caption{$2L^2$ Lyapunov exponents (LEs) of the right-upper part $h$ of the nodal-line semimetal model ${\cal H}$ 
    in the canonical basis ($t_{\perp} = 3/10,  t^{\prime}_{\|} = 1, t_{\|} = 1/4,  \Delta = 2$)
    calculated along the $z$ direction with the quasi-1D geometry
    $L \times L \times L_z$. 
    Distributions of the $2L^2$ LEs are plotted as a 
    function of the disorder strength $W$. 
    The LEs of the nodal-line semimetal model are the sum of the $2L^2$ 
    LEs of $h$ and their opposite-sign exponents.
    }
    \label{LE_SM}
\end{figure}

\begin{figure}[pbt]
    \centering
	\subfigure[nodal-line semimetal with $\nu_z\ne 0$ and $\nu_x=\nu_y=0$ ($g= 0.22$)]{
		\begin{minipage}[t]{0.5\linewidth}
			\centering
			\includegraphics[width=1\linewidth]{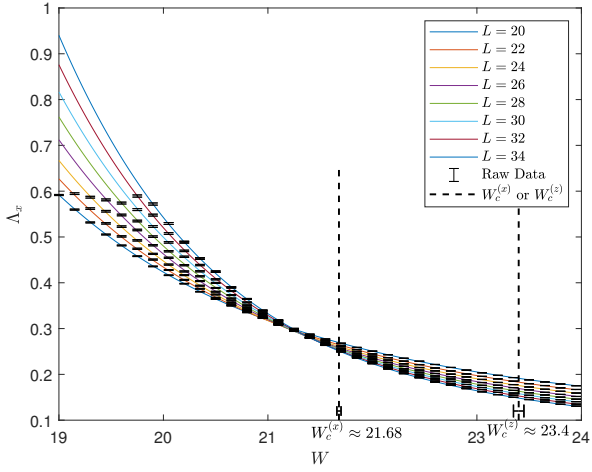}
		\end{minipage}%
	}%
	\subfigure[nodal-line semimetal with $\nu_z\ne 0$ and $\nu_x=\nu_y=0$ ($g= 1$)]{
		\begin{minipage}[t]{0.5\linewidth}
			\centering
			\includegraphics[width=1\linewidth]{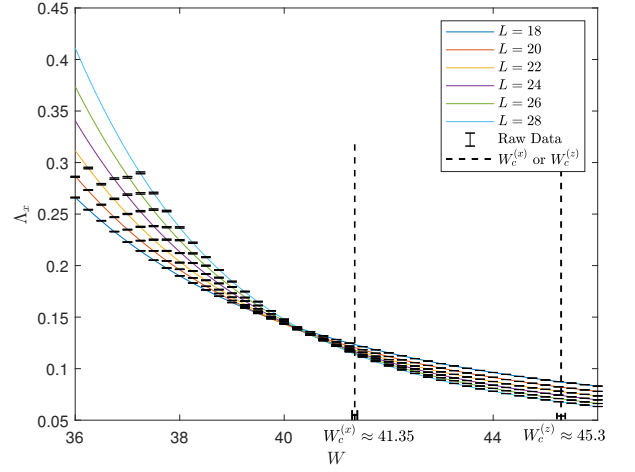}
			\label{lambda_x_g_1}
		\end{minipage}%
	}%
	
	\subfigure[nodal-line semimetal with  with $\nu_y = \nu_z \ne 0$ and $\nu_x=0$]{
		\begin{minipage}[t]{0.5\linewidth}
			\centering
			\includegraphics[width=1\linewidth]{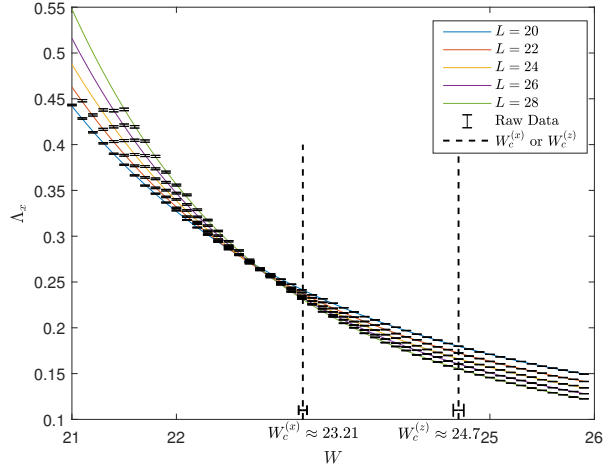}
		\end{minipage}%
	}%
	\caption{
    Normalized localization length $\Lambda_x \equiv \xi_x/L$ along 
    the $x$ direction as a function of the disorder strength $W$ for different system sizes $L$. 
    (a, b)~Nodal-line semimetal model in Eq.~(\ref{nodal-line-h}) with $\Delta = 0$, $t_{\|} = \sinh g, t_{\|}^{\prime} = \cosh g$ [(a)~$g=0.22$ and (b)~$g=1$], $ t_{\perp} = 1$, and 
	(c)~nodal-line semimetal model in Eq.~(\ref{nodal-line-h2}) with  $\Delta = 0$, $t_{\|} = \cosh g, t_{\|}^{\prime} = \sinh g$ ($g=0.2$), $ t_{\perp} = 1$. 
    The black points are the raw data with the error bars. 
    The solid lines for different $L$ and the dashed vertical line $W_c^{(x)}$ with the error bars are the results of the fitting according to Eqs.~(\ref{fit1}) and (\ref{fit2}) with $(m,n) = (2,3)$.
    The dashed line $W_c^{(z)}$ is evaluated by the fitting of the Lyapunov exponents along the $z$ direction by Eqs.~(\ref{3-0a}) and (\ref{3-0b}).}
	\label{SM_lambda_x}
\end{figure}

\begin{table}[bt]
\centering
\caption{Polynomial fitting results of the normalized localization length 
$\Lambda_x \equiv \xi_x/L$ in the $x$ direction around its scaling invariant point 
of the different nodal-line semimetal models ${\cal H}$ [Eq.~(\ref{nodal-line-h})] 
and ${\cal H}_2$ [Eq.~(\ref{nodal-line-h2})]. 
Both models are characterized by the parameters 
$\Delta = 0$, $t_{\|} = \sinh g, t_{\|}^{\prime} = \cosh g$, $ t_{\perp} = 1$. 
The column ``parameter'' specifies the value of $g$. 
The data of $\Lambda_x$ in these models are calculated with the 
quasi-1D geometry ($L\times L \times L_x$, $L_x\gg L$). The range of the system size $L$, critical disorder strength $W^{(x)}_c$, critical exponent $\nu$, scaling dimension $-y$ of the least irrelevant variable, critical localization length $\Lambda_c$, the goodness of fitting (GOF), and the fitting order 
$(m,n)$
in Eqs.~(\ref{fit1}) and (\ref{fit2}) 
are shown. 
The square 
brackets are the 95\% confidence error bars determined by the synthetic data.
Note that $\Lambda_c$ is non-universal because of the spatial anisotropy.}
\begin{tabular}{cccccccccc}
\hline \hline
model & parameter & $L$ & $m$ & $n$  & GOF  &$W^{(x)}_c$& $\nu$ & $y$ & $\Lambda_c$ \\ \hline
${\cal H}$ &$g =0.22$ & 20-34&  2 & 3 & 0.61 & 21.688[21.647,21.729] & 0.866[0.792,0.916] &  1.840[1.608,2.130] & 0.239[0.232,0.246]  \\
${\cal H}$ &$g =0.22$ & 20-34 &3 & 3 & 0.58 & 21.730[21.662,21.803] & 0.860[0.679,0.918] &  1.619[1.284,2.031] & 0.232[0.218,0.243] \\
${\cal H}$  &$g = 1$ & 18-28 & 2 & 3 & 0.14 & 41.284[41.209,41.365] & 0.787[0.754,0.818] &  2.256[2.013,2.516] & 0.110[0.108,0.113] \\
${\cal H}$ &$g = 1$ & 18-28 & 3 & 3 & 0.24 & 41.358[41.264,41.463] & 0.785[0.751,0.817] &  2.088[1.816,2.381] & 0.108[0.104,0.111] \\
${\cal H}_2$ &$g = 0.2$& 20-28 &2 & 3 & 0.32 & 23.214[23.146,23.334] & 0.857[0.712,0.920] &  2.595[1.845,3.300] & 0.223[0.208,0.231]  \\
${\cal H}_2$ &$g = 0.2$ & 20-28 & 3 & 3 & 0.31 & 23.229[23.131,23.422] & 0.855[0.691,0.921] &  2.513[1.593,3.418] & 0.222[0.197,0.232]  \\
\hline
\end{tabular}
\label{fitting_table_SM}
\end{table}

\begin{table}[tb]
    \centering
    \caption{Comparison of the two critical disorder strengths, $W^{(x)}_c$ and $W^{(z)}_c$, in the nodal-line semimetal models ${\cal H}$ [Eq.~(\ref{nodal-line-h})] and ${\cal H}_2$ [Eq.~(\ref{nodal-line-h2})]. 
    The parameters of these models are the same as in Table~\ref{fitting_table_SM}. 
    Both models are characterized by the parameters $\Delta = 0$, $t_{\|} = \sinh g, t_{\|}^{\prime} = \cosh g$, $ t_{\perp} = 1$. The column ``parameter'' specifies the value of $g$. In the two models, the non-localized regions extend from $W=0$ to $W=W^{(z)}_{c}$, and the quasi-localized phases extend from $W^{(x)}_{c}$ to $W^{(z)}_{c}$.}
    \begin{tabular}{cccccc}
    \hline
    \hline
\makecell[c]{symmetry \\ class} & ~model~ & ~parameter~  & $W_c^{(x)}$ & $W_c^{(z)}$ \\
    \hline
      BDI & ${\cal H}$ & $g = 0.22$    & ~21.73[21.66,21.80]~ & ~23.4[23.3,23.5]~ \\
       BDI & ${\cal H}_2$ &$g = 1$    & 41.36[41.26,41.46] & 45.3[45.2,45.4]\\
       BDI&${\cal H}_2$ &%
       $g =0.2$ &  23.23[23.13,23.42] & 24.7[24.6,24.8]\\
     \hline
    \end{tabular}
    \label{W_x_z_1}
\end{table}

For the directions with non-zero weak topological indices 
[i.e., $z$ direction in Eq.~(\ref{NDSM-NH}) and $y,z$ directions in 
Eq.~(\ref{NDSM-NH2})], 
we calculate the LEs of the right-upper part 
[i.e., $h$ in Eq.~(\ref{NDSM-NH}) and $h_2$ in Eq.~(\ref{NDSM-NH2})] 
as a function of the disorder strength $W$. 
In all these models, the distributions of the LEs for large $L$ show  
the $W$-dependence described in Fig.~\ref{LE_SM}. For $W = 0$, the $2L^2$ LEs 
form a continuum spectrum in the large $L$ limit, including zero $\gamma = 0$. When $W$ increases, the spectrum splits into 
the two continuous spectra. For $W>W^{(z)}_{c}$, all the $L^2$ LEs in the lower 
spectrum become negative. 
The non-localized region extends from $W=0$ to $W=W^{(z)}_{c}$, while the Anderson 
insulator phase appears in $W>W^{(z)}_{c}$. 
$W^{(z)}_{c}$ is determined by the finite-size scaling of $\gamma^{(1)}_{\rm max}$ as in Sec.~\ref{supplement-sec-Lyapunov-scaling}, which is summarized in Table~\ref{W_x_z_1}. 

For the directions with zero weak topological indices [i.e., $x,y$ directions 
in Eq.~(\ref{NDSM-NH}) and $x$ direction in Eq.~(\ref{NDSM-NH2})], we calculate the localization 
length $\xi_x$ in the quasi-1D geometry ($L_{x}\times L \times L$, 
$L_x \gg L$). 
Figure~\ref{SM_lambda_x} shows the normalized localization length 
$\Lambda_x \equiv \xi_x/L$ as a function of the disorder strength $W$. 
In all these models, $\Lambda_x$ shows scale-invariant behavior inside the 
non-localized region. 
Fitting $\Lambda_x$ around the scale-invariant points by the scaling functions of 
Eqs.~(\ref{fit1}) and (\ref{fit2}), we evaluate the critical exponent $\nu$ 
and the critical disorder strength $W_c^{(x)}$, 
as summarized in Table~\ref{fitting_table_SM}.  
The critical disorder strength $W_c^{(x)}$ is far below $W_c^{(z)}$, 
demonstrating the presence of the quasi-localized phases for 
$W_c^{(x)} < W  <W_c^{(z)}$ in these three models (Table~\ref{W_x_z_1}). 
The evaluated critical exponents are consistent with 
the critical exponent $\nu = 0.820[0.783,0.846]$ shown 
in the main text for the nodal-line semimetal in class BDI.
This consistency suggests that
all the phase transitions between the metal and quasi-localized phases in 
the 3D models in symmetry class BDI are of the same nature.

Note also that $\nu$ in the nodal-line semimetal model of Eq.~(\ref{NDSM-NH}) with $t_{\|} = \cosh g, t_{\|}^{\prime} = \sinh g$ ($g=0.22$) 
shows the larger error bars in their fitting results (see Table~\ref{fitting_table_SM}). 
These larger error bars may stem from a severe crossover effect. 
For $ t_{\|}^{\prime} = \sinh g = 0$, the nodal-line semimetal model has an extra unitary symmetry 
${\cal H} =\sigma_z {\cal H} \sigma_z$. 
The Hamiltonian can be 
block-diagonalized into two parts, and each block
belongs to 
orthogonal class.
For nonzero but small $g$, 
this unitary symmetry is only weakly broken. 
Thus, the finite-size 
systems with smaller $g$ must suffer from a stronger crossover effect.

\subsubsection{Weak topological insulators and ordinary insulators in class BDI}
Zero-energy states of the nodal-line semimetal model in Eq.~(\ref{nodal-line-h}) can be either in a topological 
insulator state with $(\nu_x,\nu_y,\nu_z) = (0,0,1)$ or in 
an ordinary insulator state with $(\nu_x,\nu_y,\nu_z) = (0,0,0)$, 
depending on its tight-binding parameters. For simplicity, 
we assume $\Delta, t_{\perp}, t_{\|} > 0 $ in Eq.~(\ref{nodal-line-h}). 
For $\Delta + 4t_{\perp} < 2|t_{\|}^{\prime}|$ ($\Delta - 4t_{\perp} > 2|t_{\|}^{\prime}|$), 
the zero-energy states of ${\cal H}$ in Eq.~(\ref{nodal-line-h}) are in 
the topological (ordinary) insulator state in the clean limit ($W=0$).

In Ref.~\cite{Luo20}, the localization lengths of Eq.~(\ref{nodal-line-h}) 
along the $x$ direction were calculated with the quasi-1D geometry 
($L_x \times L \times L$, $L_x\gg L$), 
and the two consecutive 
disorder-driven phase transitions were identified for the following set of parameters:
\begin{equation}
    t_{\perp} = 3/10,\quad  
    t^{\prime}_{\|} = 1,\quad
    t_{\|} = 1/4,\quad
    \Delta = 1/2\quad
    \text{(topological insulator in the clean limit).}
    \label{para-0.5}
\end{equation}
The two phase transitions are 
i) a transition from the topological insulator phase to the 
diffusive metal phase at $W^{(x)}_{c,1}=3.135 \!\ [3.132,3.138]$
and ii) a transition from 
the diffusive metal phase to the Anderson insulator phase 
at $W^{(x)}_{c,2}=11.96 \!\ [11.92, 12.02]$, respectively.
In addition, Ref.~\cite{Luo20} studied another parameter set, 
\begin{align}
    t_{\perp} = 3/10,\quad  
    t^{\prime}_{\|} = 1,\quad 
    t_{\|} = 1/4,\quad  
    \Delta = 4\quad 
    \text{(ordinary insulator in the clean limit),}
    \label{para-4.0}
\end{align}
where a disorder-driven phase transition from the 3D ordinary 
band insulator phase to the diffusive metal phase 
[Fig.~\ref{NDSM_p2}] was found at 
$W^{(x)}_{c,3}=4.76 \!\ [4.75,4.77]$. 
The normalized localization length $\Lambda_x \equiv \xi_x/L$ 
shows scale-invariant behavior at these critical 
disorder strengths. From the finite-size scaling analyses, 
it was clarified that these phase transitions are universally 
characterized by the same critical exponent $\nu=0.82 \pm 0.04$~\cite{Luo20}, which is consistent with the evaluations of the disordered nodal-line semimetal models studied in the main text.

\begin{figure}
    \centering
	\subfigure[$t_{\perp} = 3/10,  t^{\prime}_{\|} = 1, t_{\|} = 1/4,  \Delta = 1/2$]{
		\begin{minipage}[t]{0.5\linewidth}
			\centering
			\includegraphics[width=1\linewidth]{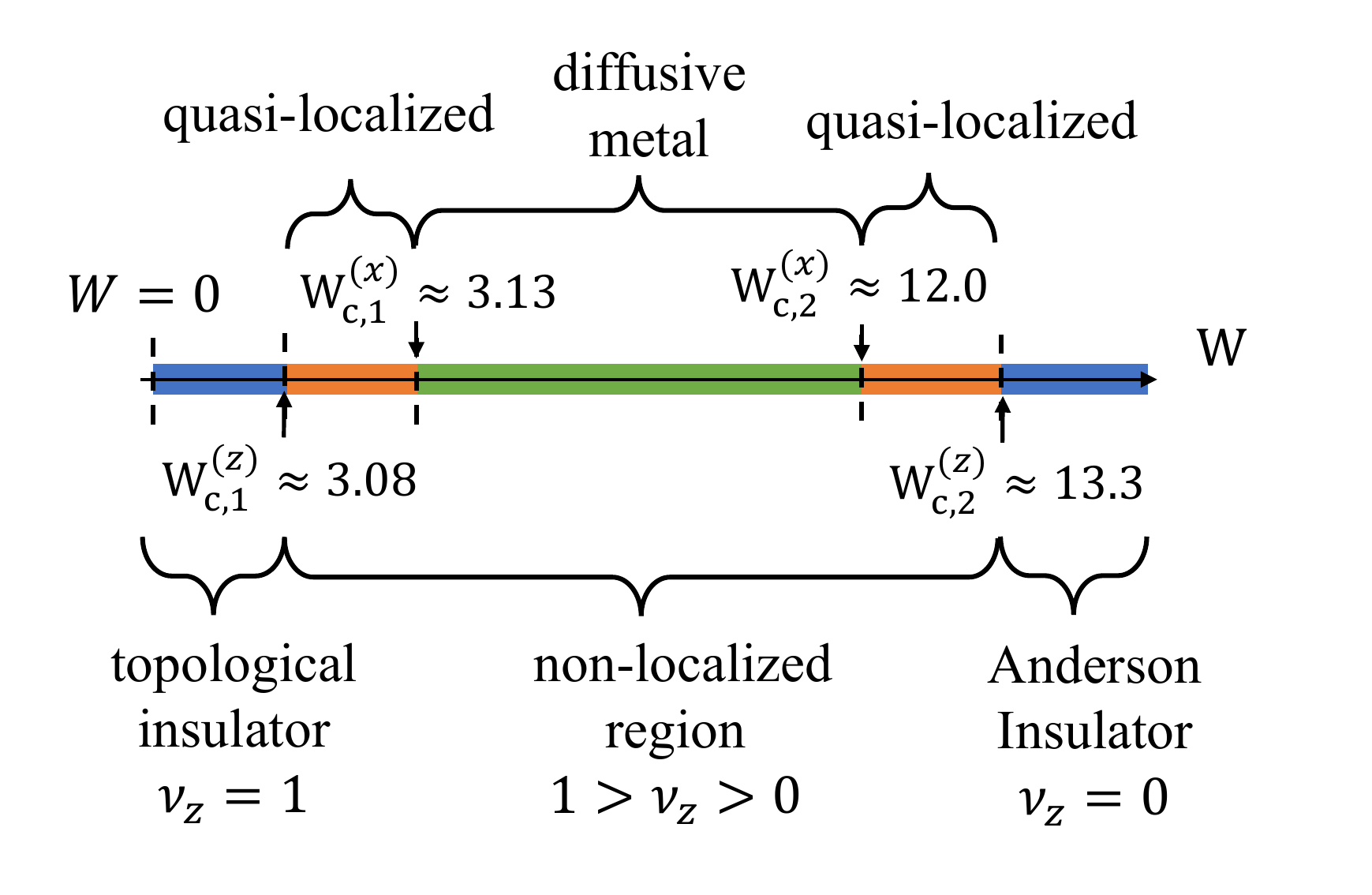}
        \label{NDSM_p1}
		\end{minipage}%
	}%
    \subfigure[$t_{\perp} = 3/10,  t^{\prime}_{\|} = 1, t_{\|} = 1/4,  \Delta = 4$]{
		\begin{minipage}[t]{0.5\linewidth}
			\centering
			\includegraphics[width=1\linewidth]{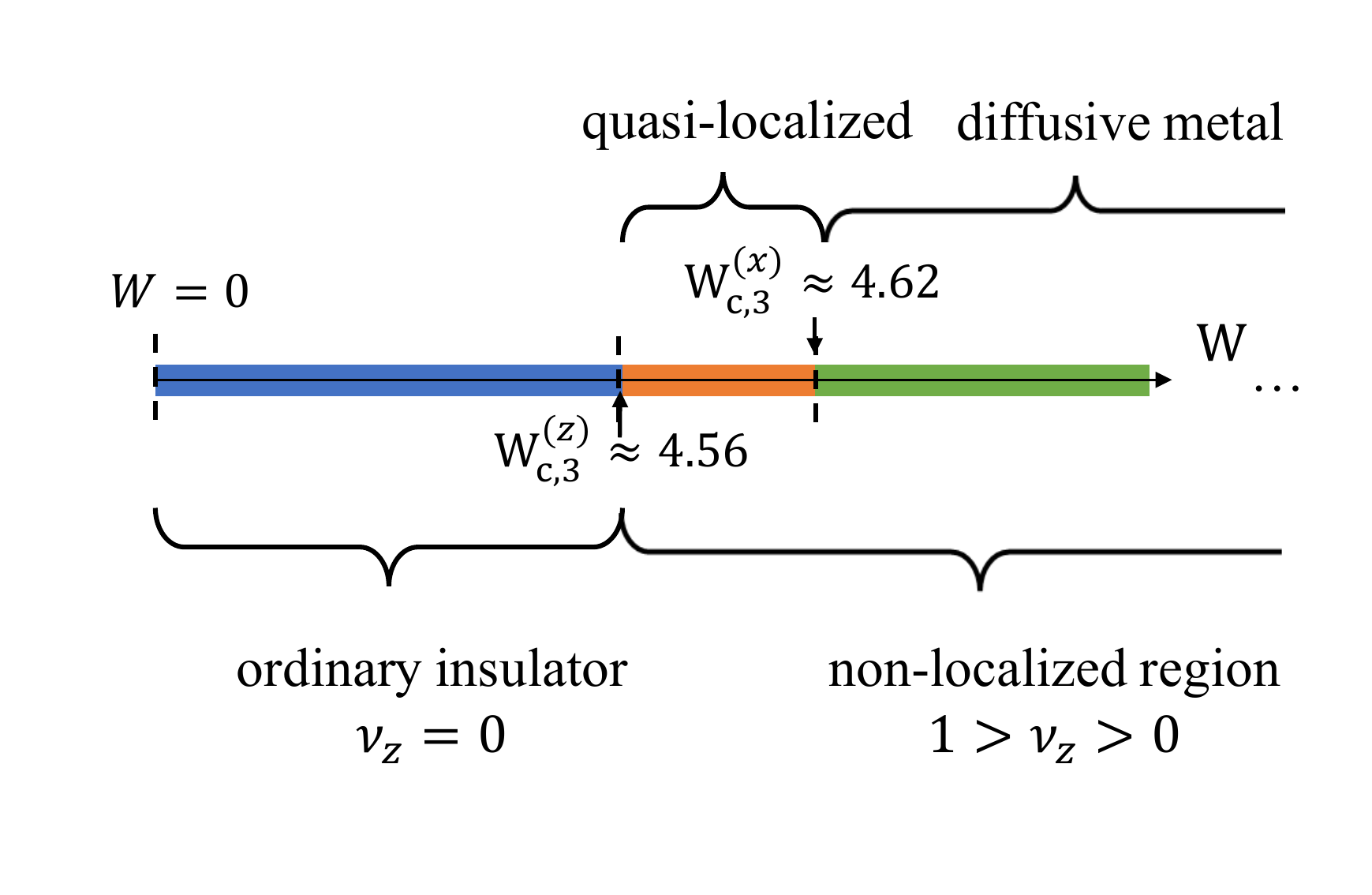}
		\label{NDSM_p2}
		\end{minipage}%
	}%

    \subfigure[$\gamma-W$ in (a)]{
		\begin{minipage}[t]{0.5\linewidth}
			\centering
			\includegraphics[width=1\linewidth]{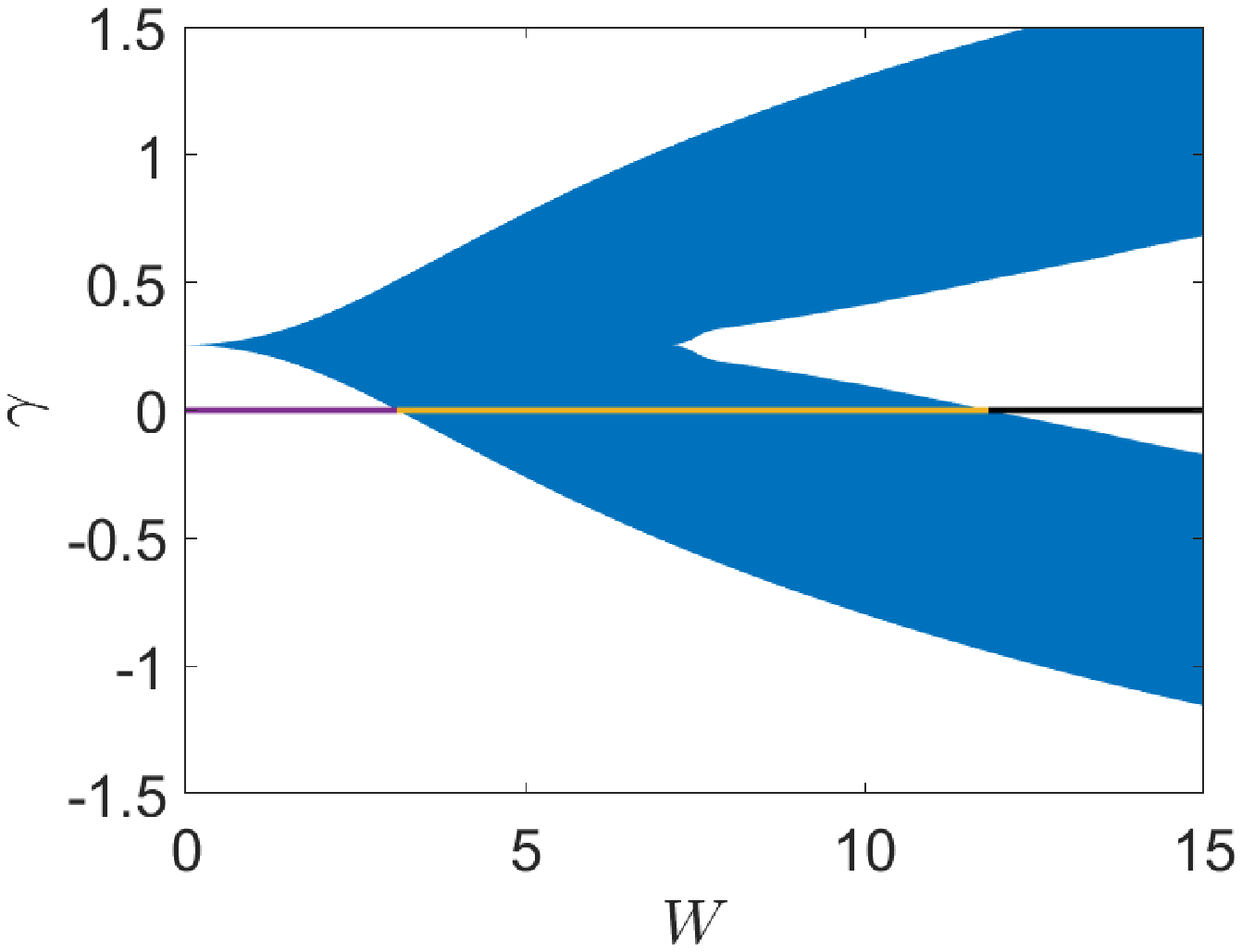}
        \label{LE_M1}
		\end{minipage}%
	}%
    \subfigure[$\gamma-W$ in (b)]{
		\begin{minipage}[t]{0.5\linewidth}
			\centering
			\includegraphics[width=1\linewidth]{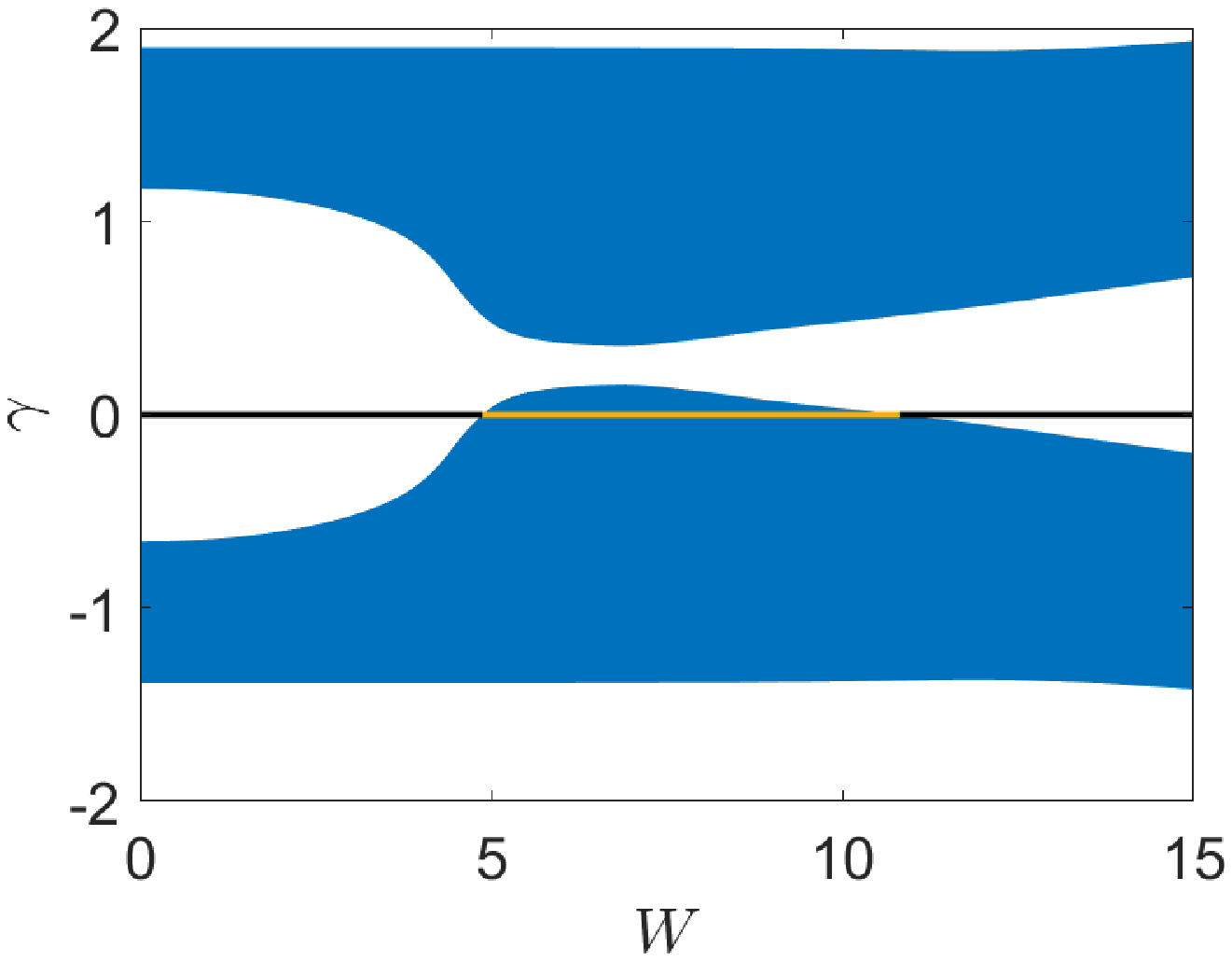}
		\label{LE_M2}
		\end{minipage}%
	}%
	\caption{(a,b)~Schematic phase diagrams of ${\cal H}$ in different parameter sets, (a) weak topological and (b) ordinary insulator sides. $W_{c,i}^{(x)} (i = 1,2,3)$ stand for critical points of the Anderson transitions 
    in the $x$ or $y$ direction~\cite{Luo20}.
    The weak topological index $\nu_z$ is finite for  $W_{c,1}^{(z)}<W<W_{c,2}^{(z)}$ 
    of (a) and for $W^{(z)}_{c,3}<W$ of (b). The $2L^2$ Lyapunov exponents of $h$ in the same parameter 
    sets as (a) and (b) are shown in (c) and (d), respectively.
    }
    \label{LE_phase}
\end{figure}

For these two sets of parameters, we study the localization length and the winding 
number along the $z$ direction. 
We calculate the LEs of the right-upper part 
$h$ of ${\cal H}$ in the canonical basis [i.e., Eq.~(\ref{NDSM-NH})]
with the quasi-1D geometry ($L\times L \times L_z$, $L_z\gg L$). 
For the parameters in Eq.~(\ref{para-0.5}), the LEs show a 
$W$-dependence described as Fig.~\ref{LE_M1}. 
The topological insulator is stable under weak disorder. 
For $W<W^{(z)}_{c,1}$, all the $2L^2$ LEs are positive, and
the localization length is finite. 
The winding number 
$w_{z}$ along the $z$ direction is $L^2$, giving rise to
$\nu_z = 1$. 
The non-localized region appears 
from $W=W^{(z)}_{c,1}$ to $W=W^{(z)}_{c,2}$ ($>W^{(z)}_{c,1}$), where  
a continuous spectrum of LEs includes zero $\gamma=0$. When $W$ increases 
from $W^{(z)}_{c,1}$ to $W^{(z)}_{c,2}$, $L^2$ positive LEs cross 
zero and become negative; $\nu_z$ changes from 1 to 0. 
For $W>W^{(z)}_{c,2}$, the $L^2$ LEs are positive and the other $L^2$ LEs 
are negative, leading to $\nu_z = 0$. 
For the parameters in Eq.~(\ref{para-4.0}), 
the LEs shows a $W$-dependence described as Fig.~\ref{LE_M2}, where 
the non-localized region appears from $W=W^{(z)}_{c,3}$ to $W=W^{(z)}_{c,4}$ 
($>W^{(z)}_{c,3}$). 
In the non-localized region, the number of positive LEs is 
greater than the number of negative ones, leading to $\nu_z > 0$.

From the finite-size scaling analyses of the minimal or 
maximal LEs by Eqs.~(\ref{3-0a}), (\ref{3-0b}), or (\ref{3-0c}), 
we determine the phase boundaries of the non-localized regions as 
$W^{(z)}_{c,1}=3.08 \!\ [3.07,3.09]$, $W^{(z)}_{c,2}=13.3 \!\ [13.2,13.4]$, and 
$W^{(z)}_{c,3}=4.56 \!\ [4.55,4.57]$ (see Table~\ref{gamma_m_fit2}).  
From a comparison of these numbers with $W^{(x)}_{c}$ 
obtained in Ref.~\cite{Luo20} (see Table~\ref{W_x_z_3}), we conclude 
that the quasi-localized phases 
appear inside the non-localized regions: 

\begin{table}[t]
    \centering
    \caption{
    Finite-size scaling analyses of $\gamma^{(1)}_{\rm max}(W,L)$ or $\gamma^{(1)}_{\rm min}(W,L)$ for the several disorder strength $W$ around $W = W_c^{(z)}$  
    for the disordered topological insulator model
    with the parameters in Eq.~(\ref{para-0.5}) (shown as ``P1" in the ``parameter set"), 
    and the disordered ordinary insulator model with the parameters in Eq.~(\ref{para-4.0}) (shown as ``P3" in the ``parameter set"). The square brackets are the 95\% confidence error 
    bars determined by the Monte Carlo analyses.}
    \centering
    
    \begin{minipage}[t]{1\linewidth}
    \begin{tabular}{cccccccccc}
    \hline \hline
    ~parameter set~ & $W$ & $L$ & $\gamma_{\rm min}(W,L)$ & $a$ & ~GOF~ \\ \hline
    P1  & ~3.07~ & ~24 - 60~ & ~0.0013 [0.0012,0.0014]~ & ~0.050 [0.046,0.053]~ & 0.64\\
    P1  & 3.08 & 24 - 60 & -0.0002 [-0.0003,-0.0000] & 0.051 [0.047,0.054] & 0.90\\
    P1  & 3.09 & 24 - 60 & -0.0015 [-0.0016,-0.0014] & 0.048 [0.045,0.051] & 0.46\\
    \hline
    \hline
    \quad \quad
    \end{tabular}
    \end{minipage}
    
    \begin{minipage}[t]{1\linewidth}
    \begin{tabular}{cccccccccc}
    \hline \hline
    ~parameter set~ & $W$ & $L$ & $\gamma^{(1)}_{\rm max}(W,L)$ & $a$ & ~GOF~ \\ \hline
    P1  & ~13.2~ & ~14 - 28~ & ~0.006 [0.002,0.010]~ & ~-0.967 [-1.048,-0.889]~ & 0.15\\
    P1  & 13.3 & 14 - 28 & 0.0007 [-0.003,0.004] & -0.954 [-1.027,-0.880] & 0.74\\
    P1  & 13.4 & 14 - 28 & -0.006 [-0.010,-0.002] & -0.951 [-1.026,-0.870] & 0.11\\
    P3  & 4.60 & 14 - 28 & -0.007 [-0.011,-0.0028] & -1.177 [-1.257,-1.097] & 0.74  \\  
    P3  & 4.62 & 14 - 28 & -0.003 [-0.007,0.001] & -1.091 [-1.163,-1.009] & 0.19 \\
    P3  &4.63 & 14 - 28 & 0.007 [0.002,0.0109] & -1.150 [-1.231,-1.064] & 0.79    \\
    \hline
    \hline
    \quad \quad
    \end{tabular}
    \end{minipage}
    \label{gamma_m_fit2}
\end{table}

\begin{table}[t]
    \centering
    \caption{Comparison of the critical disorder strengths, $W^{(z)}_c$ and $W^{(x)}_c$, 
    in the disordered topological insulator model with the parameters 
    in Eq.~(\ref{para-0.5}) (shown as ``T1" and ``T2" in the ``transition"), and the disordered ordinary insulator 
    model with the parameters in Eq.~(\ref{para-4.0}) (shown as ``T3" in the ``transition"). 
    The square 
    brackets are
    the 95\% confidence error 
    bars determined by 
    the synthetic data. 
    }
  \begin{minipage}[t]{\linewidth}
    \begin{tabular}{ccccc}
    \hline
    \hline
    \makecell[c]{symmetry \\ class} & ~transition~ & $W_c^{(x)}$ & $W_c^{(z)}$ \\
    \hline
    BDI &     T1  & 3.135[3.132,3.138]\footnote{from Ref.~\cite{Luo20}}& ~3.08[3.07,3.09]~ \\
      BDI&  T2 & 11.96[11.92,12.02]\footnotemark[1]&13.3[13.2,13.4] \\
      BDI &  T3 &  4.62[4.60,4.63]\footnotemark[1]& 4.56[4.55,4.57]\\
     \hline
    \end{tabular}
  \end{minipage}
    \label{W_x_z_3}
\end{table}

\noindent
\begin{align}
    \left\{\begin{array}{lccl} 
    W<W^{(z)}_{c,1} & & & (\text{topological insulator phase}), \\
     W^{(z)}_{c,1}<W<W^{(x)}_{c,1}& & & (\text{quasi-localized phase}), \\
     W^{(x)}_{c,1}<W<W^{(x)}_{c,2}& & & (\text{diffusive metal phase}), \\
     W^{(x)}_{c,2}<W<W^{(z)}_{c,2}& & & (\text{quasi-localized phase}), \\
     W^{(z)}_{c,2}<W& & & (\text{Anderson insulator phase}), \\
    \end{array}\right. 
\end{align}
for the parameters in Eq.~(\ref{para-0.5}) and 
\begin{align}
    \left\{\begin{array}{lccl} 
    W<W^{(z)}_{c,3} & & & (\text{ordinary insulator phase}), \\
     W^{(z)}_{c,3}<W<W^{(x)}_{c,3}& & & (\text{quasi-localized phase}), \\
     W^{(x)}_{c,3}<W<... & & & (\text{diffusive metal phase}), \\ 
\end{array}\right.
\end{align}
for Eq.~(\ref{para-4.0}). 
Here, ``...'' means that when $W$ is further increased, 
the system undergoes a transition from the diffusive metal phase to the quasi-localized phase at $W^{(x)}_{c,4}$, and a transition from the quasi-localized phase to the Anderson insulator phase at $W^{(z)}_{c,4}$, but the respective critical disorder strengths $W^{(x)}_{c,4}$ and $W^{(z)}_{c,4}$ are not determined.  
The phase diagrams of the 
disordered topological insulator and ordinary insulator
are shown in Figs.~\ref{NDSM_p1} and \ref{NDSM_p2}.

\subsection{Anderson transitions in chiral-symmetric models with no weak topological indices} 
    \label{supplement-sec-Q1D-triv}

For comparison, we study three-dimensional (3D) chiral-symmetric models 
in symmetry classes BDI and AIII, where statistical symmetries 
enforce all the three topological indices to be zero. 
We refer 
to these models as non-topological models.
Notably, the disordered 
ordinary insulator model in Eq.~(\ref{nodal-line-h}) with the parameters in Eq.~(\ref{para-4.0}) 
is a topological model because non-zero $\nu_z$ is induced by the disorder 
[see also Figs.~\ref{LE_M2} and \ref{NDSM_p2}]. Non-topological models have the 
following three features that are distinct from the topological models 
in the same chiral symmetry classes: 
\begin{enumerate}
    \item In the quasi-1D geometry ($L \times L\times L_{\mu}$, $L_{\mu}\gg L$), 
    the localization length along any spatial direction 
    is always finite with finite $L$. 
    In the topological models with $\nu_{z}\ne 0$, the localization length along the $z$ 
    direction can diverge for finite $L$ when the 1D winding $w_{z}$ changes. 
    \item In the thermodynamic limit $L \rightarrow \infty$, the localization lengths along all 
    the spatial directions diverge at the same critical point, which implies no quasi-localized phase. 
    In the topological models, the localization length along the $z$ direction and 
    those along the other two directions diverge at different critical points in the thermodynamic limit, which gives rise to  
    the quasi-localized phase. 
    \item  The divergence of the localization length along all the directions is characterized by the same critical exponent in the non-topological models. 
    In the topological models, the divergence of the localization length along the $z$ direction and those along the other directions are characterized by the different critical exponents. 
    The two exponents in the topological models are also different from 
    the exponents in the non-topological models in the same symmetry class. 
\end{enumerate}

\subsubsection{Three-dimensional non-topological models in classes BDI and AIII}
Let us introduce the following non-topological chiral-symmetric model that belongs to symmetry class BDI 
or AIII,  
\begin{equation}
    {\cal H}_{0} = \sum_{\bm r=(x,y,z)} \left\{ (\Delta + \epsilon_{\bm{r}}) c^{\dagger}_{\bm{r}} \sigma_z c_{\bm{r}} + \epsilon^{\prime}_{\bm r} c^{\dagger}_{\bm r} \sigma_y c_{\bm r} + \left[ 
    \sum_{\mu = x,y} \left(  t_{\perp}c^{\dagger}_{\bm r+\bm{e_{\mu}}} \sigma_0 c_{\bm{r}}  \right)   
     +t_{\|}   c^{\dagger}_{\bm r+\bm{ e_z}}  \sigma_z c_{\bm{r}} + {\rm i}t_{\|}^{\prime}   c^{\dagger}_{\bm r+\bm{ e_z}}  \sigma_y c_{\bm{r}} +\text{H.c.} \right]  \right\}, 
     \label{h1-sup}
\end{equation}
where $c_{\bm{r}}$ is a two-component annihilation operator at the cubic-lattice 
site ${\bm r} \equiv (r_x,r_y,r_z)$, ${\bm e}_{\mu}$'s ($\mu=x,y,z$) are the unit vectors 
connecting the nearest neighbor cubic-lattice sites, 
$\sigma_{\mu}$'s ($\mu = 0,x,y,z$) are the two-by-two unit matrix and Pauli matrices for the two orbitals, 
$\Delta, t_{\perp}, t_{\|}, t_{\|}^{\prime}$ are the real parameters, 
and $\epsilon_{\bm r}$ and $\epsilon^{\prime}_{\bm r}$ 
are the real-valued on-site random potential. 
We choose the parameters to be $\Delta = 0, t_{\perp} = 1, t_{\|} = 13/12, t_{\|}^{\prime} = 5/12$. 
The model in the clean limit ($\epsilon_{\bm r} \equiv \epsilon^{\prime}_{\bm r} \equiv 0$) 
has a finite density of states at $E = 0$. 
${\cal H}_{0}$ respects chiral symmetry, 
${\cal H}_{0} =-\mathcal{C}^{\dagger}{\cal H}_{0}^{\dagger} \mathcal{C}$ 
with the chiral operator
\begin{align}
\mathcal{C}_{ {\bm r , \bm r^{\prime}}} \equiv (-1)^{x+y}\delta_{{\bm r , \bm r^{\prime}}} \sigma_x, 
\end{align}
satisfying ${\cal C} = {\cal C}^T$.
For $\epsilon^{\prime}_{\bm r} = 0$, ${\cal H}_0$ 
respects time-reversal symmetry ${\cal H}_{0} = {\cal H}^*_{0}$ and hence belongs to the chiral orthogonal class 
(class BDI). 
For $\epsilon^{\prime}_{\bm r} \ne 0$, time-reversal symmetry is broken, and ${\cal H}_0$ belongs to 
the chiral unitary class (class AIII). 
For ${\cal H}_0$ in class BDI, we choose $\epsilon_{\bm r}$ 
to be uniformly distributed in $[-W/2,W/2]$. 
For ${\cal H}_0$ in class AIII, on the other hand, we choose $\epsilon_{\bm r}$ 
and $\epsilon^{\prime}_{\bm r}$ to be uniformly distributed 
for $\epsilon^2_{\bm r} + {\epsilon^{\prime}_{\bm r}}^2 \leq W^2$.

Following Eq.~(\ref{relation}), we decompose ${\cal H}_0$ into the block-off-diagonal structure in a basis that diagonalizes the chiral operator ${\cal C}$.
The right-upper part $h_0$ of ${\cal H}_0$ 
is regarded as a single-orbital tight-binding model on the cubic lattice, 
\begin{align}
        h_{0} &= \sum_{{\bm r}} \left[ (\Delta + \epsilon_{\bm r} + {\rm i } \epsilon_{\bm r}^{\prime}) f^{\dagger}_{\bm r} f_{\bm r} + \sum_{\mu = x,y}  \left(t_{\perp} f^{\dagger}_{{\bm r}+{\bm e}_{\mu}}  f_{\bm{r}}  +\text{H.c.}  \right) 
        +\left(t_{\|} - (-1)^{r_x+r_y} t_{\|}^{\prime} \right)  f^{\dagger}_{{\bm r}+{\bm e}_z}  f_{\bm r} +
         \left(t_{\|} + (-1)^{r_x+r_y} t_{\|}^{\prime} \right)  f^{\dagger}_{\bm r}  f_{{\bm r}+ {\bm e}_z}
         \right] \, . \label{h3-nh} 
\end{align}
Transposition exchanges $t_{\|} - (-1)^{r_x+r_y} t^{\prime}_{\|}$ and $t_{\|} + (-1)^{r_x+r_y} t^{\prime}_{\|}$ in $h_0$. 
Thus, 
as a unitary transformation in Eqs.~(\ref{statistical-symmetry}) and (\ref{unitary}), 
we apply a spatial translation along the $x$ or $y$ direction  
by ${\bm e}_{x}$ or ${\bm e}_y$, 
\begin{align}
    {\cal U}_{(r_x,r_y,r_z|r^{\prime}_x,r^{\prime}_y,r^{\prime}_z)} 
    = \left\{\begin{array}{c}
    \delta_{r_x,r^{\prime}_x+1} \delta_{r_y,r^{\prime}_y} \delta_{r_z,r^{\prime}_z}, \\
    \delta_{r_x,r^{\prime}_x} \delta_{r_y,r^{\prime}_y+1} \delta_{r_z,r^{\prime}_z}. \\
    \end{array}\right. \label{unitary3-0}
\end{align}
Instead of Eq.~(\ref{unitary3-0}), we can also use 
a mirror operation with respect to 
the $r_x=1/2$ plane or the $r_y =1/2$ plane 
\begin{align}
    {\cal U}_{(r_x,r_y,r_z|r^{\prime}_x,r^{\prime}_y,r^{\prime}_z)} 
    = \left\{\begin{array}{c}
    \delta_{r_x+r^{\prime}_x,1} \delta_{r_y,r^{\prime}_y} \delta_{r_z,r^{\prime}_z}, \\ 
    \delta_{r_x,r^{\prime}_x} \delta_{r_y+r^{\prime}_y,1} \delta_{r_z,r^{\prime}_z}. \\ 
    \end{array}\right. \label{unitary3}
\end{align}
Since $\epsilon_{\bm r}$ ($\epsilon^{\prime}_{\bm r}$) at different lattice points ${\bm r}$
is statistically equivalent, an ensemble of $h_0$ defined in 
Eq.~(\ref{h3-nh}) is statistically invariant under 
the combination of transposition and any of these unitary transformations.
These statistical symmetries require the LEs of $h_0$ along all the 
directions to come in opposite-sign pairs, leading to $\nu_x=\nu_y=\nu_z=0$.

For ${\cal H}_0$ in symmetry class BDI, we calculate the localization lengths $\xi_x,\xi_z$ along the $x,z$ directions with the quasi-1D geometry $L^2 \times L_{\mu}$ ($\mu = x,z$; $L_{\mu} \gg L$).
Because of chiral symmetry, it is sufficient to calculate 
the product of the transfer matrices of  $h_{0}$ [see Eq.~(\ref{TMv1})]. 
Because of the statistical symmetries, the LEs of $h_{0}$ come in opposite-sign pairs. 
Both normalized localization length $\Lambda_{x} \equiv \xi_x/L $ and $\Lambda_{z} \equiv \xi_z/L$ show scale-invariant behavior around the same critical 
disorder strength $W \approx 23$ (see Fig.~\ref{non_topo_BDI}). From the 
fitting by the polynomial expansion of the finite-size scaling function 
[Eqs.~(\ref{fit1}) and (\ref{fit2})], we determine the critical disorder 
strength and the critical exponent 
(see the fourth to seventh
rows of Table~\ref{fitting_table_SM_0}). 
The critical disorder strength and 
exponent determined from $\Lambda_z$ and those determined 
from $\Lambda_x$ are consistent with each other. 
The critical exponent is $\nu = 1.089 \, [1.005,1.128]$, 
and different 
from the two exponents ($\nu = 0.820\,[0.787,0.848]$ and $\nu^{\prime} = 1$) 
of the topological model in the same symmetry class (i.e., class BDI). 
Reference~\cite{Wang21} studied the localization length of ${\cal H}_{0}$ along the $z$ direction with the different parameters 
$\Delta = t_{\perp} =  t_{\|}^{\prime} = 1, t_{\|} = 1/2$ 
and evaluated the critical exponent to be $\nu = 1.119 [0.973.1.241]$, which is consistent with our evaluation.  

For ${\cal H}_0$ in symmetry class AIII, we calculate 
the normalized localization length $\Lambda_{z} = \xi_z/L$ with the quasi-1D geometry 
$L^2 \times L_{z}$ and $L_z \gg L$. 
$\Lambda_z$ shows scale-invariant behavior around the critical disorder strength $W \approx 8$ (see Fig.~\ref{non_topo_AIII}). From the fitting 
by the polynomial expansion of the finite-size scaling function 
[Eqs.~(\ref{fit1} and (\ref{fit2})], we determine the critical 
disorder strength and critical exponents. The critical 
exponent is $\nu = 1.024 \, [0.973,1.070]$ (see the last row of Table~\ref{fitting_table_SM_0})
and different from the two critical exponents 
$\nu = 0.824 \, [0.776,0.862] $ and $\nu^{\prime} = 1$ of the topological model in the same symmetry class (i.e., class AIII).

\begin{figure}[t]
    \centering
	\begin{minipage}[t]{0.9\linewidth}
			\centering
			\includegraphics[width=1\linewidth]{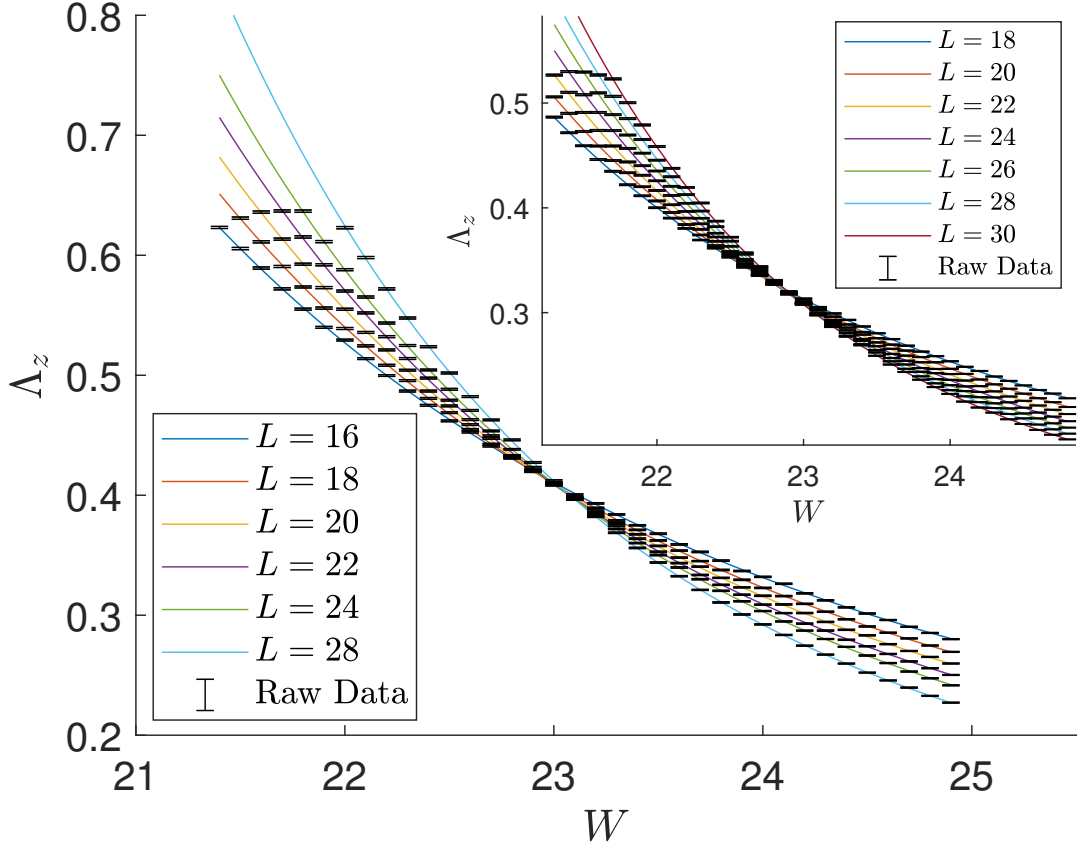}
		\end{minipage}%
			\caption{
            Normalized localization length $\Lambda_z \equiv \xi_z/L$ along the $z$ direction
            as a function of the disorder strength $W$ in the non-topological model 
            ${\cal H}_{0}$ ($\Delta = 0, t_{\perp} = 1, t_{\|} = 13/12, t_{\|}^{\prime} = 5/12$) 
            in class BDI [Eq.~(\ref{h1-sup})] with the quasi-1D geometry ($L\times L\times L_z$). 
            The black points are the raw data with the error bars. The solid lines for different $L$ 
            are the results of the fitting according to 
            Eqs.~(\ref{fit1}) and (\ref{fit2}) with 
            $(m,n) = (2,3)$.
            Inset: $\Lambda_x \equiv \xi_x/L$ as a function of $W$ in the same model.}
			\label{non_topo_BDI}
\end{figure}

\begin{figure}[bt]
    \centering
	\begin{minipage}[t]{0.8\linewidth}
			\centering
			\includegraphics[width=1\linewidth]{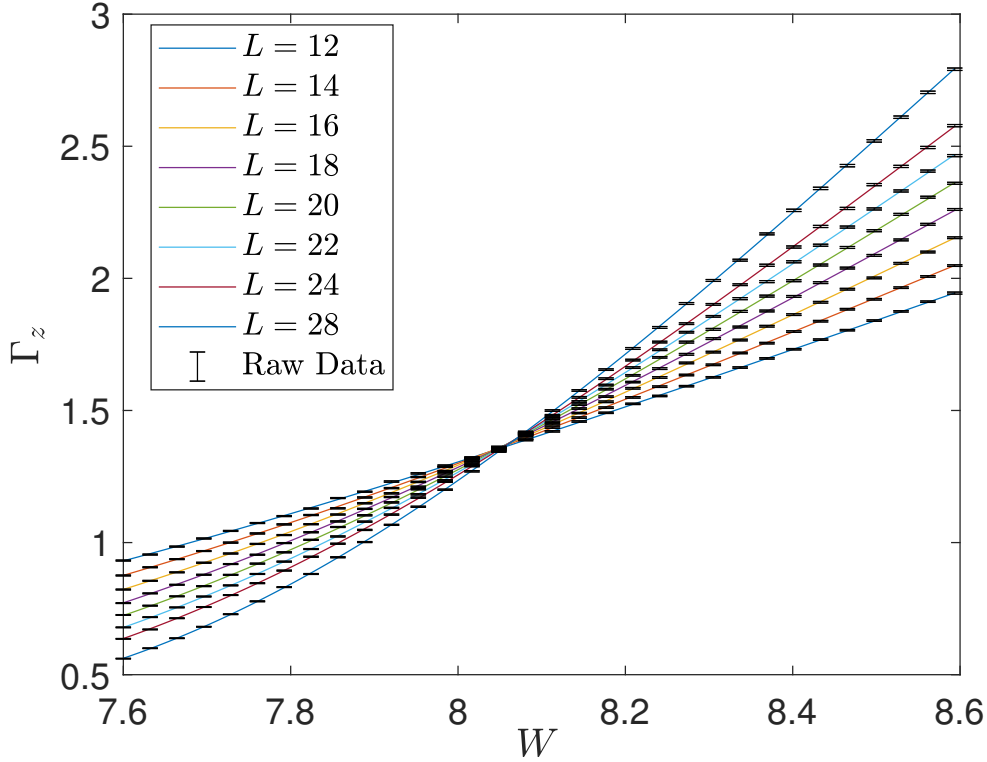}
		\end{minipage}%
			\caption{
            Inverse $\Gamma_z \equiv 1/ \Lambda_z \equiv \xi_z/L$ of the normalized localization length 
            along the $z$ direction as a function of the disorder strength $W$ in the non-topological model ${\cal H}^{\prime}$ ($\Delta = 0, t_{\perp} = 1, t_{\|} = 13/12, t_{\|}^{\prime} = 5/12$) in class AIII [Eq.~(\ref{h1-sup})] with the quasi-1D geometry ($L\times L\times L_z$). 
            The black points are the raw data with the error bars. 
            The solid lines for different $L$ are the results of the fitting according 
            to Eqs.~(\ref{fit1} and (\ref{fit2}) with $(m,n) = (2,3)$.
            }
			\label{non_topo_AIII}
\end{figure}

\subsubsection{Another three-dimensional non-topological model in class AIII} 
Reference~\cite{Wang21} introduced another 3D non-topological model 
in symmetry class AIII, 
\begin{equation}
    {\cal H}_{0}^{\prime}= \sum_{\bm r=(r_x,r_y,r_z)} \left\{ (\Delta + \epsilon_{\bm r}) 
c^{\dagger}_{\bm r} \sigma_z c_{\bm r} +  \left[ c^{\dagger}_{{\bm r}+{\bm e}_{x}} \left( t_1 \sigma_z + {\rm i}t_{\perp}\sigma_x \right) c_{\bm r}     
 +  c^{\dagger}_{{\bm r}+{\bm e}_y} \left( t_2 \sigma_0 + {\rm i}t_{\perp}\sigma_y \right) c_{\bm r}   +  t_{\|} c^{\dagger}_{{\bm r}+{\bm e}_z} \sigma_0 c_{\bm r} + \text{H.c.} \right]  \right\},  \label{non-topo-aiii}
\end{equation}
where the disorder potential $\epsilon_{\bm r}$ distributes uniformly in $[-W/2,W/2]$, and the parameters are chosen to be $\Delta = 0,t_{\perp}=3/5,t_{\|}=2/5,t_1 = t_2 = 1/2$. 
${\cal H}_{0}^{\prime}$ respects chiral symmetry 
${\cal H}_{0}^{\prime} =-\mathcal{C}^{\dagger}{{\cal H}^{\prime}_{0}}^{\dagger} \mathcal{C}$ with a chiral operator ${\cal C}$
\begin{align}
\mathcal{C}_{ {\bm r , \bm r^{\prime}}} \equiv (-1)^{y+z}\delta_{{\bm r , \bm r^{\prime}}} \sigma_y. 
\end{align}

In terms of Eq.~(\ref{relation}), ${\cal H}^{\prime}_0$ is decomposed into 
the block off-diagonal structure in a basis that diagonalizes the chiral operator. 
The right-upper part $h^{\prime}_0$ of ${\cal H}^{\prime}_0$ in this 
basis is given by a single-orbital tight-binding model on the cubic lattice site,  
\begin{align}
    h_{0}^{\prime}&= \sum_{\bm r=(r_x,r_y,r_z)} \left[ (\Delta + \epsilon_{\bm r}) 
    f^{\dagger}_{\bm r} f_{\bm r} +    \left( t_{1} + (-1)^{r_y+r_z} t_{\perp}\right)f^{\dagger}_{{\bm r}+{\bm e}_{x}}  f_{\bm r} +  \left( t_{1} - (-1)^{r_y+r_z} t_{\perp}\right)f^{\dagger}_{\bm r}  f_{{\bm r} +{\bm e}_{x}}   \right. \nonumber \\
     &\left. +\left( t_{2} - (-1)^{r_y+r_z} {\rm i} t_{\perp} \right)  f^{\dagger}_{{\bm r}+{\bm e}_y}  f_{\bm r} +
     \left( t_{2} - (-1)^{r_y+r_z} {\rm i} t_{\perp} \right)  f^{\dagger}_{\bm r}  f_{{\bm r}+ {\bm e}_y} + \left( t_{\|} 
     f^{\dagger}_{{\bm r} + {\bm e}_z} f_{\bm r} + \text{H.c.} \right)
     \right] \, .\label{h3-nh2}
    \end{align}
Hermitian conjugation exchanges $t_1 + (-1)^{r_y+r_z} t_{\|}$ 
and $t_1 - (-1)^{r_y+r_z} t_{\|}$, but transforms 
$t_2 - (-1)^{r_y+r_z} {\rm i} t_{\perp}$ into $t_2 + (-1)^{r_y+r_z} {\rm i} t_{\perp}$. 
Thus, as a unitary transformation in Eqs.~(\ref{unitary}) and (\ref{statistical-symmetry-2}), 
we apply a spatial translation along the $y$ or $z$ direction 
by ${\bm e}_{y}$ or ${\bm e}_z$, 
\begin{align}
    {\cal U}_{(r_x,r_y,r_z|r^{\prime}_x,r^{\prime}_y,r^{\prime}_z)} 
    = \left\{\begin{array}{c}
    \delta_{r_x,r^{\prime}_x} \delta_{r_y,r^{\prime}_y+1} \delta_{r_z,r^{\prime}_z}, \\
    \delta_{r_x,r^{\prime}_x} \delta_{r_y,r^{\prime}_y} \delta_{r_z,r^{\prime}_z+1} . \\
    \end{array}\right. \label{unitary2-0}
\end{align}
Instead of Eq.~(\ref{unitary2-0}), we can also 
use a mirror operation with respect to 
the $r_y = 1/2$ plane and the $r_z = 1/2$ plane,
\begin{align}
    {\cal U}_{(r_x,r_y,r_z|r^{\prime}_x,r^{\prime}_y,r^{\prime}_z)} 
    = \left\{\begin{array}{c}
    \delta_{r_x,r^{\prime}_x} \delta_{r_y+r^{\prime}_y,1} \delta_{r_z,r^{\prime}_z}, \\ 
    \delta_{r_x,r^{\prime}_x} \delta_{r_y,r^{\prime}_y} \delta_{r_z+r^{\prime}_z,1}. \\ 
    \end{array}\right. \label{unitary2}
\end{align}
Since $\epsilon_{\bm r}$ is statistically equivalent 
for different lattice points ${\bm r}$, an ensemble of $h^{\prime}_0$ defined in 
Eq.~(\ref{h3-nh2}) is statistically invariant under 
the combination of Hermitian conjugation and any of these unitary transformations.
These symmetries 
require the LEs of 
$h^{\prime}_0$ along all the directions 
to come in opposite-sign pairs, leading to $\nu_x=\nu_y=\nu_z=0$. 
Reference~\cite{Wang21} evaluated the critical exponent to be  
$1.059\,[1.022,1.100]$, which is consistent with our evaluation of the exponent in 
the non-topological model in symmetry class AIII. 
 
\subsection{Conductance 
and 
weak topological indices}
    \label{conductance}
    
In this section, we provide detailed numerical results of the two-terminal dimensionless conductance $g$ in the quasi-localized, metallic, and Anderson-localized phases. 
We show that 
the quasi-localized phase is characterized by the finite conductance along the direction with the divergent localization length and the vanishing conductance along the other directions. 
By contrast, the metallic and Anderson-localized phases 
exhibit the finite and vanishing conductance along all the directions, respectively. 
We also demonstrate large sample fluctuations of the conductance along the direction with the divergent length. 
These unique properties in the quasi-localized phase are of direct relevance to transport experiments. 
In fact, $g$ is directly related to the electric conductance $G$ and thermal conductance $G_T$ 
by $h G=e^2 g$ and $h G_T\propto k^2_BT g$, respectively, 
where $h$ is the Planck constant, $e$ is the elementary charge, $k_B$ is the Boltzmann constant, and $T$ is the temperature. 

  \begin{figure}[bt]
    \centering
	\subfigure[metallic phase ($W=20$)]{
		\begin{minipage}[t]{0.3\linewidth}
			\centering
			\includegraphics[width=1\linewidth]{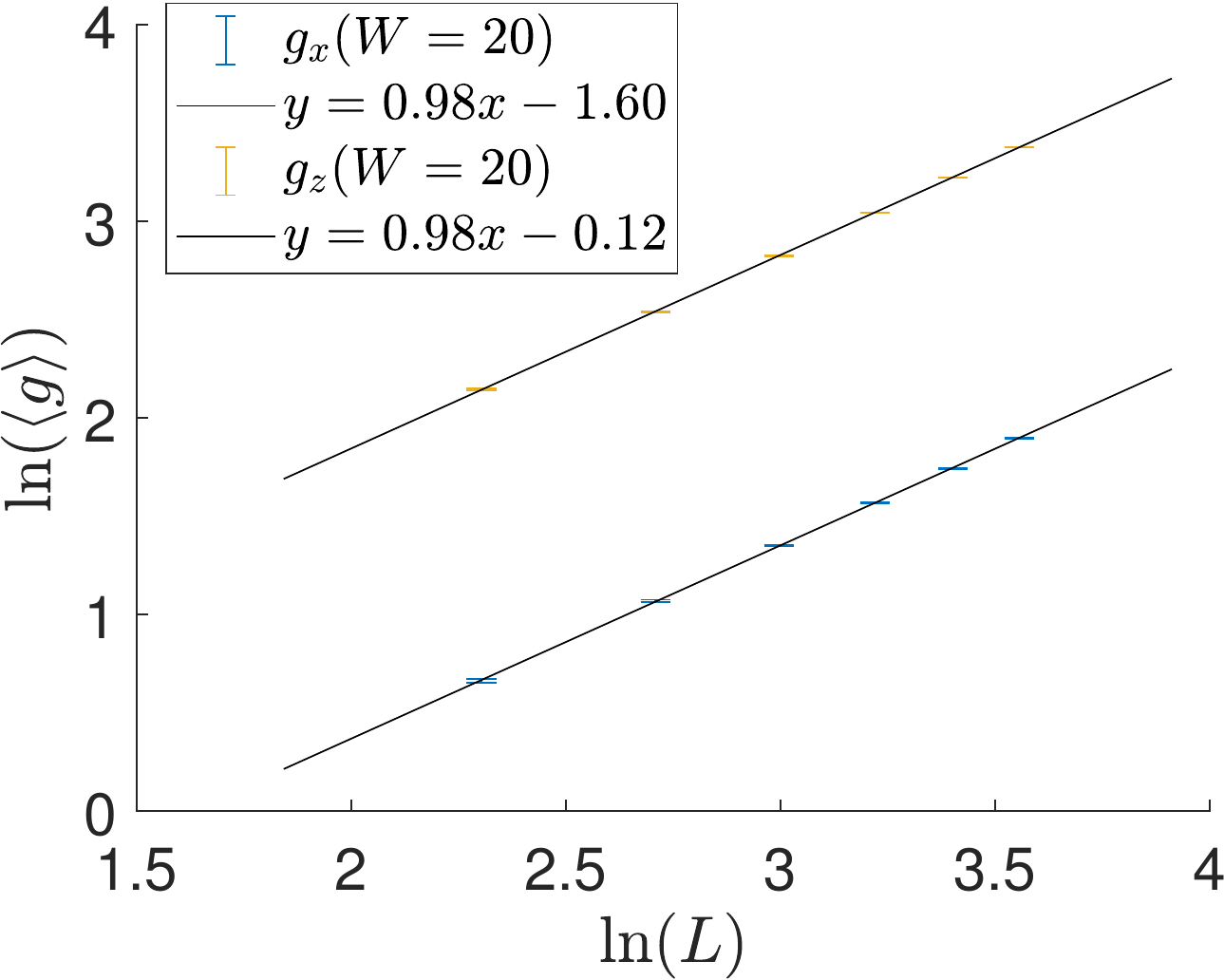}
            \label{metal_gxz}
		\end{minipage}%
	}%
	\subfigure[quasi-localized phase ($W=43.3$)]{
		\begin{minipage}[t]{0.3\linewidth}
			\centering
			\includegraphics[width=1\linewidth]{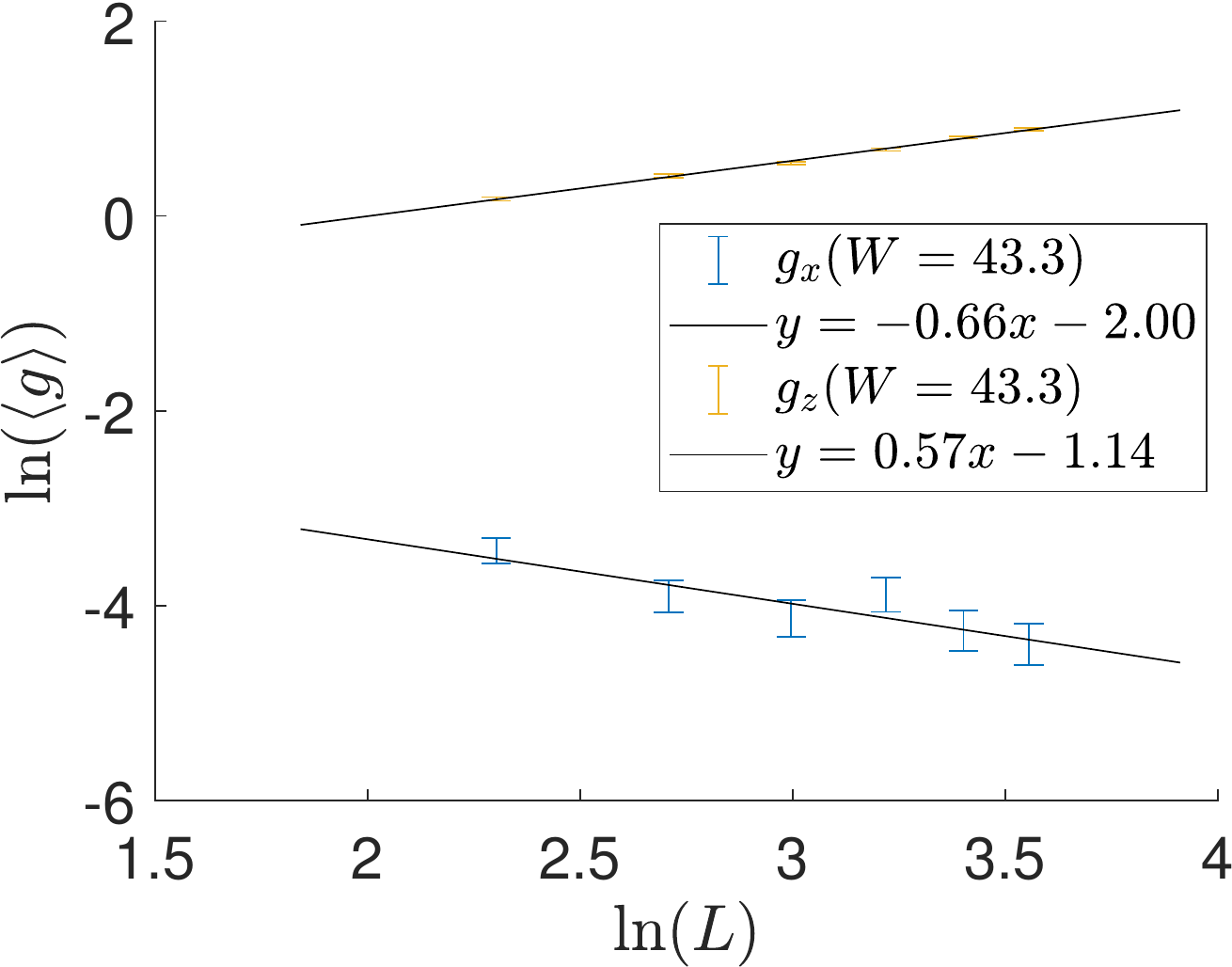}
		\label{ql_gxz}
		\end{minipage}%
	}%
	\subfigure[quasi-localized phase ($W=43.3$) and two transition points ($W=41.3$, $45.3$)]{
		\begin{minipage}[t]{0.3\linewidth}
			\centering
			\includegraphics[width=1\linewidth]{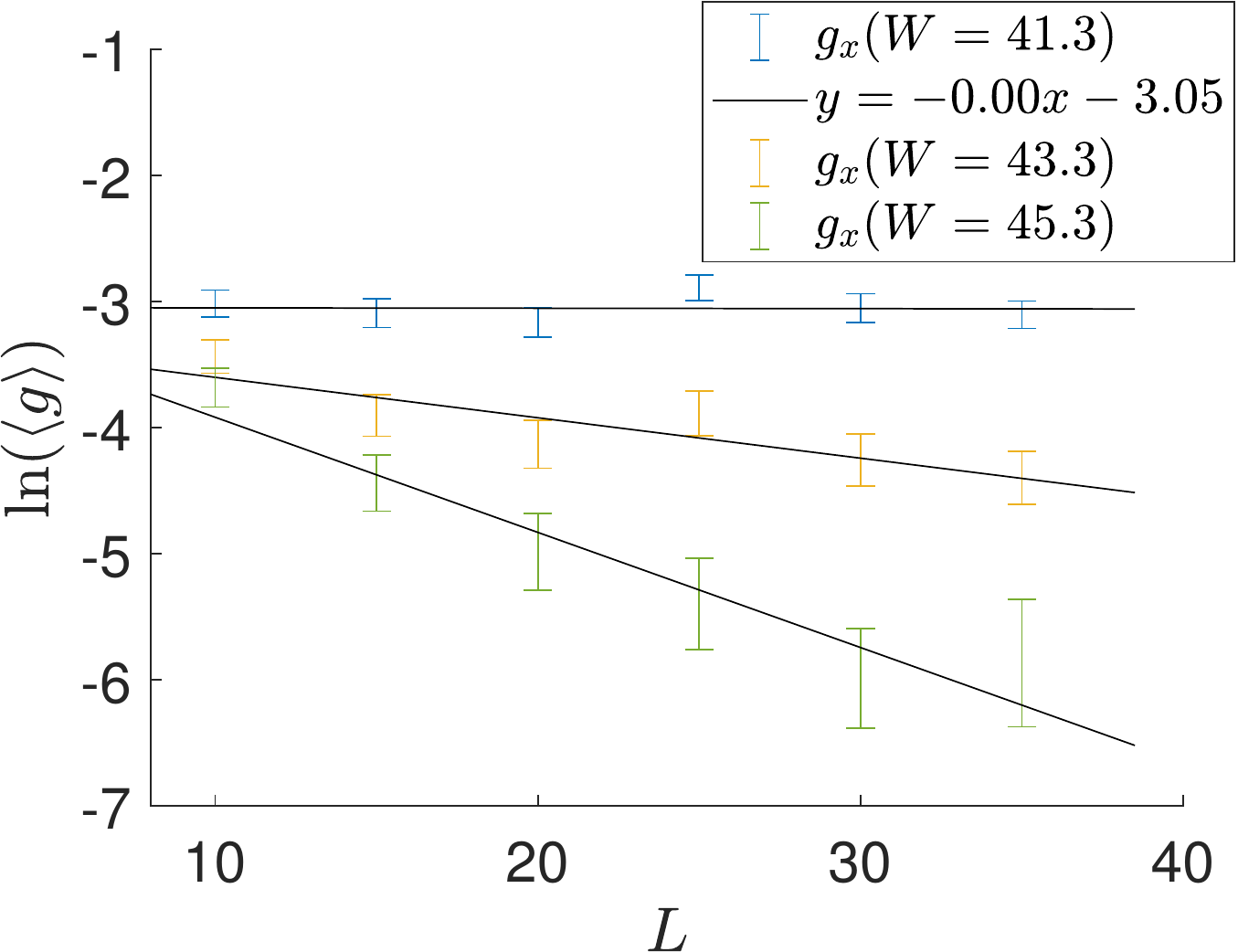}
		\label{ql_x}
		\end{minipage}%
	}%
    
    \subfigure[quasi-localized phase ($W=41.3$, $42.3$, $43.3$, $44.3$)]{
		\begin{minipage}[t]{0.3\linewidth}
			\centering
			\includegraphics[width=1\linewidth]{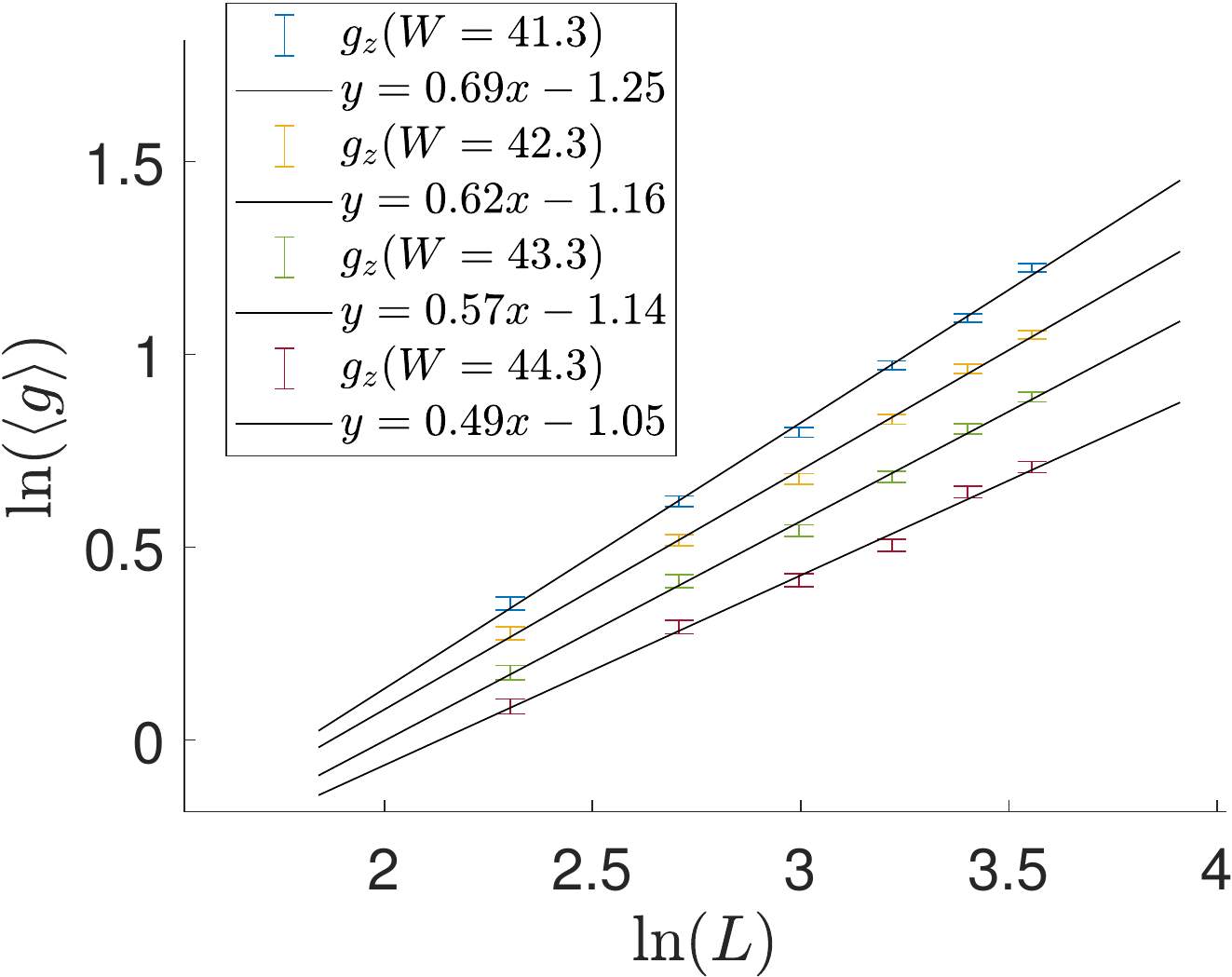}
			\label{ql_z}
		\end{minipage}%
	}%
    \subfigure[Anderson-localized phase ($W=60$)]{
		\begin{minipage}[t]{0.3\linewidth}
			\centering
			\includegraphics[width=1\linewidth]{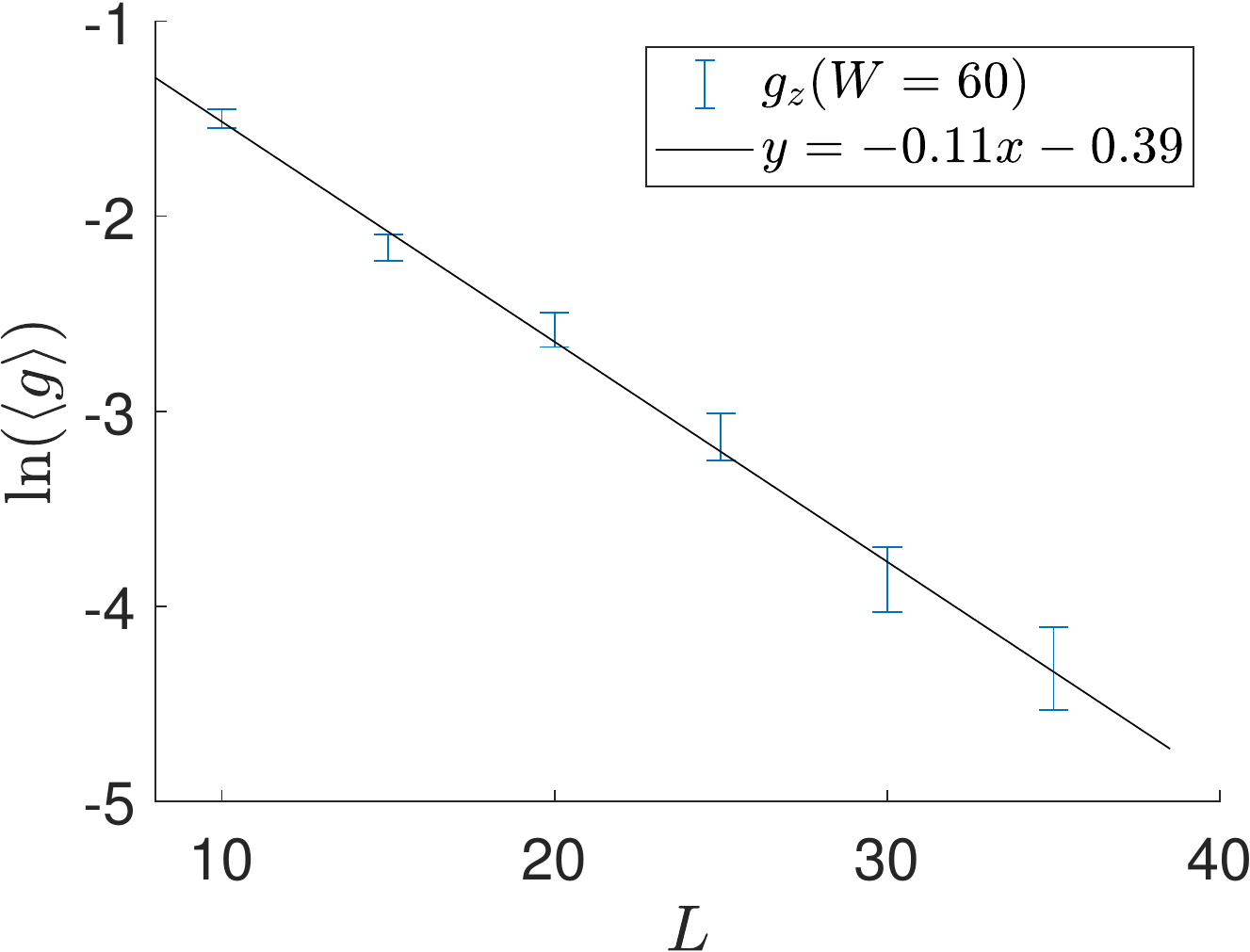}
			\label{AI_z}
		\end{minipage}%
	}%
    \subfigure[distributions of $g_z$ in the quasi-localized phase ($W=43.3$) and around the transition point ($W=41.3$)]{
		\begin{minipage}[t]{0.3\linewidth}
			\centering
			\includegraphics[width=1\linewidth]{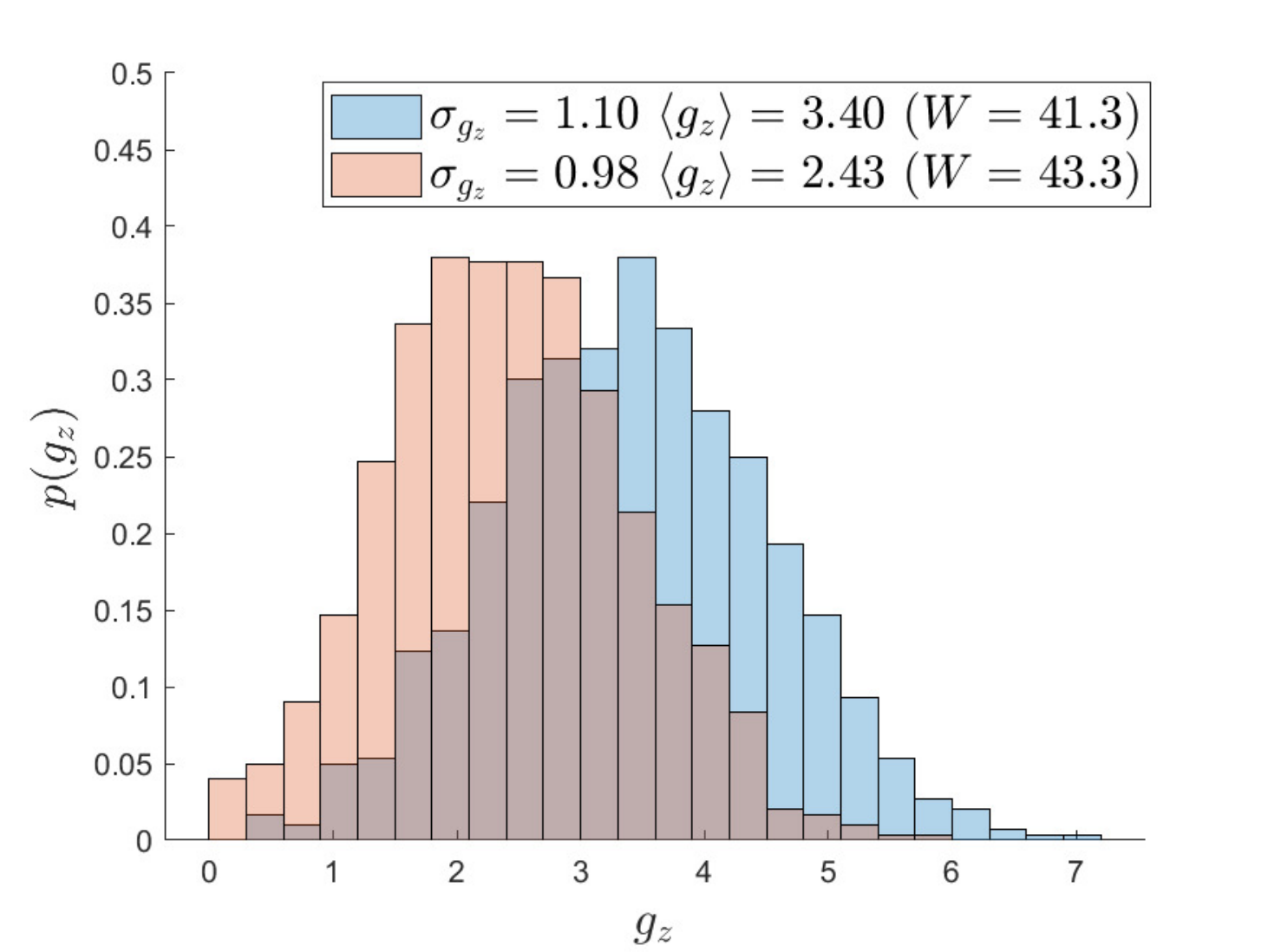}
			\label{hist_gz}
		\end{minipage}%
	}%
	\caption{(a)-(e)~System-size dependence of the mean conductance $\mg{\mu}$ $(\mu = x,z)$ of the nodal-line Hamiltonian [Eq.~(\ref{nodal-line-h})] with $\nu_{x} = \nu_y=0$ and $\nu_z\ne 0$. 
    The mean conductance 
    for the different cubic system sizes $L^3$ 
    is plotted as a function of $L$ (maximal size $L=35$) 
    (a)~in the metallic phase ($W=20$), (b)~in the quasi-localized phase ($W=43.3$), (c)~in the quasi-localized phase ($W=43.3$) and around the two transition points ($W=41.3 \approx W^{(x)}_{c}$, $W=45.3 \approx W^{(z)}_c$), (d)~in the quasi-localized phase ($41.3<W<45.3$), (e)~in the Anderson-localized phase ($W=60$). 
    We calculate the conductance of $N=1000$ samples for each cubic system size and disorder strength $W$. 
    The error bars of $\mg{\mu}$ are estimated as $\sqrt{( \langle g_{\mu}^2  \rangle -  \mg{\mu}^2 )/N}$.
    (f)~Distributions of $g_z$ in the quasi-localized phase ($W=43.3$) and around the transition point ($W=41.3 \approx W^{(x)}_{c}$). 
    The distributions are evaluated for the maximal system size $L = 35$.  
    $\sigma_{g_z} =\sqrt{ \langle g_{z}^2  \rangle -  \mg{z}^2 }$ is the standard deviation of $g_z$.}
\end{figure}

We calculate the two-terminal conductance $g$ of the nodal-line Hamiltonian in Eq.~(\ref{nodal-line-h}) with $\nu_x=\nu_y=0$ and $\nu_z \ne 0$ for the different disorder strength $W$. 
Here, 
$g$ is obtained from the transmission matrix $t$ by the transfer matrix method~\cite{pendry90, kramer05}, $g = \Tr(t^{\dagger} t)$. 
To compute the transmission matrix $t$ along the $\mu$ direction ($\mu=x,z$), we take the nodal-line Hamiltonian on a cubic lattice of size $L^3$
and couple it with two leads at its two ends. 
Each lead is composed of decoupled 1D metal wires along the $\mu$ direction,
\begin{align} 
    \h_{\rm lead} =  - t_{\rm lead} \sum_{{\bm r} = (r_x,r_y,r_z)}( c_{\bm r+ \bm{e_{\mu}}}^{\dagger} \sigma_z c_{\bm r} + {\rm H.c.}), \label{lead}
  \end{align}
which respects the same time-reversal and chiral symmetries as the nodal-line Hamiltonian.  
We choose the parameters of the nodal-line Hamiltonian in Eq.~(\ref{nodal-line-h})
as $\Delta = 0$, $t_{\|} = \sinh 1, t_{\|}^{\prime} = \cosh 1$, $ t_{\perp} = 1$. 
The phase diagram of the Hamiltonian was obtained by the localization lengths along the $x$ and $z$ directions [see Fig.~\ref{lambda_x_g_1}]. 
As the disorder strength $W$ increases, the zero-energy states undergo the phase transitions: 
\begin{equation} 
    \begin{cases}
     W < W_c^{(x)} \approx 41.3 &  \text{(metallic phase);}  \\
      W_c^{(x)} <W < W_c^{(z)} \approx 45.3 &  \text{(quasi-localized phase);}  \\
  W_c^{(z)} <W &     \text{(localized phase).}  \\
    \end{cases}
  \end{equation}
In the metallic phase ($W<41.3$), both $g_x$ and $g_z$ show Ohm's law, $\mg{\mu} \propto L$ for $\mu=x,z$ [Fig.~\ref{metal_gxz}]. 
Around the transition point between the metallic and quasi-localized phases ($W = 41.3$), $\mg{x}$ becomes scale-invariant [Fig.~\ref{ql_x}], which is consistent with the scale-invariant behavior of the localization length along the $x$ direction. 
In the quasi-localized phase ($41.3<W<45.3$), $\mg{x}$ decays exponentially with the system size $L$ [see Fig.~\ref{ql_x}]. 
In contrast, $\langle g_z \rangle$ grows with $L$ in the power law, $\langle g_z \rangle \propto L^\alpha$, characterized by a non-universal exponent $\alpha$ 
($0<\alpha<1$) [Figs.~\ref{ql_gxz} and \ref{ql_z}]. 
Notably, $g_z$ exhibits large sample fluctuations, and its standard deviation is comparable to the mean value [Fig.~\ref{hist_gz}]. 
In the Anderson-localized phase ($45.3<W$), both $g_x$ and $g_z$ decay exponentially with $L$ [Fig.~\ref{AI_z}]. 



          \end{widetext}
 

\end{document}